\documentclass[reprint,nofootinbib,showkeys,showpacs]{revtex4-1} 

\usepackage[utf8]{inputenc}
\usepackage[greek,british]{babel}
\languageattribute{greek}{ancient}
\usepackage[LGR, T1]{fontenc}
\usepackage[babel=true]{csquotes}
\usepackage{lmodern}
\usepackage{graphicx,pstricks-add} 
\usepackage{listings}
\usepackage[colorlinks=true,linkcolor=blue,citecolor=blue]{hyperref}
\usepackage{amsmath,amssymb,mathrsfs,xspace,enumitem,mathtools,cases}  
\usepackage{amsfonts}

\DeclareMathOperator{\expo}{e}
\DeclareMathOperator{\imag}{i}
\DeclareMathOperator{\diag}{diag}
\DeclareMathOperator{\tr}{tr}
\DeclareMathOperator{\id}{id}
\DeclareMathOperator{\spn}{span}
\DeclareMathOperator{\rank}{rank}
\DeclareMathOperator{\mat}{\mathsf M}
\DeclareMathOperator*{\argmin}{arg\,min}
\DeclareMathOperator*{\argmax}{arg\,max}
\newcommand{\rep}[1]{\mathsf #1}

\newcommand*\diff{\mathop{}\!\mathrm{d}}
\newcommand{\module}[1]{\ensuremath{\lvert #1 \rvert}}
\newcommand{\ket}[1]{\ensuremath{\lvert #1\rangle}\xspace}
\newcommand{\bra}[1]{\ensuremath{\langle #1\rvert}\xspace}
\newcommand{\bk}[2]{\ensuremath{\langle #1| #2\rangle}\xspace}
\newcommand{\kb}[2]{\ensuremath{\lvert #1\rangle\langle #2\rvert}\xspace}

\makeatletter
\newsavebox\myboxA
\newsavebox\myboxB
\newlength\mylenA

\newcommand*\xoverline[2][0.75]{%
    \sbox{\myboxA}{$\m@th#2$}%
    \setbox\myboxB\null
    \ht\myboxB=\ht\myboxA%
    \dp\myboxB=\dp\myboxA%
    \wd\myboxB=#1\wd\myboxA
    \sbox\myboxB{$\m@th\overline{\copy\myboxB}$}
    \setlength\mylenA{\the\wd\myboxA}
    \addtolength\mylenA{-\the\wd\myboxB}%
    \ifdim\wd\myboxB<\wd\myboxA%
       \rlap{\hskip 0.5\mylenA\usebox\myboxB}{\usebox\myboxA}%
    \else
        \hskip -0.5\mylenA\rlap{\usebox\myboxA}{\hskip 0.5\mylenA\usebox\myboxB}%
    \fi}
\makeatother

\newcounter{theorem}
\newcounter{prop}
\newcounter{cor}
\newcounter{lem}

\begin{document}

\title{\vspace{1em}On the parallel transport in quantum mechanics\\with an application to three-state systems}

\author{Rapha\"el Leone}

\email{raphael.leone@univ-lorraine.fr} 

\affiliation{Laboratoire de Physique et Chimie Th\'eoriques (UMR CNRS 7019)}

\affiliation{Laboratoire d'Histoire des Sciences et de Philosophie -- Archives Henri Poincar\'e (UMR CNRS 7117), F-54501 Nancy Cedex, Universit\'e de Lorraine, France}

\date{\today}

\begin{abstract}
The aim of this article is to give a rigorous although simple treatment of the geometric notions around parallel transport in quantum mechanics. I start by defining the teleparallelism (or generalized Pancharatnam connection) between $n$-dimensional vector subspaces (or $n$-planes) of the whole Hilbert space. It forms the basis of the concepts of parallel transport and of both cyclic and non-cyclic holonomies in the Grassmann manifold of $n$-planes. They are introduced in the discrete case (broken lines) before being rendered \enquote{continuous} (smooth curves) and the role of the geodesics is stressed. Then, I discuss the interest of such a construction to geometrize a part of the dynamics when a (quasi-)dynamical invariant is known, especially in the adiabatic limit. Finally, I illustrate the general theory with a three-state toy model allowing for non-Abelian adiabatic transports.

\end{abstract}



\begin{small}
\noindent\textsl{Abracadabra, Manel Tekel Phares, Pape Satan Pape Satan Aleppe, le vierge le vivace et le bel aujourd'hui, ogni vol­ta che un poeta, un predicatore, un capo, un mago hanno emesso-borborigmi insignificanti, l'u­manit\`a spende secoli a decifrare il loro messaggio.}

\hfill Umberto Eco, \textsl{Il Pendolo di Foucault}
\end{small}

\maketitle 

\section{Introduction}

It has been known for a long time \cite{BornFock} that, under the non-crossing condition, a quantum state initially in an eigenlevel of a time-dependent Hamiltonian $H(t)$ keeps clinging to that level in the adiabatic limit whereby $H(t)$ varies infinitely slowly. This is a natural consequence of the quantum adiabatic theorem \cite{BornFock,Kato,Messiah,Teufel}. However, infinite slowness remains an unattainable horizon and the original true purpose of an adiabatic theory is essentially to quantify the discrepancy between this ideality and the actual solution of the Schr\"odinger equation \cite{BerryHistories}. Concretely, it consists in estimating the transition probabilities between the level which is supposed to be occupied by the state in the adiabatic regime and the others. In general, it is far from being a simple task but some helpful formulas exist under certain conditions often realized --- or almost realized --- in practice; one may for example cite the archetypal Landau-Zener formula \cite{Landau,Zener,Wittig,JoyeLZ} which is widely used, especially in molecular quantum chemistry within the Born-Oppenheimer approximation \cite{BO,Malhado}. 

Overall phase factors have no incidence on transition probabilities. This is certainly one of the reasons why the question of the \enquote{phase adiabatically accumulated} after a cyclic evolution of the Hamiltonian, by a wavefunction in a non-degenerate energy level, was ignored for half a century. Another reason is --- rather paradoxically --- the fact that the consequences of the \enquote{insignificance} of the quantum phase regarding the physical state were not thoroughly investigated. However, at the turn of the 1930s, all was in place for a proper treatment of the phase issue: the 1927 fifth Solvay conference \cite{Bacciagaluppi} had provided a solid foundation for the \enquote{new quantum theory},\footnote{By opposition to the \enquote{old quantum theory} of Bohr, Sommerfeld, and Kramers, that was beautifully synthesized by Born in his masterpiece \textit{The Mechanics of the Atom} (\textit{Vorlesungen \"uber Atommechanik}) \cite{Atommechanik}.} the geometric notion of parallel transport initiated by Levi-Civita \cite{LeviCivita1917} had been generalized by Weyl and Cartan who pioneered the key-concepts of connection and holonomy \cite{Scholz},\footnote{The works of Levi-Civita, Weyl, and Cartan (amongst other geometers), at the turn of the 1920s, were to a large extent motivated by the question of the nature of the \enquote{world geometry} initiated by Einstein.} Weyl, again, had introduced the idea of gauge\footnote{One can find the translation in english of the seminal papers of Weyl on this idea in reference \cite{ORaifeartaigh}. In particular, he shaped in his famous 1929 article a gauge invariance implying the phase of a two-component spinor.} and coined the term \enquote{ray space},\footnote{Initially: \enquote{ray field} (\textit{Strahlenk\"orper}) \cite{Weyl1927}. The term \enquote{ray space} stuck in the quantum physics community whereas mathematicians prefer call that space as it is: a projective space.} and the question of the \enquote{non-integrability of phase} had been considered in a profound work of Dirac \cite{Dirac}.\footnote{Thouless estimated that this paper \enquote{shows Dirac thinking in ways that were unfamiliar to the general community of physicists, and using arguments which would not become part of the general equipment of theorists for many years to come} \cite{Thouless}. It should be stressed that Dirac was influenced by Weyl's 1929 article.} Even if some attention had been occasionally paid to the physical meaning of a non-integrability of phase in certain precise contexts --- one may mention the paradigmatic Aharonov-Bohm effect \cite{EhrenbergSiday,AharonovBohm} or, in quantum chemistry, the phase accumulated by the electronic wavefunction due to a circuit in the nuclear coordinates space within the Born-Oppenheimer approximation \cite{LonguetHiggins,Stone,MeadTruhlar} --- it was not until the 1980s that the phase issue became considered per se in the seminal work of Berry \cite{Berry1984}. He discovered that, if the time-dependence of the Hamiltonian is entirely realized through slow parameters $x=(x_i)$, and if the parameter-point $x$ traces out an oriented loop $\mathcal C$ in the parameter space $\mathscr P$, then, in the adiabatic limit and under the non-crossing condition, a wavefunction initially in a non-degenerate eigenlevel (the $k$-th, say) acquires after the journey along $\mathcal C$ an extra phase (besides the usual dynamical one) which does not depend on the initial arbitrary choice of eigenstate in the $k$-th level but only on $\mathcal C$. This phase $\gamma_k[\mathcal C]$ is thus a purely geometric property associated with the oriented contour $\mathcal C$ in the parameter space and the considered non-degenerate eigenlevel. It is in particular non-dynamical by its independence on the parametrization of $\mathcal C$, or equivalently on how $\mathcal C$ is traced out in time (provided it is very slowly), in contrary to the well-named dynamical phase. It depends only on the chronological order, i.e.\ the orientation of $\mathcal C$ (it naturally changes sign if the orientation is reversed). The discovery of the geometrical \enquote{extra} phase allowed Berry to subsume the above cited phase effects under the same thought pattern.

Simon immediately saw in Berry's geometric phase factor the holonomy of a natural Hermitian connection in a so-called spectral line bundle over $\mathscr P$ \cite{Simon1983}. Without going into too much detail, let me clarify the last sentence in an informal manner. The bundle in question is formed by the collection of $x$-dependent eigenspaces $V_k(x)$ associated with the $k$-th level: at each point $x$ in $\mathscr P$ (the base space) is anchored the complex line $V_k(x)$ (the fibre over $x$ which is an element of the ray space). As for the connection, it is, roughly speaking and as the name suggests, a structure which connects or glues together the \enquote{clump} of fibres over infinitesimal regions of $\mathscr P$.\footnote{In a domain of $\mathscr P$ coordinatized by $(x_i)$, two points $x=(x_i)$ and $x'=(x'_i)$ are infinitely close if their difference is an infinitesimal quantity, a property which is independent of the chosen coordinates. (Indeed, $x'-x$ is in this case an infinitesimal contravariant vector or, more properly, the representation in coordinates of an infinitesimal (absolute) contravariant vector.) In the ray space, the closeness of two rays $V_1$ and $V_2$ can for example be measured by the distance $\inf\|\ket{v_2}-\ket{v_1}\|$ taken over the unit vectors $\ket{v_1}\in V_1$ and $\ket{v_2}\in V_2$, where $\|\cdot\|$ is the Hermitian norm induced by the inner product in the Hilbert space. The \enquote{clump} of fibres over an infinitesimal region of $\mathscr P$ are themselves infinitely close in the ray space under the tacit assumption whereby the Hamiltonian depends smoothly on the parameters (an infinitesimal alteration of $H$ produces an infinitesimal modification of the $n$-th eigenspace).} Actually, it defines a one-to-one linear correspondence between pairs of fibres over infinitely close points which is called an equipollence \enquote{in the infinitesimal}: while the parameter point passes from $x$ to $x'=x+\diff x$, the connection makes any state $\ket{v}$ in the fibre $V_k(x)$ \enquote{transits} to the state in $V_k(x')$ which is said equipollent to $\ket{v}$. In the present context, the epithet \enquote{natural} attributed to the connection puts the emphasis on the fact that it does not call upon any extra-structure than the Hermitian one which pre-exists in the Hilbert space $\mathcal H$ (this adjective will be frequently used throughout the paper). Indeed, the transition is the orthogonal projection $V_k(x)\to V_k(x')$ induced by the inner product in $\mathcal H$. Alternatively stated, the transmission by equipollence from $V_k(x)$ to $V_k(x')$ is the mechanism which takes as input a vector $\ket{v}\in V_k(x)$ and returns as output the vector $\ket{v'}\in V_k(x')$ characterized by the property of being the closest element of $V_k(x')$ to $\ket{v}$ with respect to the ambient Hermitian norm or by the equivalent property of being the element of $V_k(x')$ which overlaps maximally with $\ket{v}$. This transmission is norm-preserving by the infinite closeness of these two spectral lines (see the body of this paper). 
One thereby obtains, over a path $\mathcal C$ in $\mathscr P$, a unitary evolution of the adiabatic wavefunction in the fibres (putting aside the dynamical phase) which is called a parallel transport. When $\mathcal C$ is closed, i.e.\ when the parameter point returns to its initial position, the final wavefunction can be compared with the initial one inside their common fibre: the former differs from the latter by a phase factor which does not depend on the initial phase choice of the wavefunction since the transport is unitary and therefore preserves the phase differences. This phase depends only on the loop $\mathcal C$ and, obviously, on the considered non-degenerate level. It is an example of holonomy in the language of mathematicians (see below) which is here the Berry phase factor associated with the $n$-th energy level over the contour $\mathcal C$ in the parameter space. 

The above construction is on many aspects similar to the parallel transport (\textit{trasporto per parallelismo}) elaborated by Levi-Civita \cite{LeviCivita1917} in the early twentieth century when regarding specifically manifolds embedded in Euclidean spaces. Here, the fibres are the tangent spaces which are connected, over infinitesimal regions of the manifold, through the orthogonal projections induced by the inner product in the ambient space. It is worth noting that Levi Civita's connection plays already a central role in adiabatic phenomena arising in classical mechanics. It suffices to think of the celebrated Foucault pendulum whose natural frequency is four orders of magnitude greater than the sidereal rotation rate of the Earth. The standard dynamical analysis of this problem, within the usual hypothesis (long massless rod, small angle, frictionless pivot, bob sufficitently massive to neglect air resistance, uniform rotation of the Earth considered as spherical), can be found by browsing through innumerable treatises or textbooks in classical mechanics (e.g.\ in Ref.\ \onlinecite{Sommerfeld} to cite but one example). It consists in recognizing that, beyond the \enquote{real} forces exerted on the bob, the Coriolis force due to Earth's sidereal rotation is by far the dominating \enquote{fictitious} force which manifests itself in the laboratory frame of reference (e.g. the \textit{Panth\'eon}, Paris), albeit sufficiently weak to be treated as a small perturbation of the problem that would arise if the laboratory were an inertial frame. If the pendulum is carefully launched \cite{Foucault}, the sensible effect of that correction is to make the \enquote{oscillation plane} precess around the vertical as seen from the laboratory, with a frequency equal to Earth's angular velocity multiplied by the sinus of the latitude where the experiment takes place. (Actually, within the above hypothesis, the trajectory of the bob is more precisely an horizontal rotating ellipse which is so flat\footnote{The ratio between the semi major axis and the semi minor one is equal to the ratio between the precession period $T_{\text p}$ and the natural period of the pendulum $T_0$. In some old references certainly influenced by Appell's treatise \cite{Appell2}, this result is known as the theorem of Chevilliet. Actually, Chevilliet published only a small number of works and none of them concerns this point. It seems that this theorem was communicated by him to R\'esal while he was proofreading the \textit{Trait\'e de M\'ecanique G\'en\'erale} \cite{Resal}. R\'esal included this \enquote{curious theorem} as a note in his treatise (p. 112). In the case of the pendulum suspended in 1851 from the dome of the Panth\'eon, Paris, one has $T_{\text p}\simeq 32\,\text{h}$ and $T_0\simeq 16\,\text{s}$ so that the ratio is approximatively $7200$.} that its major axis appears as an instantaneous oscillation plane.) 

Geometric interpretations of the precession phenomenon flourished as soon as the experimental demonstration of Foucault. Let me insist on the fact that by \enquote{geometric} must be understood a description which do not call upon amounts of time or time rates, but only on the chronological order. The first interpretation of this kind is due to Foucault himself. In his \textit{compte rendu} at the french \textit{Acad\'emie des Sciences}, he notably said that his \enquote{law of the sinus of the latitude} can be determined by having recourse to either calculus or mechanical and geometrical considerations \cite{Foucault}.\footnote{\textit{Pour d\'eterminer la loi suivant laquelle varie ce mouvement sous les diverses latitudes, il faut recourir soit \`a l'analyse, soit \`a des consid\'erations m\'ecaniques et g\'eom\'etriques que ne comporte pas l'\'etendue restreinte de cette Note [\dots]}} Foucault's insight is confirmed by Binet \cite{Binet1} and developed in a draft of a letter (addressed to an unknown recipient) found in his papers.\footnote{The draft is reproduced in reference \cite{oeuvres}.} It is wholly contained in this excerpt:\footnote{I reproduce Romer's translation \cite{Romer}.}
\begin{quote}
I begin by boldly adopting as a postulate the following. When the vertical, always of course lying in the plane of oscillation, changes its direction in space, the successive positions of the plane of oscillation are determined by the condition that the angles between successive positions shall be minimized. To state this in ordinary language: when the vertical direction changes from its initial position, the plane of oscillation follows, while remaining as  parallel as possible to the initial plane of oscillation.\footnote{\textit{Je commence par poser effront\'ement un postulatum tel que celui-ci : Quand la verticale, toujours comprise dans le plan d'oscillation change de direction dans l'espace, les positions successives du plan d'oscillation, sont déterminées par la condition de faire entre elles des angles minima. Autrement dit et en langue vulgaire : lorsque la verticale sort du plan d'impulsion primitive, le plan d'oscillation la suit en restant aussi parall\`ele que possible.}}
\end{quote}  
That breadcrumb trail allowed him to infer heuristically the law of the sinus\footnote{Recall that Foucault could not arrive at his law by the experience alone since he based all his reflections on the observations made at Paris. A pure experimental proof would have been to install pendula at different latitudes (that was obviously done in many places afterwards to corroborate the law). It gives an idea of how inspired he was.} and was later confirmed rigorously by Bertrand \cite{Bertrand}. Using a unit vector $\mathbf u$ that orients continuously the intersection $\Delta$ between the oscillation plane and the horizontal one, the \textit{postulatum effront\'e} of Foucault can be rephrased as the condition that the distance between two successive orientations $\mathbf u$ shall be minimized, or equivalently that their overlap shall be maximized (in the sense of the ambient scalar product). The property of being displaced while remaining \enquote{as parallel as possible} with respect to the absolute space, under the constraint of confinement to the horizontal plane, is exactly the property of remaining strictly parallel with respect to the immobile geometrical sphere $S$ occupied by Earth's surface, in the sense of Levi-Civita's parallel transport. Several equivalent ways of describing geometrically the phenomenon exist. The most direct one relies on the cone drawn by the tangent to the meridian towards the north while the locus $x$ of the experiment traces out in one sidereal day a circle $\mathcal C$ on $S$. The straight line $\Delta$ is at any point $x$ of $\mathcal C$ tangent to the cone. The cone touching $S$ on $\mathcal C$, the parallel transport along $\mathcal C$ with respect to $S$ can be equivalently thought with respect to the cone which is a developable surface, and the geometric analysis becomes child's play (see e.g. Refs \onlinecite{Tardieu,Young}).\footnote{Levi-Civita, in his \textit{Lezoni di calcolo differenziale absoluto} \cite{LeviCivita1927}, defines intuitively the parallel transport along a path on developable surfaces, then extends it to any surface by using the developable envelope of the family of tangent planes along the considered path, and finally obtains an intrinsic formulation (i.e.\ with nothing but the metric induced on the surface) from which follows readily an extension to any Riemannian manifold. Although he had research interests in the adiabaticity in classical mechanics (but we must say that it chiefly concerned the adiabatic invariants), I did not find in his works an explicit application of his geometric conceptions to the adiabatic theory (but maybe I missed some references). As pointed out by Arnold \cite{Arnold2}, Levi-Civita's 1917 paper inspired to Radon a \textit{Gedankenexperiment} to characterize the geometry of a surface through slow displacements of an oscillating system over it \cite{Klein}.}  

Let me return to the \enquote{quantal phase factors accompanying adiabatic changes}, as Berry named it. As Simon wrote in the introduction of his paper \cite{Simon1983}, the concept of fibre bundle endowed with a connection was becoming familiar to physicists because it was recognized as the basic framework of gauge theories \cite{WuYang}. Using the common terminology of those theories, Berry's phase can be computed in two ways: as the circulation of a $U(1)$ gauge field $\mathcal A$ along $\mathcal C$ or, thanks to Stokes' theorem, as the integral of the gauge-invariant curvature $\diff\mathcal A$ on a surface bounded by $\mathcal C$. The first way suggests that the wavefunction \enquote{records} an information of the followed path in its memory while the second that it feels a region which it has not visited. According to Berry in the conclusion of his seminal article, 
\begin{quote}
[t]he remarkable and rather mysterious result of this paper is that [\dots] the system records its history in a deeply geometrical way, whose natural formulation [in terms of the gauge curvature] involves phase functions hidden in parameter-space regions which the system has not visited. 
\end{quote}
The semantic similarity between Berry's assertion and the Aharonov-Bohm effect is manifest. The gauge freedom is associated with the arbitrariness in choosing a unit eigenstate of reference along $\mathcal C$ for concrete computations. Berry's phase is by construction an invariant of the deployed gauge structure. Returning to Foucault's pendulum, note that the precession angle after one sidereal day can be formally expressed as a Berry phase: it suffices to identify with $\mathbb C$ each tangent plane along the circle $\mathcal C$ traced out on $S$ --- through a (gauge) choice of an orthonormal frame along $\mathcal C$ --- and to convert the Euclidean inner product into a Hermitian one.\footnote{Foucault's pendulum and the adiabatically driven two-state quantum system share the same parameter space, to wit the sphere. In both cases, the angle or phase is connected to the solid angle sustained by the contour $\mathcal C$ \cite{Berry1984,Phase}. It is not surprising, what else but that solid angle would be naturally associated with a contour on the sphere? Indeed, the length is irrelevant since it has a nonzero value along a closed curve consisting in a path followed by its inverse.} A more conceptually-based analogy lies in the fact that the precession angle is expressible as a Hannay angle \cite{Khein1993,Saletan}, the classical homologue of the Berry's phase emerging in cyclically and adiabatically perturbed integrable systems \cite{Hannay,BerryHannay}.

It is a well-known anecdote that Simon's paper appeared before Berry's one. The reason for this lies in the fact that they met up in Australia between the receipt, by the editors, of the first version of Berry's article and the publication of the final one \cite{Berrycomment} which was in particular augmented by some references to Simon's insight (one referee confessed to have lost the first manuscript \cite{Berrycomment}). Meanwhile, Simon wrote and sent his paper in which he coined the name \enquote{Berry's phase} in the title. It is possibly Simon who brought to Berry's attention the sense given by the geometers to the word \enquote{holonomy}. Seemingly, Berry will never use this term again, considering it as a barbarism (in his own words) \cite{Berryanticipations} for it reverses the sense of the initial adjective \enquote{holonomic} coined by Hertz in 1894 \cite{Hertz}.\footnote{In the paragraph 123 of his \textit{Prinzipien der Mechanik in Neuem Zusammenhange Dargestellt} \cite{Hertz}, Hertz gave the following definition (this is the translation from the English edition):
\begin{quote}
A material system between whose possible positions all conceivable continuous motions are also possible motions is called a holonomous system. The term means that such a system obeys integral (\textgreek{ὅλος}) laws (\textgreek{νομός}) whereas material systems in general obey only differential conditions. 
\end{quote}
The Greek root \textgreek{ὅλος} may also be translated as \enquote{whole} or \enquote{complete}. The latter draws a parallel with the term \enquote{complete differential} (\textit{diff\'erentielle complette}, in the written form of Classical French) introduced by Clairaut in 1740 \cite{Clairaut} to designate those first order differentials which admit some functions as integrals. Clairaut also used as a synonym, in the same article, the term \enquote{exact differential} (\textit{diff\'erentielle exacte}), a term which stuck.}
 Indeed, in the tradition of classical mechanics \cite{Appell,Whittaker},\footnote{In the introduction of the second edition of his celebrated \textit{Trait{\'e} de M{\'e}canique Rationnelle} \cite{Appell}, Appell wrote (free translation)
\begin{quote} 
[W]e have introduced, according to the physicist Hertz, the important distinction between two classes of systems: the \emph{holonomic} systems,
for which all the constraints can be expressed by \emph{finite} relations between the coordinates, and the \emph{non-holonomic} ones,
such as the hoop or the bicycle, for which
some constraints are expressed \emph{through
non-integrable differential relations}. 
\end{quote}
} the adjective \enquote{holonomic} refers to an integrable constraint over the configuration space whereas, since Cartan,\footnote{In the first part of a long memoir \cite{Cartan1923}, Cartan borrows to mechanics the adjective \enquote{holonomic} (\textit{holonome}) and introduces a measure of the degree of \enquote{non-holonomy} of a space (to be understood in the modern language as a fibre bundle equipped with a connection) through the displacements in this space caused by infinitesimal contours in the base manifold (\textit{Ce[s] d{\'e}placement[s] mesure[nt] en quelque sorte la non-holonomie du mouvement de l'espace}).} a \enquote{holonomy} is \textit{a contrario} a measure of the degree of non-integrability. Berry (and others) makes the choice of remaining faithful to the origins and thus to the etymology.\footnote{I remember a discussion I had more than one decade ago with a mathematical physicist specialized in geometrical and topological methods in quantum and semi-classical physics. When I asked him some explanations about Berry's \enquote{anholonomies}, he exclaimed that this term is a total nonsense\dots{} a barbarism of a barbarism, so to speak.}  

Shortly after the publications of Berry and Simon, Wilczek and Zee \cite{WZ} generalized the adiabatic phase to the case of degenerate eigenspaces. It led naturally to a non-Abelian $U(n)$ gauge formulation \textit{\`a la} Yang-Mills, where $n$ is the degree of degeneracy. As they wrote in the title of their paper, it is an appearance of a gauge structure in ordinary quantum mechanical problems although they have \textit{a priori} nothing to do with gauge theories. Actually, as soon as a system acquires a description in terms of a fibre bundle, its dynamics is likely to be encoded (at least partly) in a suitable connection. The part of the dynamics which is thereby encoded in the connection gains a geometric interpretation with respect to the bundle structure whereas the dynamics boils down to its remaining part, if any. 

Berry's parallel transport is a continuous projection from fibre to fibre over the parameter space. Since the fibres are rays, it is also a continuous projection from ray to ray over the ray space. This picture, over the ray space, is by far more general and gets rid of the adiabatic assumption. It was formalized in 1987 by Aharonov and Anandan \cite{AA} and the phases induced by cyclic evolutions in the ray space were subsequently called Aharonov-Anandan phases (incidentally, that formalism allowed to give a proper geometric description of the Aharonov-Bohm effect which is not fundamentally an adiabatic phenomenon). The relevance of the parameter space in Berry's settings stems, in fact, from the knowledge a priori of a submanifold of the ray space, coordinatized by $x$, in which the state ray is constrained to live (but nothing prevents us from coordinatizing the ray space itself, e.g.\ through two angles on the Bloch-Poincar\'e sphere for two-state systems). The Aharonov-Anandan phase received shortly afterwards a non-Abelian generalization from Anandan \cite{Anandan1988}: the base space is in this context the Grassmann manifol $\text{Gr}(n,\mathcal H)$, i.e.\ the collection of $n$-dimensional vector subspaces (or $n$-planes) of $\mathcal H$. With respect to the general framework implemented by Aharonov and Anandan, the holonomies of Berry, Wilczek and Zee became particular examples which manifest themselves in the adiabatic limit. 

Let me go back, once more, to the comparison between Foucault's precession angle and the quantum geometric phase. Beyond their above-mentioned analogies, there is a strong difference between them. Indeed, in the former phenomenon, there exists another natural way to transfer vectors on the sphere. This is due to the local identification between the tangent plane and the sphere which allows to determine the precession angle at each instant. It suffices to draw initially (small) vectors on the floor and to let Earth's rotation dragging them along the circle of latitude. By construction, this transfer is unitary and circles back to itself. Hence, an initial orthonormal basis $(\mathbf e_1,\mathbf e_2)$ drawn on the floor (e.g.\ towards the east and the north) generates along the circle a preferred reference frame field with respect to which the orientation $\mathbf u$ of the oscillation plane is compared, instant by instant. Obviously, there is no such identification between the ray space and the fibres in the quantum adiabatic context. However, the following question arises: is there a natural way of defining preferred reference frame fields, along a path in the ray space, in Aharonov-Anandan's settings? The answer is mostly yes. It came from the Raman Research Institute (RRI) of Bangalore, India. Samuel and Bhandari \cite{SamuelBhandari1988} exploited an idea that Pancharatnam (a former physicst of the RRI) developed in a paper of 1956 about the classical polarization states of light beams \cite{Pancharatnam}. They identified vectors belonging to any two rays $V$ and $W$ (provided that these rays are not orthogonal) by connecting the pairs of vectors $\ket v\in V$ and $\ket w\in W$ which have the same norm ($\|v\|=\|w\|$) and are in phase ($\bk{v}{w}>0$, their overlap is maximal). Such a (reflexive and symmetric) relation between non-orthogonal rays can be seen as a distant equipollence (or teleparallelism). Clearly, it coincides with the equipollence in the infinitesimal previously discussed when the two rays are infinitely close. Then, with respect to a natural line element, they characterized a geodesic $\mathcal C$ in the ray space, from $V$ to $W$, such that the parallel transport along $\mathcal C$ realizes exactly the teleparallel transfer from $V$ to $W$: an initial vector of $V$ ends at its teleparallel equivalent in $W$ after the parallel transport along $\mathcal C$. Consequently, a two-step teleparallel transfer $V\to W\to X$ amounts to a parallel transport along the geodesic triangle $VWX$ and, if the triangle encloses a nonzero area, the geometric phase accumulated along it is in general nonzero. This is a geometric proof \textit{\`a la} Pancharatnam of the non-transitivity of the equipollence: the two-step teleparallel transfer $V\to W\to X$ is in general distinct from the one-step one $V\to X$.\footnote{In Pancharatnam's paper \cite{Pancharatnam}, two light beams are in-phase if they interfer maximally. The geodesic triangles are traced out on the Poincar\'e sphere built on the polarization states of light. The phase difference between two routes $\text P\to\text Q\to \text R$ and $\text P\to\text R$ of \enquote{in-phase transfers} is measured by the solid angle sustained by the geodesic triangle PQR.} Following Samuel and Bhandari, all is in place to define the geometric phase accumulated by a time-dependent ray $V(t')$ from $t'=0$ to $t'=t$ as the Aharonov-Anandan phase accompanying the loop composed by the path $V(t')$ in the ray space ended by a geodesic from $V(t)$ to $V(0)$ --- provided that $V(0)$ and $V(t)$ are not orthogonal. Choosing a smooth vector field $\ket{v(t')}\in V(t')$, this so-called non-cyclic (or open path) geometric phase is more simply the phase difference between $\ket{v(t)}$ and the vector $\ket{v^\parallel(t)}\in V(t)$ which is the teleparallel equivalent of $\ket{v(0)}\in V(0)$. Alternatively stated, $\ket{v(0)}$ generates a preferred reference vector field $\ket{v^\parallel(t)}\in V(t)$ with respect to which is determined the instantaneous geometric phase accumulated by the parallel transport \cite{Pati1995}. When the ray traces out a loop, the reference field circles back to itself (since the equipollence is reflexive) and the geometric phase coincides with the usual Aharonov-Anandan phase. 

The connection between pairs of non-orthogonal rays through the in-phase relation (i.e.\ the teleparallelism) is known today as the Pancharatnam connection. The work of Samuel and Bhandari, which revealed its role in quantum mechanics, was motivated by two of their colleagues of the RRI, namely Ramaseshan and Nityananda, who were the first to recognize in the Pancharatnam phase an early example of geometric phase \textit{\`a la} Berry \cite{Ramaseshan1986}.\footnote{The references cited in this paragraph and the preceeding one demonstrate how active were the researches on the geometric phase in India. See also the detailed treatment of Mukunda and Simon \cite{SimonMukunda1993} (Rajiah Simon, not to be confused with Barry Simon who was cited previously).} But we can take the opposite view and see Berry's phase as an example of Pancharatnam's phase for which the teleparallel transfer is made continuously along curves (this is the angle chosen in this article). The Pancharatnam connection was extended to the non-Abelian case by Anandan and Pines \cite{Anandan1989} through the polar decomposition of their mutual projection.\footnote{The possibility of establishing a natural unitary correspondance between two $n$-planes (provided that some conditions are fulfilled) is today commonly known but it is not trivial and some authors were unaware of it in the past. Let me illustrate this point with an example. In a famous paper of 1958 on the perturbation theory, Bloch \cite{Bloch} started from a $g$-fold degenerate eigenenergy $E_0$ of an unperturbed Hamiltonian $H_0$, and designated by $\Omega_0$ the corresponding eigenspace. Then, he \enquote{switched on} a small perturbation. The Hamiltonian became $H=H_0+\lambda V$ ($\lambda\ll1$) and the unperturbed energy $E_0$ split into $g$ energies (not necessarily distinct) $E_0+E_\alpha$ ($\alpha=1,\dots,g$) associated with orthonormal eigenstates $\ket{\alpha}$ such that $H\ket{\alpha}=(E_0+E_\alpha)\ket{\alpha}$. His aim was basically to deduce the $E_\alpha$'s and $\ket{\alpha}$'s from an eigenproblem formulated in $\Omega_0$. To this end, he needed to choose a basis $(\ket{\alpha}_{\text{np}})_{\alpha=1,\dots,g}$ of $\Omega_0$. It amounted to define a \enquote{bridge} between $\Omega=\spn(\ket{\alpha})$ and $\Omega_0$ through $\ket{\alpha}\leftrightarrow\ket{\alpha}_{\text{np}}$. Then, he wrote (free translation)
\begin{quote}
In the usual perturbation theory, one tries to restrict the arbitrariness of the $\ket{\alpha}_{\text{np}}$'s by imposing on them to form an orthonormal system. In fact, this condition produces the opposite effect since it is clear from a geometric viewpoint that it is in general impossible to find in $\Omega_0$ a system of orthonormal wavefunctions which is naturally related to the $\ket{\alpha}$'s. To see it, it suffices to imagine in the three-dimensional ordinary space two non-parallel planes $\Omega$ and $\Omega_0$, and an orthogonal axes system in $\Omega$. None of the orthogonal axes systems of $\Omega_0$ is naturally related to the one of $\Omega$.
\end{quote}
Bloch preferred abandoning the orthogonality condition. Two bases of $\Omega_0$ emerged naturally: the system $\{\ket{\alpha}_0\}$ formed by the (orthognal) projections of the $\ket{\alpha}$'s into $\Omega_0$, and the system $\{\ket{\bar\alpha}_0\}$ formed by the vectors whose projections into $\Omega_0$ were the $\ket{\alpha}$'s (these two systems were well-defined since $\Omega$ and $\Omega_0$ are supposed sufficiently close so that the mutual projections $\Omega\leftrightarrow\Omega_0$ were bijective). The latter led essentially to Kato's perturbation theory \cite{Kato1949} while Bloch's theory was based on the former. It still remains that Bloch's statement is retrospectively erroneous: there exists an orthonormal basis of $\Omega_0$ which is naturally related to the $\ket{\alpha}$'s through the generalized Pancharatnam connection (or teleparallelism). But, to characterize this basis, one must perform a polar decomposition whereas the systems $\{\ket{\alpha}_0\}$ and $\{\ket{\bar\alpha}_0\}$ are more simply defined (note that they are dual in the sense that ${}_0\bk{\bar\alpha}{\beta}_0=\delta_{\alpha\beta}$).}

The concept of geometric phase has to do with the mathematical formalism of quantum mechanics. It is thus observable in a wide variety of systems, has been theoretically studied and experimentally measured in many systems for more than three decades now, and has many applications, especially in quantum information theory and solid state physics (many reviews exist; for books, see references \onlinecite{Phase,Chruscinski,Niuetal,Vanderbilt}). Its fundamental importance justifies its treatment in most of the modern textbooks on quantum mechanics at any level (see e.g.\ references \onlinecite{Griffiths,Sakurai,Shankar,Ballentine,Weinberg}), although the discussion is generally restricted to the ubiquitous Berry phase (a notable exception is B\"ohm's book \cite{Bohm} which contains two chapters devoted to the geometric phase and some of its applications, but B\"ohm was particularly interested in that subject). Furthermore, by its mathematical nature, it admits analogues in other domains provided that they are suitably formalized (here and there in this introduction were for example evoked the classical Hannay's angle and Pancharatnam's phase). To remain in quantum mechanics, it also received interesting generalizations, especially the so-called off-diagonal phases \cite{Manini,Kult2007}, and the geometric phases of mixed states \cite{Uhlmann1989,Uhlmann1991,Filipp} which paved the way for investigations of geometric effects at finite temperature and in open quantum systems. However, we will not deal with these generalizations in this article. We will be concerned with the generally non-Abelian non-cyclic geometric phase factors associated with evolutive $n$-planes. 

The paper is organized as follows. Section \ref{sec:2} is devoted to the parallelism and the induced holonomies regarding $n$-planes. I begin by defining the teleparallelism before detailing its principal characteristics, especially in terms of orthonormal frames. Then, the multistep teletransfers are rendered continuous to derive the parallel transport mechanism and the holonomies, and the emerging gauge description is discussed in few words. The reason to proceed in this order is that it is more natural since the various primary objects are once and for all defined as absolute quantities, whereas the gauge field is only a derived quantity. Otherwise, unless we are comfortable with the representation-free description of gauge theories \cite{Bleecker}, if we start from the gauge picture through representations, then we must verify a posteriori that the considered quantities represent absolute ones to be interpreted. It would lead to unnecessary complications since, in the present context, the existence of an ambient Hilbert space allows for a very simple representation-free description of the geometry. Section \ref{sec:2} continues with formulas to calculate holonomies, the roles of the geodesics are stressed, and the interest of the geometric settings in the presence of a (quasi-)dynamical invariant is discussed, especially in the adiabatic limit. The whole theory is applied in section \ref{sec:3} to a three-state toy model possessing a twofold degenerate level. It is the lowest dimensional framework allowing for such a nontrivial degeneracy. It had already been considered by Mostafazadeh \cite{Mostafazadeh1997,Mostafazadeh1999}, with possible realizations through a spin-1 interacting with a magnetic or electric field. The largest parameter space of the problem is coordinatized by two positive real numbers $r,s$ and two angles $\vartheta,\varphi$. Fixing $r$ and $s$, one obtains a torus over which non-Abelian adiabatic teleparallelism and parallel transports along toroidal helices are studied in some details. Finally, the results are compared with the exact dynamics.

\section{Parallelism and holonomies}\label{sec:2}

\subsection{Teleparallelism}\label{subsec:teleparallelism}

\subsubsection{General settings}\label{generalities}

Let us consider a Hilbert space $\mathcal H$. Thanks to the Hermitian structure, any vector subspace $V$ of $\mathcal H$ can be identified with the (orthogonal) projection operator $P_V$ into $V$. Now, suppose that we focus our attention on the collection of $n$-dimensional vector subspaces (in short: $n$-planes) for a given positive number $n<\dim\mathcal H$. In the mathematical literature, this set is known as the Grassmann manifold $\text{Gr}(n,\mathcal H)$ \cite{Kobayashi-Nomizu2}. When $n=1$, one recovers the ray space. For any $n$-planes $V$ and $W$, there is a natural linear map $\Pi_{WV}\colon V\to W$ --- identifiable\footnote{A linear map $F\colon X\to Y$ between two subspaces of $\mathcal H$ is identified with the operator in $\mathcal H$ which realizes $F$ on $X$ and is null on $X^\perp$.} with the operator $P_WP_V$ in the ambient space $\mathcal H$ --- which projects $V$ into $W$. Since $(P_WP_V)^\dag=P_VP_W$, the maps $\Pi_{VW}$ and $\Pi_{WV}$ are mutually adjoint (with respect to the inner products induced in $V$ and $W$). Consequently, they have the same rank whose value is an intrinsic property of the couple $V$ and $W$. We say that $V$ and $W$ are anti-orthogonal \cite{Anandan1989} if this rank is $n$, i.e.\ if no nonzero vector of $V$ is orthogonal to $W$ ($V\cap W^\perp=\{0\}$) or equivalently if no nonzero vector of $W$ is orthogonal to $V$ ($V^\perp\cap W=\{0\}$). Let me emphasise that the anti-orthogonality is not a transitive relation in the Grassmann manifold: if $V$, $W$, and $X$, are three $n$-planes such that $V$ and $W$ are anti-orthogonal as well as $W$ and $X$, then $V$ and $X$ are not necessarily so.\footnote{A counterexample in the case $n=1$ is easily found by taking two orthogonal nonzero vectors $\ket{v}$ and $\ket{x}$ in $\mathcal H$, and building $V=\spn(\ket{v})$, $W=\spn(\ket{v}+\ket{x})$, and $X=\spn(\ket{x})$.}

Now, let $V$ and $W$ be two $n$-planes and $r\leqslant n$ be the rank of their mutual projections. By the polar decomposition theorem reviewed in appendix \ref{PD}, there exist unitary maps $U,U'\colon V\to W$ and positive semidefinite self-adjoint maps $R\colon V\to V$, $R'\colon W\to W$, such that $\Pi_{WV}=U R=R' U'$, the product $UR$ (resp.\ $R'U'$) being called a right (resp.\ left) polar decomposition of $\Pi_{WV}$. The maps $R$ and $R'$ are unique and will be respectively denoted by $R_{WV}$ and $R'_{WV}$. Furthermore, the set of unitary maps entering the right polar decompositions coincide with the set of unitary maps entering the left polar ones. This common set will be denoted by $\mathcal U_{WV}$. Since $\Pi_{VW}$ is the adjoint of $\Pi_{WV}$, one has $R_{VW}=R'_{WV}$, $R'_{VW}=R_{WV}$ and the set $\mathcal U_{VW}$ is formed by the inverses of the elements of $\mathcal U_{WV}$. Hence, the Hermitian structure of $\mathcal H$ allows to define preferred unitary correspondences between $V$ and $W$.

Technically, $R_{WV}$, $R_{VW}$ and the elements of $\mathcal U_{WV}$ are constructed as follows (here we merely apply the general theory exposed in appendix \ref{PDbasic}). First, one knows from basic linear algebra that the two endomorphisms $A=\Pi_{VW}\Pi_{WV}$ and $B=\Pi_{WV}\Pi_{VW}$ of $V$ and $W$, respectively, have the same eigenvalues with the same multiplicities (see the lemma \ref{lem1}). Then, $\Pi_{WV}$ and $\Pi_{VW}$ being mutually adjoint, $A$ and $B$ are positive semidefinite self-adjoint maps, thus diagonalizable with nonnegative eigenvalues and orthogonal eigenspaces. Now, let $\sigma_1^2\leqslant\sigma_2^2\leqslant\dots\leqslant\sigma_p^2$ be their distinct nonzero eigenvalues, where the $\sigma_i$'s are positive numbers known as the singular values of $\Pi_{WV}$ (and of $\Pi_{VW})$. One has the decompositions into orthogonal direct sums
\begin{align*}
V=V_0\oplus\underbrace{V_1\oplus\dots\oplus V_p}_{V_*}\;\;\text{and}\;\; W=W_0\oplus\underbrace{W_1\oplus\dots\oplus W_p}_{W_*}
\end{align*}
where $V_1,\dots,V_p$ (resp.\ $W_1,\dots,W_p$) are the eigenspaces of $A$ (resp.\ $B$) associated with the eigenvalues $\sigma_1^2,\dots,\sigma_p^2$ while $V_0$ (resp.\ $W_0$) is the kernel of $\Pi_{VW}$ (resp.\ $\Pi_{WV}$). Concretely, $V_0=V\cap W^\perp$ (resp.\ $W_0=V^\perp\cap W$) is the vector subspace of $V$ (resp.\ $W$) which is orthogonal to $W$ (resp.\ $V$). Hereafter, the Latin index $i$ will cover the range $1,\dots,p$ and the Greek index $\mu$ the range $0,\dots,p$. We will set $\sigma_0=0$ for later convenience and $n_\mu$ the dimension shared by $V_\mu$ and $W_\mu$. By the lemma \ref{lem1}, one knows that $\Pi_{WV}$ realizes on each $V_i$ an isomorphism $V_i\to W_i$ which is clearly the projection $\Pi_{W_iV_i}$. Hence, the projection of $V$ into $W$ is \enquote{block-diagonalized} as the sum
\begin{align}
P_WP_V=P_{W_1}P_{V_1}+\dots+P_{W_p}P_{V_p}\,.\label{PP}
\end{align}
Then, according to the theorem \ref{thm3}, $R_{WV}$ and $R_{VW}$ are the respective square roots of $A$ and $B$. They decompose as (see the theorem \ref{thm1})
\begin{align}
R_{WV}=\sum_{i=1}^p\sigma_iP_{V_i}\quad\text{and}\quad R_{VW}=\sum_{i=1}^p\sigma_iP_{W_i}\,.\label{RR}
\end{align}
As for the elements of $\mathcal U_{WV}$, they are the maps having the form $U_0+U_*$ where $U_0$ is an arbitrary unitary map from $V_0$ to $W_0$ while
\begin{align}
U_*=\sum_{i=1}^p\sigma_i^{-1}P_{W_i}P_{V_i}\label{Ustar}
\end{align}
is a fixed unambiguous isomorphism from $V_*$ to $W_*$. Actually, each term $U_i=\sigma_i^{-1}P_{W_i}P_{V_i}$ in the above sum is a unitary isomorphism from $V_i$ to $W_i$ (see the theorem \ref{thm2}).

Suppose that $V$ and $W$ are anti-orthogonal. In this case, $V=V_*$, $W=W_*$, the polar decompositions are unique and $\mathcal U_{WV}$ contains as single element the unitary isomorphism $U_*$ that will be denoted by $\Gamma_{WV}$ (with $\Gamma_{VW}=U_*^{-1}$ as inverse). Hence, we have a natural unitary correspondence between $V$ and $W$ that will be called an \textit{equipollence} or \textit{teleparallelism}: two vectors $\ket{v}\in V$ and $\ket w\in W$ are equipollent if and only if (iff) $\ket w=\Gamma_{WV}\ket v$ or equivalently $\ket v=\Gamma_{VW}\ket w$. The map $\Gamma_{WV}$ will be called the \emph{teletransporter} from $V$ to $W$. By construction, it preserves the inner product, i.e.\ the norm (and the phase differences). 

When $V$ and $W$ are not anti-orthogonal, we have only a natural unitary correspondence between the two anti-orthogonal $r$-subspaces $V_*$ and $W_*$, realized by $U_*$ and its inverse. Actually, it is the teleparallelism between $V_*$ and $W_*$ for which $U_*$ is the teletransporter $\Gamma_{W_*V_*}$. In finer details, each map $U_i$ is the teletransporter $\Gamma_{W_iV_i}$. Completing naturally $U_*$ with the null map on $V_0$, one obtains a map from $V$ to $W$ that can be also denoted by $\Gamma_{WV}$ and that remains expressible as in the right-hand side of \eqref{Ustar}. By construction, $\Gamma_{WV}$ and $\Gamma_{VW}$ are mutually pseudoinverse maps in the Moore-Penrose sense.\footnote{See the definition given in appendix \ref{PDbasic}.} The correspondence they realize between $V$ and $W$ is a \enquote{partial teleparallelism}: $V_*$ and $W_*$ are unitarily related by $U_*$ and its inverse whereas $V_0$ and $W_0$ are disregarded. The map $\Gamma_{WV}$ is thus a \enquote{partial teletransporter} from $V$ to $W$.

Let us derive some basic properties of the above construction, in the general case. Hereafter, the vector of $W_*$ (resp.\ $V_*$) which is equipollent to a vector $\ket{v}$ of $V_*$ (resp.\ $\ket w$ of $W_*$) will be denoted by $\ket{v^\parallel}$ (resp.\ $\ket{w^\parallel}$), and one has clearly $\bk{v^\parallel}{w}=\bk{v}{w^\parallel}$ for any $\ket{v}\in V_*$, $\ket{w}\in W_*$. (i) As it is clear from \eqref{PP}, $V_i$ and $W_{i'}$ are orthogonal if $i\ne i'$. (ii) A direct consequence of (i) is that, on each $V_i$, the teletransport from $V$ to $W$ is merely realized by the projection $V\to W$ rescaled by the multiplicative factor $\sigma_i^{-1}$ (and vice versa by exchanging the letters $V$ and $W$). (iii) Let $\ket{v_i}$ and $\ket{\tilde v_i}$ be two vectors of $V_i$. Using the idempotence of $P_W$, one obtains the relationships 
\begin{align*}
\bk{v_i^{\,\parallel}}{\tilde v_i}=\bk{v_i}{\tilde v_i^{\,\parallel}}=\sigma_i\bk{v_i^{\,\parallel}}{\tilde v_i^{\,\parallel}}=\sigma_i\bk{v_i}{\tilde v_i}.
\end{align*}
In particular, $\ket{v_i}$ is orthogonal to $\ket{\tilde v_i}$ iff $\ket{{v_i}^\parallel}$ is so. Furthermore, taking $\ket{v_i}=\ket{\tilde v_i}\ne 0$, we see that $\ket{v_i}$ and $\ket{{v_i}^\parallel}$ are in phase (i.e.\ $\bk{v_i}{{v_i}^\parallel}>0$). Then, using the norm preservation, we can write
\begin{align*}
\sigma_i=\frac{\bk{v_i}{{v_i}^\parallel}}{\|v_i\|\cdot\|{v_i}^\parallel\|}=\frac{\Re\big[\bk{v_i}{{v_i}^\parallel}\big]}{\|v_i\|\cdot\|{v_i}^\parallel\|}=\frac{\module{\bk{v_i}{{v_i}^\parallel}}}{\|v_i\|\cdot\|{v_i}^\parallel\|}\,.
\end{align*}
Alternatively stated, $\sigma_i$ is simultaneously the cosine of the complex, the Euclidean, and the Hermitian angles between $\ket{v_i}$ and $\ket{{v_i}^\parallel}$, whatever the nonzero vector $\ket{v_i}\in V_i$ be. Let $\phi_i\in[0,\tfrac{\pi}{2}]$ be this single angle. We will also set $\phi_0=\tfrac{\pi}{2}$ so that $\cos\phi_0=0=\sigma_0$. The geometric angles $\phi_i$ (if $n_0=0$) or $\phi_\mu$ (if $n_0\ne 0$) are called the \emph{principal angles} between $V$ and $W$. Actually, each $\phi_{i/\mu}$ is the minimal angle between $V_{i/\mu}$ and $W_{i/\mu}$ in the sense of Dixmier.\footnote{Dixmier \cite{Dixmier} defined the minimal angle between two closed vector subspaces $Y$ and $Z$ of $\mathcal H$ as the angle $\phi\in[0,\tfrac{\pi}{2}]$ whose cosine is the supremum of the cosines of the Hermitian angles between the vectors of $Y$ and the vectors of $Z$. In particular, $\phi=0$ (resp.\ $\tfrac{\pi}{2}$) iff $Y$ and $Z$ intersect non trivially (resp.\ are orthogonal).} The \enquote{in-phase} relation between equipollent vectors is not reserved to the elements of the subspaces $V_{i}$ and $W_{i}$. Indeed, if $\ket{v}$ and $\ket{\tilde v}$ are two vectors of $V_*$ which decompose as $\ket{v_1}+\dots+\ket{v_p}$ and $\ket{\tilde v_1}+\dots+\ket{\tilde v_p}$ with $\ket{v_i},\ket{\tilde v_i}\in V_i$, one has
\begin{align}
\bk{v^\parallel}{\tilde v}=\bk{v}{\tilde v^\parallel}=\sum_{i=1}^p\sigma_i\bk{v_i^\parallel}{\tilde v_i^\parallel}=\sum_{i=1}^p\sigma_i\bk{v_i}{\tilde v_i}.\label{scalaire}
\end{align}
In particular, $\bk{v^\parallel}{v}>0$ for nonzero vectors $\ket{v}\in V_*$.\footnote{Let $\ket{v}=\ket{v_0}+\dots+\ket{v_p}\in V$ and $\ket{w}=\ket{w_0}+\dots+\ket{w_p}\in W$ with $\ket{v_\mu}\in V_\mu$, $\ket{w_\mu}\in W_\mu$. One has
\begin{align*}
\bk{v}{w}=\sum_{i=1}^p\bk{v_i}{w_i}=\sum_{i=1}^p\sigma_i\bk{v_i}{w_i^{\,\parallel}}\leqslant\sigma_p\sum_{i=1}^p\|v_i\|\cdot\|w_i\|.
\end{align*}  
It implies $\bk{v}{w}\leqslant \sigma_p\|v\|\cdot\|w\|$ by the Cauchy-Schwarz inequality. Since $\bk{v_p}{w_p}=\sigma_p\|v_p\|\cdot\|w_p\|$, one deduces that $\phi_p$ is the minimal angle between $V$ and $W$. Then, $\phi_{p-1}$ is the minimal angle between $V_0\oplus\dots\oplus V_{p-1}$ and $W_0\oplus\dots\oplus W_{p-1}$, etc.} (iv) From $\|v_i\|=\|{v_i}^\parallel\|=\sigma_i^{-1}\|P_{W_i}\ket{v_i}\|$, one has
\begin{align*}
\sigma_i^2=\frac{\|P_{W_i}\ket{v_i}\|^2}{\|v_i\|^2}=\left\langle\frac{v_i}{\|v_i\|}\right|P_{W_i}P_{V_i}\left|\frac{v_i}{\|v_i\|}\right\rangle
\end{align*}
for any nonzero vector $\ket{v_i}\in V_i$, and it is clear that $n_i\sigma_i^2=\tr(P_{W_i}P_{V_i})$. (v) It is evident that $\sigma_p$ is equal to 1 iff $V$ and $W$ have a nontrivial intersection, in which case $V_p=W_p=V\cap W$ and the teletransport reduces to the identity on $V\cap W$. It is certainly true when $\mathcal H$ has a finite dimension and $2n>\dim\mathcal H$.\footnote{Since $\dim\mathcal H\geqslant \dim(V+W)=\dim V+\dim W-\dim(V\cap W)$ one has $\dim(V\cap W)\geqslant 2n-\dim\mathcal H>0$.} 

As an example, let us consider the case $n=1$. Let $V$ and $W$ be two non-orthogonal rays spanned by two nonzero vectors $\ket{v}$ and $\ket{w}$, respectively. The unique singular value of $\Pi_{WV}$ is the invariant $\tr(P_WP_V)^{1/2}$ and the teletransporter from $V$ to $W$ is thus
\begin{align*}
\Gamma_{WV}=\frac{P_WP_V}{\sqrt{\tr(P_WP_V)}}=\expo^{\imag\gamma_{w,v}}\left|\frac{w}{\|w\|}\right\rangle\left\langle\frac{v}{\|v\|}\right|,
\end{align*}
where the quantity $\gamma_{w,v}\coloneqq\arg\bk{w}{v}$ is the so-called Pancharatnam phase between $\ket v$ and $\ket w$ (in this order since one has the antisymmetry $\gamma_{v,w}=-\gamma_{w,v}$ which reflects the relation $\Gamma_{VW}=(\Gamma_{WV})^{-1}$ in the case $n=1$). Whatever the reference vector $\ket{w}\in W$ be, the vector of $W$ equipollent to a vector $\ket{v}\in V$ is given by $\ket{v^\parallel}=\|v\|\expo^{\imag\gamma_{w,v}}\ket{w/\|w\|}$. Here, the in-phase relation becomes $\gamma_{v,v^\parallel}=0$. The case $n=2$ can also be treated exhaustively in a straightforward way.

\subsubsection{Teleparallelism and orthonormal bases}\label{telebases}

It is interesting to look upon the teleparallelism between $V$ and $W$ through the lens of their (ordered) orthonormal bases. From now on, the Latin indices $j$ and $k$ will cover the range $1,\dots,n$. In any orthonormal bases $\mathcal V=(\ket{v_j})_j$ and $\mathcal W=(\ket{w_j})_j$ of $V$ and $W$, respectively, the map $\Pi_{WV}$ is represented by the $n\times n$ overlap matrix $\rep S(\mathcal W,\mathcal V)\coloneqq(\bk{w_j}{v_k})_{jk}$ between $\mathcal V$ and $\mathcal W$ (in this order). Now, fix an orthonormal basis $\mathcal V$ of $V$. For any unitary map $U\colon V\to W$, $U(\mathcal V)=(U\ket{v_j})_j$ is an orthonormal basis of $W$, and reciprocally, for any basis $\mathcal W$ of $W$, there exists a unique unitary map $U\colon V\to W$ such that $\mathcal W=U(\mathcal V)$. By the choice of a basis of $V$, the set of orthonormal bases of $W$ is thereby identified with the set of unitary maps $V\to W$.  

Theorems \ref{thm4} and \ref{thm5} in appendix \ref{subsec:matchar} bring the two following characterizations of $\mathcal U_{WV}$: amongst all the unitary maps $U\colon V\to W$, the elements of $\mathcal U_{WV}$ are the ones which (i) render the overlap matrix $\rep S(U(\mathcal V),\mathcal V)$ positive semidefinite and self-adjoint, and (ii) maximize the real part of the trace of $\rep S(U(\mathcal V),\mathcal V)$. As a corollary of (i), $\rep S(U(\mathcal V),\mathcal V)$ is diagonal and nonnegative iff $U\in\mathcal U_{WV}$ and $\mathcal V$ is, up to a reordering of its elements, adapted\footnote{See the definition given in the appendix \ref{PDbasic}.} to the decomposition $V_0\oplus\dots\oplus V_p$ of $V$ (see the corollary \ref{cor1}). Now, as a corollary of (ii), the elements of $\mathcal U_{WV}$ are also the unitary maps $U\colon V\to W$ which minimize the Frobenius distance between $\rep S(U(\mathcal V),\mathcal V)$ and the $n\times n$ identity matrix \cite{Mead1991} (see the corollary \ref{cor2}). In this sense, the bases $U(\mathcal V)$ such that $U\in\mathcal U_{WV}$ are the orthonormal bases of $W$ which maximally overlap with $\mathcal V$. Moreover, since $V$ and $W$ live in the same ambient space $\mathcal H$, the point (ii) furnishes a last characterization of $\mathcal U_{WV}$ as the set of unitary maps $U\colon V\to W$ which minimize the \enquote{root mean-square distance} \cite{Kult2006}
\begin{align*}
d(U(\mathcal V),\mathcal V)=\bigg(\frac1n\sum_{j=1}^n\|U\ket{v_j}-\ket{v_j}\|^2\bigg)^{1/2}
\end{align*}
between $\mathcal V$ and $U(\mathcal V)$.\footnote{\label{Stiefel}This distance is defined on the space of (ordered orthonormal) $n$-frames of $\mathcal H$ known as the Stiefel manifold $\text{St}(n,\mathcal H)$. In this space, there is a natural equivalence relation $\sim$ : two $n$-frames are equivalent if they span the same $n$-plane. There is also a free action of the unitary group $U(n)$ on $\text{St}(n,\mathcal H)$: any element $\rep T\in U(n)$ \enquote{rotates} any $n$-frame $\mathcal V$ to give the equivalent $n$-frame $\mathcal V\rep T$ (see the appendix \ref{subsec:matchar} for this notation). The Grassmann manifold $\text{Gr}(n,\mathcal H)$ is both the quotient space $\text{St}(n,\mathcal H)/\sim$ and the orbit space $\text{St}(n,\mathcal H)/U(n)$. The mean-square distance has the following compatibility with the action of $U(n)$: for any element $\rep T\in U(n)$ and $n$-frames $\mathcal V$, $\mathcal W$, one has $d(\mathcal V\rep T,\mathcal W\rep T)=d(\mathcal V,\mathcal W)$. We can also define another distance between two $n$-frames $\mathcal V$ and $\mathcal W$ as the quantity $\module{\det\rep S(\mathcal V,\mathcal W)}$. The latter is fully compatible with $\sim$ (or the action of $U(n))$: if $\mathcal V\sim\mathcal V'$ and $\mathcal W\sim\mathcal W'$ then $\module{\det\rep S(\mathcal V,\mathcal W)}=\module{\det\rep S(\mathcal V',\mathcal W')}$. It allows to define a distance $\delta$ between $n$-spaces $V$ and $W$ as $\delta(V,W)=\module{\det\rep S(\mathcal V,\mathcal W)}$ for an arbitrary choice of bases $\mathcal V$ and $\mathcal W$ of $V$ and $W$ respectively. Clearly, $\delta(V,W)=(\cos\phi_0)^{n_0}(\cos\phi_1)^{n_1}\dots(\cos\phi_p)^{n_p}$. The geometric angle $\phi\in[0,\tfrac{\pi}{2}]$ such that $\cos\phi=\delta(V,W)$ is known as the Fubini-Study distance between $V$ and $W$. It is equal to $\tfrac{\pi}{2}$ iff $V$ and $W$ are not anti-orthogonal. In the cane $n=1$, the Fubini-Study distance is also known as the statistical (or Wootters \cite{Wootters}) distance.} Indeed, after an expansion of the norms, the squared distance
\begin{align*}
d(U(\mathcal V),\mathcal V)^2=2-\frac{2}{n}\,\Re\big[\tr\rep S(U(\mathcal V),\mathcal V)\big]
\end{align*}
attains its minimum iff $U\in\mathcal U_{WV}$. Finally, it must be said that the various extrema presented in this paragraph do not depend on the choice of $\mathcal V$; they are thus intrinsic to the couple $V$ and $W$. In particular, taken over all the bases $\mathcal V$ of $V$ and $\mathcal W$ of $W$, one has
\begin{align*}
\max_{\mathcal V,\mathcal W}\Re\big[\tr\rep S(\mathcal W,\mathcal V))\big]=\sum_{i=1}^pn_i\cos\phi_i=\tr(P_WP_V),
\end{align*}   
and this maximum is attained iff $\mathcal W$ is the image of $\mathcal V$ by an element of $\mathcal U_{WV}$.

\subsubsection{Discrete transports and (non-)cyclic discrete holonomies}

The teleparallelism is not a transitive relation in the following sense: if $V$, $W$, and $X$, are three $n$-planes, $\Gamma_{XV}$ is in general different from $\Gamma_{XW}\Gamma_{WV}$. In particular, supposing these three spaces mutually anti-orthogonal, if $\ket v\in V$ and $\ket w\in W$ are equipollent as well as $\ket w\in W$ and $\ket x\in X$ then $\ket v\in V$ and $\ket x\in X$ are generally not. For $n=1$, it amounts to say that one has generally $\gamma_{x,v}\ne\gamma_{x,w}+\gamma_{w,v}$.\footnote{Consider for example two orthogonal vectors $\ket v$ and $\ket z$, then define $\ket w=\ket v+\imag\,\ket z$ and $\ket x=\ket v+\ket z$: these three vectors are such that $0=\gamma_{x,v}\ne\gamma_{x,w}+\gamma_{w,v}=\tfrac{\pi}{4}+0$.} The non transitivity says that if we want to transfer by teleparallelism the vectors from $V=V_0$ to $W=V_N$ by passing through $N-1$ auxiliary $n$-planes $V_1,\dots,V_{N-1}$ then the net result is the transporter 
\begin{align*}
\Gamma(V_N,V_{N-1},\dots,V_1, V_0)\coloneqq\Gamma_{V_NV_{N-1}}\dots\Gamma_{V_2V_1}\Gamma_{V_1V_0}
\end{align*}
which crucially depends on the sequence $V_1,\dots,V_{N-1}$. The succession of steps $V\to V_1\to\dots\to V_{N-1}\to W$ can be seen as a discrete path in the Grassmann manifold. In this viewpoint, $\Gamma(V_N,\dots, V_0)$ is a purely geometric quantity associated with the \enquote{broken line} $V_0V_1\dots V_N$ in $\text{Gr}(n,\mathcal H)$. In the following, we will suppose that, step by step, the transport is done between anti-orthogonal $n$-planes $V_a$ and $V_{a+1}$ ($a=0,\dots,N-1$) in order to avoid \enquote{destroying} vectors during the process. Clearly, the transporter is transformed into its inverse if the path is reversed.

One obtains a discrete loop when $V=W$. In this case, the final vectors can be compared with the initial ones inside $V$, and the transporter can be called a cyclic discrete holonomy. To characterize the comparison, it suffices to choose an initial basis $\mathcal V$ of $V$ and to compare with it the final basis $\mathcal V'$ obtained by the transport of $\mathcal V$. This is done by finding the coordinates of the vectors of $\mathcal V'$ in $\mathcal V$, that is, by evaluating the matrix of the transporter in the basis $\mathcal V$. A holonomy being a unitary automorphism of $V$, it is unitarily diagonalizable and the argument of its eigenvalues define the geometric phases $\gamma_j$ of the transport along the discrete loop.  Any vector $\ket{v_j}$ belonging to the eigenspace associated with $\expo^{\imag \gamma_j}$ picks up, after the discrete loop, the phase $\gamma_j$. Note that the geometric phases change sign if the loop is reversed whereas they remain invariant if the $n$-planes are cyclically permuted\footnote{Suppose that the base point is displaced into $V_1$. The holonomy $\Gamma$ along the loop $V_0V_1\dots V_N$ becomes $\Gamma'=\Gamma(V_1,V_0,V_N,\dots,V_1)=\Gamma_{V_1V_0}\Gamma(\Gamma_{V_1V_0})^{-1}$ along the loop $V_1\dots V_NV_0V_1$. Hence, if $V_\gamma$ is the eigenspace of $\Gamma$ associated with an eigenvalue $\expo^{\imag\gamma}$ then $V'_\gamma=\Gamma_{V_1V_0}(V_\gamma)$ is the eigenspace of $\Gamma'$ associated with the same eigenvalue, and one has $\dim V_\gamma=\dim V'_{\gamma}$.} (i.e.\ if the base point is transferred to another point on the loop without changing the orientation). In the case $n=1$, for example, there is (mod $2\pi$) a single geometric phase $\gamma$ given by
\begin{align*}
\gamma&=\gamma_{v_0,v_N}+\gamma_{v_N,v_{N-1}}+\dots+\gamma_{v_2,v_1}+\gamma_{v_1,v_0}\\
&=\arg\big(\bk{v_0}{v_{N}}\bk{v_{N}}{v_{N-1}}\dots\bk{v_2}{v_1}\bk{v_1}{v_0}\big),
\end{align*}
with $\ket{v_a}$ ($a=0,\dots,N$) an arbitrary nonzero vector of $V_a$. By construction, the right-hand sides are invariant under any redefinition of the vectors $\ket{v_a}$ while the quantity inside the bracket remains invariant under any multiplication of the $\ket{v_a}$ by a (local) phase factor. When the vectors are unitary, the latter quantity is sometimes called a Bargmann invariant \cite{Simon1993} (Bargmann used invariants of this kind, with $N=2$, for his proof of Wigner's theorem on symmetry operations \cite{Bargmann}). In the case $n>1$, holonomies are said non-Abelian since, generally, transporters along two loops based at the same point do not commute, in contrary to the Abelian case $n=1$.

When $V$ and $W$ are distinct and supposed anti-orthogonal, the final vectors can obviously not be compared with the initial ones as in the previous paragraph. However, one can decide to compare the final vectors with the vectors of $W$ equipollent to the initial ones. It suffices to transport an initial basis $\mathcal V$ and to compare its image $\mathcal W$, after the transport, with the basis $\mathcal V^\parallel=\Gamma_{WV}(\mathcal V)$ of $W$ equipollent to $\mathcal V$, i.e.\ to evaluate the matrix of the transporter in the bases $\mathcal V$ and $\mathcal V^\parallel$. However, the vectors of $\mathcal W$ have in the basis $\mathcal V^\parallel$ the same coordinates as the vectors of $\mathcal W^\parallel=\Gamma_{VW}(\mathcal W)$ in $\mathcal V$: the comparison can equivalently be done inside $V$ after a closure of the discrete path by a last step $W\to V$. It amounts to evaluating the matrix of $\Gamma_{V_0V_N}\Gamma(V_N,\dots,V_0)$ in the basis $\mathcal V$. Hence, one defines 
\begin{align}
\Gamma_{\text h}(V_N,\dots,V_0)=\Gamma_{V_0V_N}\Gamma(V_N,\dots,V_0)\label{partial_holonomy}
\end{align}
as the non-cyclic discrete holonomy associated with the discrete path $V_0\dots V_N$. It is nothing but the cyclic holonomy along the discrete loop $V_0\dots V_NV_0$. Of course, in the present context, making a distinction between cyclic and non-cyclic holonomies seems artificial since any non-cyclic holonomy is a holonomy, and vice versa. However, it allows to introduce an idea of evolution of the holonomy step by step along a path (closed or not): if all the $n$-planes $V_1$, $V_2$, etc. are anti-orthogonal to $V$, one is able to define the successive intermediate holonomies $\Gamma_{\text h}(V_1,V_0)=\id_{V}$, $\Gamma_{\text h}(V_2,V_1,V_0)$, etc. In particular, one can see how evolve the geometric phases. Even if $V$ and $W$ are not anti-orthogonal, partial discrete holonomies can be defined through formula \eqref{partial_holonomy}.

\subsection{Continuous parallel transport} \label{subsec:continuous}

\subsubsection{From the discrete to the continuous}

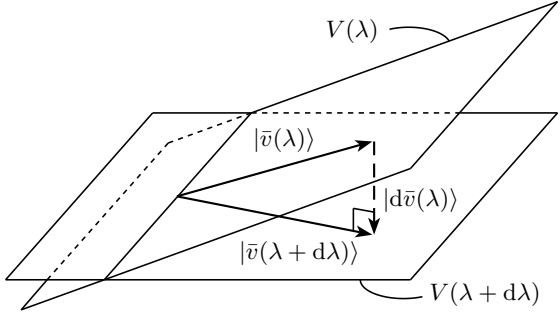
\begin{figure}
\centering
\psset{xunit=2.3cm,yunit=1.0cm,algebraic=true,dotstyle=o,dotsize=3pt 0,linewidth=0.6pt,arrowsize=3pt 2,arrowinset=0.25}
\begin{pspicture*}(1.149,3.6)(4.35,7.8)
\psline(1.15,4.05)(2,6.27)
\psline(1.15,4.05)(3.49,4.05)
\psline(2,6.27)(2.57,6.27)
\psline(3.78,6.27)(4.34,6.27)
\psline[linestyle=dashed,dash= 2pt 2pt](2.57,6.27)(3.78,6.27)
\psline(4.34,6.27)(3.49,4.05)
\psline(1.72,4.05)(2.57,6.27)
\psline[linewidth=.8pt]{->}(2.14,5.16)(3.27,5.89)
\psline(1.24,3.65)(1.39,4.05)
\psline[linestyle=dashed,dash= 2pt 2pt](1.39,4.05)(2.09,5.87)
\psline(3.49,5.53)(4.34,7.75)
\psline(2.57,6.27)(4.34,7.75)
\psline[linestyle=dashed,dash= 2pt 2pt](2.57,6.27)(2.09,5.87)
\psline(1.24,3.65)(1.72,4.05)
\psline(1.72,4.05)(3.49,5.53)
\psline[linewidth=.8pt,linestyle=dashed]{<-}(3.28,4.65)(3.28,5.89)
\psline[linewidth=.8pt]{->}(2.14,5.16)(3.27,4.65)
\psline(3.16,4.71)(3.16,5)
\psline(3.16,5)(3.28,4.95)
\rput[bl](3.33,4.95){$\ket{\diff \bar v(\lambda)}$}
\rput[bl](2.58,5.75){$\ket{\bar v(\lambda)}$}
\rput[bl](2.5,4.25){$\ket{\bar v(\lambda+\diff\lambda)}$}
\rput[bl](3.6,3.7){$V(\lambda+\diff\lambda)$}
\parametricplot{3.133620514450571}{5.16441195586637}{1*0.25*cos(t)+0*0.25*sin(t)+3.45|0*0.25*cos(t)+1*0.25*sin(t)+4.05}
\parametricplot{0.21226365904010916}{2.0518029752739864}{1*0.24*cos(t)+0*0.24*sin(t)+3.48|0*0.24*cos(t)+1*0.24*sin(t)+7.17}
\rput[bl](3,7.2){$V(\lambda)$}
\end{pspicture*}
\caption{An illustration of the parallel transport rule: while passing from $V(\lambda)$ to an infinitely close point $V(\lambda+\diff\lambda)$ in the Grassmann manifold, a vector $\ket{\bar v(\lambda)}$ in $V(\lambda)$ is orthogonally projected into the vector $\ket{\bar v(\lambda+\diff\lambda)}$ in $V(\lambda+\diff\lambda)$.}\label{transportillustration}
\end{figure}

We now consider the most interesting case where the parallel transport from $V$ to $W$ is made along a smooth path $\mathcal C$ in the Grassmann manifold $\text{Gr}(n,\mathcal H)$. By definition, this curve can be smoothly parameterized by a real variable $\lambda$ belonging to some interval $[\lambda_{\text i},\lambda_{\text f}]$. At a given value $\lambda$ of the parameter, the $n$-plane $V(\lambda)$ is reached, while $V(\lambda_{\text i})=V$ and $V(\lambda_{\text f})=W$. For short, the projection operator into $V(\lambda)$ will be denoted by $P(\lambda)$. Any initial vector $\ket{v_{\text i}}\in V$ generates a smooth vector field $\ket{\bar v(\lambda)}\in V(\lambda)$ along $\mathcal C$ through continuous infinitesimal projections of $V(\lambda)$ into $V(\lambda+\diff\lambda)$, that is, according to the rule
\begin{align}
\forall\lambda\in[\lambda_{\text i},\lambda_{\text f}]\,,\quad\ket{\bar v(\lambda+\diff\lambda)}=P(\lambda+\diff\lambda)\ket{\bar v(\lambda)},\label{transport1}
\end{align}
with $\ket{\bar v(\lambda_{\text i})}=\ket{v_{\text i}}$ as starting vector. The left-hand side can be rewritten as $\ket{\bar v}+\ket{\diff\bar v}$ and the right-hand side $(P+\diff P)\ket{\bar v}=\ket{\bar v}+\diff P\ket{\bar v}$ since $\ket{\bar v(\lambda)}$ belongs to $V(\lambda)$. Hence, equation~\eqref{transport1} is equivalent to $\ket{\diff\bar v}=\diff P\ket{\bar v}$. Furthermore, differentiating the equality $\ket{\bar v}=P\ket{\bar v}$ yields $\ket{\diff\bar v}=\diff P\ket{\bar v}+P\ket{\diff\bar v}$ so that \eqref{transport1} amounts to $P\ket{\diff\bar v}=0$. Alternatively stated, \eqref{transport1} is nothing but the requirement that $\ket{\diff\bar v(\lambda)}$ belongs to the orthogonal complement of $V(\lambda)$. Consequently, the norm of $\ket{\bar v}$ is preserved along the field:
\begin{align*}
\diff\bk{\bar v}{\bar v}=\bk{\diff\bar v}{\bar v}+\bk{\bar v}{\diff\bar v}=2\Re\bk{\bar v}{\diff\bar v}=0,
\end{align*}
and the projection of $V(\lambda)$ into $V(\lambda+\diff\lambda)$ is unitary. It proves that the infinitesimal transporter from $V(\lambda)$ to $V(\lambda+\diff\lambda)$ is $\Pi_{V(\lambda+\diff\lambda)V(\lambda)}$ itself. In other words, equation \eqref{transport1} is precisely the rule of continuous parallel transport in the Grassmann manifold (see figure \ref{transportillustration}).

The net result of the parallel transport from $V$ to $W$ along $\mathcal C$ is a unitary map $\Gamma[\mathcal C]$ which can be expressed by using the limit $N\to\infty$ of a sequence of discretizations of $\mathcal C$ into $N$ segments, as
\begin{align}
\Gamma[\mathcal C]=\lim_{N\to\infty}P(\lambda_N)P(\lambda_{N-1})\dots P(\lambda_1)P(\lambda_0),\label{PPPP}
\end{align}
where $\lambda_0=\lambda_{\text i}$ and $\lambda_N=\lambda_{\text f}$. From the parallel transport equation $\ket{\diff v}=\diff P\ket{v}$, it is clear that the above transporter can also be written
\begin{align*}
\Gamma[\mathcal C]=\mathcal P\expo^{\int_{\mathcal C}\diff P}P_V\,,
\end{align*}
where $\mathcal P$ is the path-ordering operator. By construction, and as the notation suggests, the transporter $\Gamma[\mathcal C]$ is a geometric quantity over the Grassmann manifold which depends only on the path $\mathcal C$ and not on the chosen smooth parameterization. If one introduces an orthonormal basis $\mathcal V_a=(\ket{v_j(\lambda_a)})_j$ of $V(\lambda_a)$ for $a=0,\dots,N$, one easily obtains from \eqref{PPPP} the following expression:
\begin{align*}
\Gamma[\mathcal C]=\sum_{j,k=1}^n \ket{v_j(\lambda_{\text f})}\bigg[\lim_{N\to\infty}\mathcal P\prod_{a=1}^N \rep S(\mathcal V_{a},\mathcal V_{a-1})\bigg]_{jk}\bra{v_k(\lambda_{\text i})}.
\end{align*}
The matrix of $\Gamma[\mathcal C]$ in the initial and final bases, $\mathcal V_0$ and $\mathcal V_N$ respectively, is thus
\begin{align}
\rep \Gamma[\mathcal C]=\lim_{N\to\infty}\mathcal P\prod_{a=1}^N \rep S(\mathcal V_{a},\mathcal V_{a-1}).\label{Wdiscrete}
\end{align}
Since the transporter is a linear map, this matrix does not depend at all on the arbitrary choice of intermediate bases $\mathcal V_1,\dots,\mathcal V_{N-1}$. If one rotates each basis $\mathcal V_a$ by an $a$-dependent unitary $n\times n$ matrix $\rep T_a$ ($a=0,\dots,N$) through $\mathcal V_a\to \mathcal V_a\rep T_a$\footnote{See the appendix \ref{subsec:matchar} for this notation.} then one has
\begin{align*}
\rep \Gamma[\mathcal C]\longrightarrow \rep T_N^{-1}\rep\Gamma[\mathcal C]\rep T_0\,.
\end{align*}
When $V=W$, i.e.\ when $\mathcal C$ is a loop, $\Gamma[\mathcal C]$ is a cyclic holonomy. In this case, it is natural to use the same basis at the endpoint: $\mathcal V_{0}=\mathcal V_{N}$. Then, an $a$-dependent rotation of the bases preserving the equality of the extremal ones (i.e.\ such that $\rep T_0=\rep T_N$) realizes a unitary similarity transformation of $\rep \Gamma[\mathcal C]$. The matrix $\rep \Gamma[\mathcal C]$ can thus be seen as a generalization of the Bargmann invariant (if $n>1$) in the continuum limit. 

\subsubsection{A sketch of the gauge structure}

One can derive a more elegant expression of the transporter along $\mathcal C$ by first noticing that a smooth vector field $\ket{\psi(\lambda)}\in V(\lambda)$ is parallelly transported iff it verifies the system of $n$ equations $\bk{v_j}{\diff \psi}=0$ where $\mathcal V(\lambda)=(\ket{v_j(\lambda)})_j$ is an orthonormal reference basis of $V(\lambda)$ for each value of the parameter. Then, if $\mathcal V(\lambda)$ is taken smooth with respect to $\lambda$, the system becomes $\diff\bk{v_j}{\psi}-\bk{\diff v_j}{\psi}=0$. Introducing the coordinate vector $\rep \Psi(\lambda)$ of $\ket{\psi(\lambda)}$ in the basis $\mathcal V(\lambda)$,
one deduces the parallel transport matrix equation $(\diff-\imag\mathcal A)\rep \Psi=0$, where $\mathcal A$ is the $n\times n$ matrix-valued one-form field whose entries are $\mathcal A_{jk}=\imag\,\bk{v_j}{\diff v_k}$. Taking patches along $\mathcal C$, one thereby obtains the following formulation of the matrix of the transporter:
\begin{align}
\rep \Gamma[\mathcal C]=\mathcal P\expo^{\imag\int_{\mathcal C}\mathcal A}, \label{Wilson}
\end{align}
in the initial and final bases. The field $\mathcal A$ is self-adjoint by the orthonormality of the basis:
\begin{align*}
\mathcal A_{jk}=\imag\,\bk{v_j}{\diff v_k}=-\imag\,\bk{\diff v_j}{v_k}=\big[\imag\,\bk{v_k}{\diff v_j}\big]^*=\mathcal A^*_{kj}\,.
\end{align*}
Its definition depends obviously on the chosen smooth field of orthonormal bases $\mathcal V(\lambda)$. If one performs a smooth $\lambda$-dependent rotation $\mathcal V(\lambda)\to \mathcal V(\lambda)\rep T(\lambda)$ of the reference frame then $\mathcal A$ undergoes the transformation
\begin{align}
\mathcal A\longrightarrow \mathcal A'=\rep T^{-1}\mathcal A\, \rep T+\imag \rep T^{-1}\diff \rep T.\label{transgauge}
\end{align}
One recognizes a $U(n)$ gauge structure over the Grassmann manifold, with $P\diff$ as covariant differential, its translation into coordinate being $\mathscr D=\diff-\imag\mathcal A$ where $\mathcal A$ is the gauge field.
From the transformation law \eqref{transgauge}, $\mathcal A$ is not the matrix field of a \enquote{linear map-valued} one-form field over $\text{Gr}(n,\mathcal H)$. As such, it does not represent an observable. Rather, it represents locally the connection of the gauge structure, i.e.\ the interconnection of the $n$-planes over infinitesimal regions of $\text{Gr}(n,\mathcal H)$. On the contrary, the gauge curvature field $\mathcal F=\diff A+\imag\mathcal A\wedge\mathcal A$ represents a \enquote{self-adjoint linear map-valued} two-form field over $\text{Gr}(n,\mathcal H)$. Actually, $\mathcal F$ measures the infinitesimal cyclic holonomies with respect to the choice of reference frame. It may be used to calculate non-infinitesimal cyclic holonomies through Stokes theorem (in its non-Abelian version if $n>1$ \cite{nonAbelian}).

\subsubsection{Computing transports and holonomies}

Analytically, the transporter along $\mathcal C$ is computed by using a smooth field of orthonormal bases $\mathcal V(\lambda)$ of $V(\lambda)$, as follows:
\begin{align*}
\Gamma[\mathcal C]=\sum_{j,k=1}^n\ket{v_j(\lambda_{\text f})}\bigg[\mathcal P\expo^{\imag\int_{\lambda_{\text i}}^{\lambda_{\text f}}\rep A(\lambda')\diff\lambda'}\bigg]_{jk}\bra{v_k(\lambda_{\text i})},
\end{align*}
where $\rep A(\lambda)=\imag(\bra{v_j(\lambda)}{\dot v_k(\lambda)})_{jk}$ is the smooth field of $n\times n$ self-adjoint matrices representing $\mathcal A$ along $\mathcal C$ with respect to the parameterization by $\lambda$, the overdot symbolizing always the total parameter-derivative. In particular, if $\mathcal C$ is a loop, one can express the matrix of $\Gamma[\mathcal C]$ in the initial basis as
\begin{align}
\rep\Gamma[\mathcal C]=\rep S(\mathcal V(\lambda_{\text i}),\mathcal V(\lambda_{\text f}))\,\mathcal P\expo^{\imag\int_{\lambda_{\text i}}^{\lambda_{\text f}}\rep A(\lambda')\diff\lambda'}.\label{Gloop}
\end{align}
The above overlap matrix has the meaning of a \enquote{twist} of the final basis to coincide with the initial one. As in the foregoing subsection, when $\mathcal C$ is an open path, one can close it by a teletransport from $W$ to $V$ and associate with $\mathcal C$ the non-cyclic holonomy $\Gamma_{\text h}[\mathcal C]=\Gamma_{VW}\Gamma[\mathcal C]$. Introducing the matrix $\rep\Gamma_{WV}$ of the teletransporter in the bases $\mathcal V(\lambda_{\text f})$ and $\mathcal V(\lambda_{\text i})$ one has
\begin{align}
\rep\Gamma_{\text h}[\mathcal C]=\rep\Gamma_{VW}\,\mathcal P\expo^{\imag\int_{\lambda_{\text i}}^{\lambda_{\text f}}\rep A(\lambda')\diff\lambda'}.
\end{align}
Obviously, $\rep\Gamma_{VW}$ is equal to $\rep S(\mathcal V(\lambda_{\text i}),\mathcal V(\lambda_{\text f}))$ if $\mathcal C$ is closed. Generalizing the conventional terminology introduced by Simon and Mukunda in the case $n=1$ \cite{SimonMukunda1993}, the matrix $\rep\Gamma_{VW}$ may be seen as the total \enquote{phase} factor accumulated by the frame field $\mathcal V(\lambda)$ which decomposes as a geometric part and a dynamical one according to
\begin{align*}
\underbrace{\rep\Gamma_{VW}}_{\text{total}}=\underbrace{\rep\Gamma_{\text h}[\mathcal C]}_{\text{geometric}}\underbrace{\mathcal P\expo^{-\imag\int_{\lambda_{\text i}}^{\lambda_{\text f}}\rep A(\lambda')\diff\lambda'}}_{\text{dynamical}}.
\end{align*}
Individually, the total and dynamical phase factors are invariant under a change of parametrization: they only depend on the path followed by $\mathcal V(\lambda)$ in the so-called Stiefel manifold $\text{St}(n,\mathcal H)$, i.e.\ in the collection of (ordered) orthonormal $n$-frames of $\mathcal H$ (see footnote \ref{Stiefel}). By their definition, and even if $\mathcal C$ is closed, none of them transform covariantly under arbitrary changes of frame field but, naturally, \eqref{Gloop} does. However, the terms \enquote{total}, \enquote{geometric} and \enquote{dynamical} are just a matter a convention since the three quantities are all geometric on their own as they represent linear maps associated with couple of points or curves in the Grassmann manifold.\footnote{One can write $\rep \Gamma_{VW}=\rep\Gamma_{\text{tot}}[C]$ and $\mathcal P\expo^{-\imag\int_{\lambda_{\text i}}^{\lambda_{\text f}}\rep A(\lambda')\diff\lambda'}=\rep\Gamma_{\text{dyn}}[C]$ where $C$ is the path traced out by $\mathcal V(\lambda)$ in the Stiefel manifold. Each point $C(\lambda)\in\text{St}(n,\mathcal H)$ projects into $\mathcal C(\lambda)\in\text{Gr}(n,\mathcal H)$ with respect to the quotient structure described in footnote \ref{Stiefel}. In the terminology of fibre bundles, $C$ is a section of $\text{St}(n,\mathcal H)$ over $\mathcal C$. The matrices $\rep\Gamma_{\text{tot}}[C]$ and $\rep\Gamma_{\text{dyn}}[C]$ depend on the section $C$ but the product $\rep\Gamma_{\text{tot}}[C]\rep\Gamma_{\text{dyn}}[C]^{-1}$ depends only on the projection $\mathcal C$.} The specificity of $\rep\Gamma_{\text h}[\mathcal C]$ stems from the fact that it is measurable. Suppose that $V$ and $W$ are anti-orthogonal. With a suitable choice of frame field, $\rep\Gamma_{\text h}[\mathcal C]$ can be reduced to the inverse of the dynamical phase factor if $\mathcal V(\lambda)$ is chosen such as to circle back to itself\footnote{Such a choice can always be made. For a given frame field $\mathcal V(\lambda)$, the matrix $\rep\Gamma_{VW}$ is unitary and has the form $\expo^{\imag\rep M}$ where $\rep M$ is a Hermitian matrix. If we change the frame field according to $\mathcal V(\lambda)\to \mathcal V'(\lambda)=\mathcal V(\lambda)\exp[-\imag\frac{\lambda-\lambda_{\text i}}{\lambda-\lambda_{\text f}}\rep M]$ then, clearly, $\rep\Gamma_{VW}=\rep I_n$.} or to the total phase factor if it is parallelly transported. In the case $n=1$, for example, if $V$ and $W$ are not orthogonal, the single geometric phase of a holonomy has the well-known expression \cite{SimonMukunda1993} 
\begin{align}
\gamma[\mathcal C]=\arg\bk{v(\lambda_{\text i})}{v(\lambda_{\text f})}-\Im\int_{\lambda_{\text i}}^{\lambda_{\text f}}\bk{v(\lambda')}{\dot v(\lambda')}\diff\lambda',\label{MukundaSimon}
\end{align}
where $\lambda\mapsto \ket{v(\lambda)}\in V(\lambda)$ is any smooth field of unit vectors. It is the difference between the total phase accumulated by $\{\ket{v(\lambda)}\}$ and its dynamical phase.

Let me emphasize that a naive discretization of \eqref{Wilson} can be used in numerical calculations only if $\mathcal C$ is a priori known analytically, that knowledge allowing to determine by hand a smooth orthonormal frame and the corresponding gauge field. However, in general, $\mathcal C$ is itself the result of a numerical computation, in which case the algorithm \enquote{chooses} at each point an orthonormal basis, without any idea of smoothness or continuity, and \eqref{Wdiscrete} must be used.

\subsection{Simple curves and geodesics}\label{trivialholo}

Let us consider two distinct $n$-planes $V$ and $W$. We will say that a curve $\mathcal C$ from $V$ to $W$ is simple if the transporter $\Gamma[\mathcal C]$ along $\mathcal C$ belongs to the set $\mathcal U_{WV}$. Our aim, in this part, is to show that geodesics in the Grassmann manifold offer such simple curves. To this end, we identify each $n$-plane $X$ of $\text{Gr}(n,\mathcal H)$ with the projection operator $P_X$. A point of that manifold is thus an operator of $\mathcal H$ which is idempotent, self-adjoint, and of trace $n$. Now, let $P=P_X$ be such a point and $\delta P$ be an infinitesimal increment of $P$. The operator $\delta P$ is an infinitesimal tangent vector of $\text{Gr}(n,\mathcal H)$ at the point $P$ iff $P+\delta P$ is a point of $\text{Gr}(n,\mathcal H)$. The self-adjointness condition of $P+\delta P$ amounts to $\delta P=(\delta P)^\dag$ while the idempotence condition of $P+\delta P$ amounts to the equality 
\begin{align*}
\delta P=P(\delta P)+(\delta P) P
\end{align*} 
since $\delta P$ is infinitesimal. The last equality implies the vanishings of $P(\delta P)P$ and $(\openone-P)(\delta P)(\openone-P)$. Thus, $\delta P$ is traceless and $\tr(P+\delta P)=n$. One deduces that the tangent space of $\text{Gr}(n,\mathcal H)$ at $P$ is the real vector space\footnote{If $K$ and $K'$ verify \eqref{tangence} then also $K+K'$. If $c$ is a complex number then $cK$ verifies the second equality but, if $K\ne 0$, the first one is verified iff $c$ is real. Hence, one has to restrict the scalar field to the reals.} $T_P$ formed by the operators $K$ in $\mathcal H$ verifying
\begin{align}
K=K^\dag\quad\text{and}\quad K=PK+KP,\label{tangence}
\end{align}
or equivalently
\begin{align}
K=K^\dag\quad\text{and}\quad PKP=0=(\openone-P)K(\openone-P).\label{tangence2}
\end{align}
The latter characterization says that the tangent space at $P$ is formed by the self-adjoint operators which are \enquote{anti-block-diagonal} with respect to the decomposition $X\oplus X^\perp$ of $\mathcal H$. Then, one defines the normal space $N_P$ at $P$ as the set of self-adjoint operators $N$ orthogonal to $T_P$ with respect to the Frobenius inner product, i.e.
\begin{align*}
\forall K\in T_P\,,\quad (N|K)\coloneqq\tr(N^\dag K)=\tr(NK)=0.
\end{align*}
It is also a real vector space. Now, any self-adjoint operator $A$ decomposes uniquely as the sum $T+N$ with
\begin{align*}
T&=PA(\openone-P)+\text{h.c.}\in T_P\,,\\
N&=PAP+(\openone-P)A(\openone-P)\in N_P\,.
\end{align*} 
Indeed, the fact that $T$ belongs to $T_P$ is clear from \eqref{tangence2} while, by the same characterization, $N$ belongs to $N_P$ since one has
\begin{align*}
\tr(PAPK)=\tr(APKP)&=0,\\
\tr[(\openone-P)A(\openone-P)K]=\tr[A(\openone-P)K(\openone-P)]&=0.
\end{align*}
As for the unicity, it is evident from the orthogonality of $T_P$ and $N_P$. Consider a field $K(\lambda)\in T_{P(\lambda)}$ of tangent operators over a curve $\mathcal C\colon\lambda\mapsto P(\lambda)$ in $\text{Gr}(n,\mathcal H)$. Its differential $\diff K$ at a given point $P$ can be decomposed into the sum of a tangent operator $D_TK$ and a normal one, $D_NK$. In particular, using \eqref{tangence} and \eqref{tangence2}, the latter component is given by
\begin{align*}
D_NK=P(\diff K)P+(\openone-P)(\diff K)(\openone-P)=\big[[K,P],\diff P\big]
\end{align*}
and is thus linear in $K$. As in the previous subsection, the field $K(\lambda)$ is parallelly transported along $\mathcal C$ if it is continuously projected from $T_{P(\lambda)}$ to $T_{P(\lambda+\diff\lambda)}$, i.e.\ if it obeys the rule
\begin{align*}
K(\lambda+\diff\lambda)=P(\lambda+\diff\lambda)K(\lambda)(\openone-P(\lambda+\diff\lambda))+\text{h.c.}
\end{align*} 
which amounts to $D_TK=0$, that is, to $P(\diff K)(\openone-P)=0$ since $\diff K$ is self-adjoint. The parallel transport is unitary since it preserves the Frobenius norm:
\begin{align*}
\diff(\|K\|^2)=\diff(K|K)=2(K|\diff K)=2(K|D_TK)=0.
\end{align*}
With respect to the parametrization of $\mathcal C$ by $\lambda$, the acceleration $\ddot P$ of the curve $\mathcal C$ decomposes as a tangential acceleration $D_T\dot P/\diff\lambda$ and a normal one, $D_N\dot P/\diff\lambda$. The curve is said autoparallel if its tangential acceleration and its velocity $\dot P$ are parallel. Choosing an arbitrary parametrization which is not pathologic, i.e.\ such that the velocity is (piecewise) nonvanishing, the condition of autoparallelism amounts to the existence of a real function $f(\lambda)$ such that
\begin{align*}
\frac{D_T}{\diff\lambda}\,\dot P=f(\lambda)\dot P.
\end{align*} 
This property does not depend on the chosen parameter. Indeed, any reparametrization $\lambda\to\lambda'$, with $\diff\lambda'/\diff\lambda>0$, preserves the form of the above equation inasmuch as $f$ is altered according to
\begin{align*}
f(\lambda)&\longrightarrow f'(\lambda')=f(\lambda)\,\frac{\diff\lambda'}{\diff\lambda}-\frac{\diff^2\lambda'}{\diff\lambda^2}\,.
\end{align*}
The function $f(\lambda)$ is easily deduced from
\begin{align*}
\frac{\diff}{\diff\lambda}\|\dot P\|^2=\frac{\diff}{\diff\lambda}\big(\dot P\big|\dot P)=2\bigg(\dot P\,\bigg|\frac{D_T}{\diff\lambda}\,\dot P\bigg)=2f(\lambda)\big(\dot P|\dot P\big),
\end{align*}
that is,
\begin{align*}
f(\lambda)=\frac{\diff}{\diff\lambda}\,\|\dot P\|=\ddot s,
\end{align*}
where $\diff s=\|\diff P\|$ is the line element in $\text{Gr}(n,\mathcal H)$. Introducing an orthonormal frame field $\mathcal V=\{\ket{v_j})_j$, its square can be written as
\begin{align}
(\diff s)^2=\tr\big[(\diff P)(\diff P)\big]=\sum_{j=1}^n\bra{\diff v_j}(\openone-P)\ket{\diff v_j}.\label{ds2}
\end{align}  
Hence, it is also the sum of the squared norms of the components of the differentials $\ket{\diff v_j}$ orthogonal to the $n$-plane. By construction, the right-hand side is independent of the chosen orthonormal frame. Now, the metric allows to define preferred parameters, said affine, with respect to which the curve is followed at constant speed, i.e.\ for which $f\equiv0$. Using such a parameter $\tau$, the autoparallel condition reduces to the vanishing of the tangential acceleration, a condition which can be stated in three equivalent ways:
\begin{align}
\frac{D_T}{\diff\tau}\,\dot P=0,\quad \ddot P\in N_P\,,\quad P\ddot P(\openone-P)=0.\label{auto}
\end{align}
The autoparallels are actually the geodesics since they leave stationary the length functional
\begin{align*}
\ell[\mathcal C]=\int_{\mathcal C}\diff s=\int_{\lambda_{\text i}}^{\lambda_{\text f}}\dot s(\lambda)\diff\lambda
\end{align*}
under variations $P\to P+\delta P$ vanishing at the endpoints. Indeed, from 
\begin{align*}
\delta\dot s=\frac{\big(\dot P\big|\delta\dot P\big)}{\dot s}=\frac{\diff}{\diff\lambda}\bigg(\frac{\dot P}{\dot s}\bigg|\,\delta P\bigg)-\bigg(\frac{\diff}{\diff\lambda}\bigg(\frac{\dot P}{\dot s}\bigg)\bigg|\,\delta P\bigg),
\end{align*}
one deduces that
\begin{align*}
\delta\ell[\mathcal C]=-\int_{\lambda_{\text i}}^{\lambda_{\text f}}\bigg(\frac{\diff}{\diff\lambda}\bigg(\frac{\dot P}{\dot s}\bigg)\bigg|\,\delta P\bigg)\diff\lambda,
\end{align*}
and the geodesic equation amounts to
\begin{align*}
\frac{\diff}{\diff\lambda}\bigg(\frac{\dot P}{\dot s}\bigg)\in N_P\,.
\end{align*}
It reduces to \eqref{auto} if one chooses an affine parameter $\tau$. From now on, we make such a choice and introduce a parallelly transported basis $\mathcal V(\tau)=(\ket{v_j(\tau)})_j$ along $\mathcal C$. The latter choice, together with \eqref{ds2}, imply the constancy of the trace of the matrix $\rep\Lambda=(\bk{\dot v_j}{\dot v_k})_{jk}$ and, through the third equality in \eqref{auto}, the geodesic equation becomes
\begin{align*}
\sum_{j=1}^n\ket{ v_j}\bra{\ddot v_j}(\openone-P)=0.
\end{align*}  
It is equivalent to the $n$ equations $(\openone-P)\ket{\ddot v_j}=0$ saying that the accelerations of the parallelly transported basis vectors belong to $V(\tau)$. It yields the vanishing of $\bk{\dot v_j}{\ddot v_k}$ and the constancy of the matrix $\rep\Lambda$:
\begin{align*}
\frac{\diff}{\diff \tau}\bk{\dot v_j}{\dot v_k}=\bk{\ddot v_j}{\dot v_k}+\bk{\dot v_j}{\ddot v_k}=0.
\end{align*}
Furthermore, one has
\begin{align*}
\bk{v_j}{\ddot v_k}=\frac{\diff}{\diff \tau}\bk{v_j}{\dot v_k}-\bk{\dot v_j}{\dot v_k}=-\bk{\dot v_j}{\dot v_k}=\text{cst.}
\end{align*}
and the $n$ equations $(\openone-P)\ket{\ddot v_j}=0$ become
\begin{align}
\ket{\ddot v_j}+\sum_{k=1}^n\,\ket{v_k}\bk{\dot v_k}{\dot v_j}=0.\label{neq}
\end{align}
However, since $\rep\Lambda$ is self-adjoint and constant, there exists a constant unitary matrix $\rep T$ such that $\rep T^{-1}\rep\Lambda\rep T$ is diagonal. Since $\rep T$ is constant, the rotated frame $\mathcal V'(\tau)=(\ket{v'_j(\tau)})_j=\mathcal V(\tau)\rep T$ is a parallelly transported basis which also verifies all the above equations. But now $\rep\Lambda$ is transformed into $\rep\Lambda'=\rep T^{-1}\rep\Lambda\rep T$ and \eqref{neq} reduces to the $n$ equations $\ket{\ddot v'_j}+\ket{v'_j}\|\dot v'_j\|^2=0$.

\begin{figure}
\centering
\begin{pspicture*}(-3,-2.51)(3,2.51)
\pscircle[linewidth=.6pt](0,0){2.5}
\psellipticarc[linestyle=dashed,linewidth=.6pt](0,0)(2.5,1){0}{180}
\psellipticarc[linewidth=.6pt](0,0)(2.5,1){180}{202}
\psellipticarc[linewidth=.6pt](0,0)(2.5,1){-35}{0}
\psellipticarc[linewidth=1pt,arrowsize=6pt]{->}(0,0)(2.5,1){-158}{-105}
\psellipticarc(0,0)(2.5,1){-107}{-35}
\psellipticarc[linewidth=.6pt]{->}(0,0)(1,.4){-158}{-35}
\rput(-.1,-.65){$\omega\tau$}
\psdots[dotstyle=*](0,0)
\psline[linewidth=1pt](0,0)(-1.768,-.707)
\psdots(-1.768,-.707)
\psline[linewidth=1pt](0,0)(2.5,0)
\psdots[dotstyle=*](2.5,0)
\psline[linewidth=1pt](0,0)(1.25,-.866)
\psdots(1.25,-.866)
\uput[-90](-1.768,-.707){$\rho(0)$}
\uput[-90](1.25,-.866){$\rho(\tau)$}
\uput[0](2.5,0){$\eta$}
\rput(-.1,-1.4){$\mathcal C_{\text{geo}}$}
\end{pspicture*}
\caption{A geodesic $\mathcal C_{\text{geo}}\colon\tau\mapsto\rho(\tau)$ in the case $n=1$ is determined by the initial position ray $\rho(0)$ and by the ray $\eta$ spanned by the initial velocity of a parallelly transported unit vector field $\ket{v(\tau)}$ such that $\rho(\tau)=\kb{v(\tau)}{v(\tau)}$. It can be represented on the Bloch-Poincar\'e sphere built on $\rho(0)$ and $\eta$ as shown in the figure.}\label{blochsphere}
\end{figure}
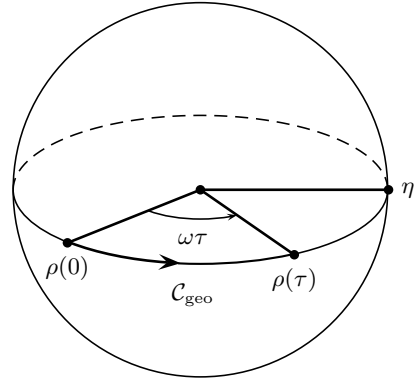

The above analysis can be resumed as follows: $\mathcal C$ is a geodesic iff (i) it admits a parallelly transported orthonormal basis $\mathcal V(\tau)$ which verifies $n$ equations of the form
\begin{align}
\ket{\ddot v_j}+\omega_j^2\ket{v_j}=0,\label{neq2}
\end{align}
where the $\omega_j$ are nonnegative constants, and (ii) $\rep\Lambda$ is the constant matrix $\diag(\omega_1^2,\dots,\omega_n^2)$. Taking the equations \eqref{neq2} alone, the general solutions have the form
\begin{subnumcases}{\label{w}\hspace{-9.5mm}\ket{v_j(\tau)}=}
\!\!\ket{v_j^0}=\text{cst.} & \hspace{-4mm}if $\omega_j=0$\label{wnul} \\
\!\!\cos(\omega_j\tau)\ket{v_j^0}+\displaystyle\frac{\sin(\omega_j\tau)}{\omega_j}\,\ket{\dot v_j^0} & \hspace{-4mm}if $ \omega_j>0$\label{wnonnul}
\end{subnumcases}
with $\ket{v_j^0}=\ket{v_j(0)}$ and $\ket{\dot v_j^0}=\ket{\dot v_j(0)}$. Now, it is readily verified that if one further imposes the initial conditions 
\begin{align}
\bk{v^0_j}{v^0_k}=\delta_{jk}\,,\quad\bk{v^0_j}{\dot v^0_k}=0,\quad\bk{\dot v^0_j}{\dot v_k^0}=\omega_j^2\delta_{jk}\,,\label{condinit}
\end{align}
then all the conditions for $\mathcal C$ to being a geodesic are fulfilled: one has at any value of the parameter, $\bk{v_j}{v_k}=\delta_{jk}$, $\bk{v_j}{\dot v_k}=0$ and $\rep\Lambda=\diag(\omega_1^2,\dots,\omega_n^2)$. The sufficient conditions \eqref{condinit} for $\mathcal C$ to being a geodesic are also clearly necessary and the geodesics are thus entirely characterized. We can make three comments on the results derived in this paragraph. \textit{Primo}, each projection $\kb{v_j(\tau)}{v_j(\tau)}$ follows separately a geodesic in the ray space (see figure \ref{blochsphere} for an illustration). \textit{Secundo}, if $2n$ is greater than $\dim(\mathcal H)$ then at least $2n-\dim(\mathcal H)$ vectors of the basis must remain constant since the $2n$ vectors $\ket{v_j^0}$, $\ket{\dot v_j^0}$ must be mutually orthogonal. \textit{Tertio}, the overlap matrix $\rep S(\mathcal V(\tau),\mathcal V(0))$ remains real and diagonal:
\begin{align}
\rep S(\mathcal V(\tau),\mathcal V(0))=\diag(\cos(\omega_1\tau),\dots,\cos(\omega_n\tau)).\label{diagoverlap}
\end{align}
In particular, if $\tau$ is such that $\cos(\omega_j\tau)\geqslant0$ for all $j$ then $\rep S(\mathcal V(\tau),\mathcal V(0))$ is obviously self-adjoint and positive semidefinite. Hence, according to the paragraph \ref{telebases}, there exists an element $U\in \mathcal U_{WV}$ such that $\mathcal V(\tau)=U(\mathcal V(0))$. Since $\mathcal V(\tau)$ results in the same time from the parallel transport of $\mathcal V(0)$ along $\mathcal C$, one concludes that $\mathcal C$ is a simple curve between $V(0)$ and $V(\tau)$. It is certainly the case for $\tau\leqslant\pi/(2\max\omega_j)$.

Now, let us determine the geodesics connecting $V$ and $W$. This task amounts to determining the $n$-frame fields $\mathcal V(\tau)$ obeying to \eqref{w} and \eqref{condinit}, such that $\mathcal V(0)$ is a basis of $V$ and $\mathcal V(1)$ of $W$, with $\rep S(\mathcal V(1),\mathcal V(0))$ a necessary real diagonal matrix. According to the paragraph \ref{telebases}, the last condition imposes the following ones: (i) up to a reordering of its elements, $\mathcal V(0)=(\ket{v_j^0})_j$ is a basis adapted to the decomposition $V_0\oplus\dots\oplus V_p$ of $V$ and (ii) there exists an element $U\in\mathcal U_{WV}$ such that each vector $\ket{v_j(1)}$ is equal or opposite to $U\ket{v_j^0}$. Hence, let $\mathcal V(0)$ be a basis formed by a concatenation of orthonormal bases $(\ket{v^0_{\mu l}})_{1\leqslant l\leqslant n_\mu}$ of $V_\mu$. The initial basis vectors generate fields $\ket{v_{\mu l}(\tau)}$ ending at final vectors $\ket{v_{\mu l}(1)}$ which have the form $\varepsilon_{\mu l}U\ket{v^0_{\mu l}}$, with $U$ a given element of $\mathcal U_{WV}$ and $\varepsilon_{\mu l}=\pm 1$. One has
\begin{align}
\bk{v_{\mu l}(1)}{v^0_{\mu' l'}}=\sigma_\mu\,\varepsilon_{\mu l}\delta_{\mu\mu'}\delta_{ll'}\,.\label{orthogonalites}
\end{align}
Formula \eqref{diagoverlap} imposes therefore $\cos\omega_{\mu l}=\varepsilon_{\mu l}\cos\phi_\mu$ with $\omega_{\mu l}=\|\dot v_{\mu l}\|$. Consider first the basis vectors of a given subspace $V_\mu$ for a value $\sigma_\mu<1$. In this case, $V_\mu$ and $W_\mu$ are non-intersecting and each $\ket{v_{\mu l}^0}$ is distinct from $\ket{v_{\mu l}(1)}$. Thus, the field $\ket{v_{\mu l}(\tau)}$ must have the form \eqref{wnonnul}. Then, since we demand $\ket{v_{\mu l}(1)}=\varepsilon_{\mu l}U\ket{v_{\mu l}^0}$, this field is
\begin{align*}
\ket{v_{\mu l}(\tau)}=\frac{\sin\big(\omega_{\mu l}(1-\tau)\big)\ket{v_{\mu l}^0}+\varepsilon_{\mu l}\sin(\omega_{\mu l}\tau)\,U\ket{v_{\mu l}^0}}{\sin\omega_{\mu l}}\,.
\end{align*}
We have $\sigma_p=1$ when $V$ and $W$ intersect nontrivially, in which case $V_p=V\cap W=W_p$ and $\omega_{pl}=k_{l}\pi$, with $k_{l}$ a nonnegative integer; it is even if $\varepsilon_{pl}=+1$ and odd if $\varepsilon_{pl}=-1$. If $k_{l}=0$ then $\ket{v_{pl}}$ remains constant as in \eqref{wnul}, otherwise it takes the form \eqref{wnonnul} with an arbitrary initial velocity, inasmuch as the $\ket{\dot v_{pl}^0}$'s are in harmony with the initial conditions \eqref{condinit}. Finally, it is an easy task to verify that the necessary conditions derived in this paragraph are also sufficient for the curve $\tau\mapsto V(\tau)=\spn(\mathcal V(\tau))$ to be a geodesic. The geodesics linking $V$ and $W$ are thus characterized. Since $\phi_\mu$ is the smallest possible value of $\omega_{\mu l}$, there is for any $U\in\mathcal U_{WV}$ one shortest geodesic $\mathcal C_{\text{min}}$ between $V$ and $W$. Its length is
\begin{align*}
\ell[\mathcal C_{\text{min}}]=\int_0^1\bigg(\sum_{\mu=0}^p\sum_{l=1}^{n_\mu}\|\dot v_{\mu l}\|^2\bigg)^{1/2}\diff\tau=\bigg(\sum_{\mu=0}^pn_\mu\phi_\mu^2\bigg)^{1/2}\!\!.
\end{align*}
This length depends only on $V$ and $W$; it may be seen as the geodesic distance between them.\footnote{When $n=1$, if $\ket v$ and $\ket w$ are two unit vectors spanning $V$ and $W$ then this length is simply the Fubini-Study distance $\arccos\module{\bk{w}{v}}$.} For the shortest geodesics, all the $\varepsilon_{\mu l}$ are equal to $+1$ and, if $\sigma_p=1$, then the $\ket{v_{pl}}$ are constant and $V(\tau)$ contains the intersection $V\cap W$.

Amongst the geodesics linking $V$ and $W$, the ones for which all the $\varepsilon_{\mu l}$ are equal to $+1$ are simple curves. Hence, one knows how to construct geodesic simple curves along which the transporter is any element of $\mathcal U_{WV}$. The shortest geodesics are such curves. They are furthermore the only geodesics for which all the portions are themselves simple curves (if one of the $\omega_{\mu l}$ is distinct from $\phi_{\mu}$ then there exist values of $\tau$ such that $\ket{v_{\mu l}(\tau)}$ and $\ket{v_{\mu l}^0}$ are in phase opposition). The case where $V$ and $W$ are anti-orthogonal is the most interesting. Indeed, any simple curve $\mathcal C_{\text{sim}}$ from $V$ to $W$ allows to express any non-cyclic holonomy along a curve $\mathcal C$ from $W$ to $V$ as the cyclic holonomy along the loop $\mathcal  C_{\text{sim}}\cup\mathcal C$ formed by the path $\mathcal C$ followed by $\mathcal C_{\text{sim}}$:
\begin{align*}
\Gamma_{\text h}[\mathcal C]=\Gamma_{WV}\Gamma[\mathcal C]=\Gamma[\mathcal C_{\text{sim}}]\Gamma[\mathcal C]=\Gamma[\mathcal C_{\text{sim}}\cup\mathcal C].
\end{align*}

\subsection{Dynamics enters the stage}\label{subsec:dynamics}

We now suppose that there exists a self-adjoint (generally time-dependent) Hamiltonian $H(t)$ governing the dynamics in $\mathcal H$ (we will set $\hbar=1$). We further assume that we have to our disposal an invariant $n$-plane \cite{Viennot2007}. By definition, it is a time-dependent $n$-plane $V(t)$ such that any initial state $\ket{\psi(0)}$ belonging to $V(0)$ generates a motion $\ket{\psi(t)}$ \enquote{remaining} inside $V(t)$ at any time $t$. It amounts to saying that the projection operator $P(t)$ into $V(t)$ verifies $P(t)=U(t)P(0)U^{-1}(t)$ or equivalently
\begin{align}
P(t)U(t)=U(t)P(0),\label{Nenciu}
\end{align}
where $U(t)$ is the time evolution operator \cite{Messiah1} from $t=0$ to $t=t$. Indeed, let $\ket{\psi(t)}$ be any motion. If $\ket{\psi(0)}$ belongs to $V(0)$ then $P(t)U(t)\ket{\psi(0)}=P(t)\ket{\psi(t)}=\ket{\psi(t)}=U(t)\ket{\psi(0)}=U(t)P(0)\ket{\psi(0)}$. Otherwise, if $\ket{\psi(0)}$ is orthogonal to $V(0)$ then, by the unitariness of the dynamics, $\ket{\psi(t)}$ remains orthogonal to $V(t)$ and one has $P(t)U(t)\ket{\psi(0)}=P(t)\ket{\psi(t)}=0=U(t)P(0)\ket{\psi(0)}$. According to appendix \ref{annexe:DI}, equality \eqref{Nenciu} characterizes the fact that $P(t)$ is a dynamical invariant. Alternatively stated, it satisfies the Schr\"odinger-von Neumann equation $\imag\dot P=[H,P]$. Of course, $\openone-P$ is also a dynamical invariant and $V(t)^\perp$ is an invariant space, too. In fact, the dynamics of the wavefunctions is \enquote{block-diagonalized} into two subdynamics inside $V(t)$ and $V(t)^\perp$. Note that, in contrary to the wavefunctions, the dynamics of $V(t)$ and $V(t)^\perp$ are insensitive to transformations $H(t)\to H'(t)=H(t)+W(t)$ of the Hamiltonian, where $W(t)$ is any operator commuting with $P(t)$, since $[H,P]=[H',P]$. In particular, nullifying the restrictions of $H(t)$ into $V(t)$ and $V(t)^\perp$, via the transformed Hamiltonian
\begin{align}
\xoverline{H}=H-PHP-(\openone-P)H(\openone-P),\label{nullification}
\end{align}
does not affect the dynamics of $V(t)$ and $V(t)^\perp$.

The determination of an invariant $n$-plane $V(t)$ can be seen as a partial resolution of the Schr\"odinger equation for the subdynamics inside $V(t)$. This information can be geometrized as follows. Let $\ket{\psi(t)}$ be a motion belonging to $V(t)$. At the instant $t$, the whole dynamics makes the state transit from $\ket{\psi(t)}$ to $\ket{\psi(t+\diff t)}$. Let $\ket{\xoverline \psi(t)}$ be the vector of $V(t)$ equipollent to $\ket{\psi(t+\diff t)}\in V(t+\diff t)$ and decompose the whole dynamics as a \enquote{geometric part} and a (remaining) \enquote{dynamical part}
\begin{align*}
\begin{array}{ccccc} V(t) & \xrightarrow{\hspace{1cm}} & V(t)& \xrightarrow{\hspace{1cm}} & V(t+\diff t)\\
& \text{dyn.} & & \text{geo.} & \\
\ket{\psi(t)} & \xmapsto{\hspace{1cm}} & \ket{\xoverline\psi(t)} & \xmapsto{\hspace{1cm}} & \ket{\psi(t+\diff t)}
\end{array}
\end{align*}
On the one hand, the dynamical part is directly measured by the covariant derivative since
\begin{align*}
P(t)\ket{\diff\psi}=P(t)\big[\ket{\psi(t+\diff t)}-\ket{\psi(t)}\big]=\ket{\xoverline\psi(t)}-\ket{\psi(t)}.
\end{align*}
On the other hand, one has, using the Schr\"odinger and the Schr\"odinger-von Neumann equation,
\begin{align}
P\ket{\diff\psi}=\ket{\diff\psi}-(\diff P)\ket{\psi}=-\imag(PHP)\ket\psi\diff t.\label{redem}
\end{align}
Therefore, equating the extreme right-hand sides of the two last equations, one obtains the mechanism of the dynamical part:
\begin{align*}
\ket{\psi(t)}\longmapsto\ket{\xoverline\psi(t)}&=\big[\openone-\imag P(t)H(t)P(t)\diff t\big]\ket{\psi(t)}\\
&=\expo^{-\imag P(t)H(t)P(t)\diff t}\ket{\psi(t)}.
\end{align*} 
It is an infinitesimal \enquote{rotation} inside the instantaneous space $V(t)$ operated by the restricted Hamiltonian. It may be interpreted geometrically as a torsion term deviating the wavefunction from the natural parallel transport. The same decomposition can of course be done for the subdynamics inside $V(t)^\perp$, too.  

By construction, the partial geometrization of the whole dynamics depends only on the path $\mathcal C$ developed by $V(t)$ in the Grassmann manifold. As was pointed out, it is insensitive to transformations $H(t)\to H'(t)=H(t)+W(t)$ of the Hamiltonian, where $W(t)$ is any operator commuting with $P(t)$. In particular, transforming $H$ into \eqref{nullification} cancels the torsions inside $V(t)$ and $V(t)^\perp$, that is, $\xoverline H$ generates the parallel transports along $t\mapsto V(t)$ and $t\mapsto V(t)^\perp$.

The reduced evolution operator $U_{\text{red}}(t)\colon V(0)\to V(t)$ of the subdynamics inside $V(t)$ is the net result of a continuous succession of infinitesimal torsions and parallel transports. Using the limit $N\to\infty$ of a sequence of discretizations of $t\mapsto V(t)$ into $N$ segments, it can be expressed as
\begin{align*}
U_{\text{red}}(t)=\lim_{N\to\infty}\mathcal T\prod_{a=0}^{N-1}P(t_{a+1})\expo^{-\imag P(t_a)H(t_a)P(t_a)(t_{a+1}-t_{a})}\!\!,
\end{align*} 
where $\mathcal T$ is the time-ordering operator. From \eqref{redem}, one has $\ket{\dot\psi}=[\dot P-\imag PHP]\ket{\psi}$ and thus
\begin{align*}
U_{\text{red}}(t)=\mathcal T\expo^{\int_0^t[\dot P(t')-\imag P(t')H(t')P(t')]\diff t'}.
\end{align*}
Now, introducing a field of orthonormal frames $\mathcal V(t)$, the reduced equation of motion \eqref{redem} amounts to the system of $n$ equations $\bk{v_j}{\diff\psi}=-\imag\bra{v_j}H\ket{\psi}\diff t$, i.e.\ to the matrix equation $(\mathscr D+\imag\rep \Omega\diff t)\rep\Psi=0$ with $\rep\Psi$ the column vector of $\ket{\psi}$ and $\rep\Omega=(\bra{v_j}H\ket{v_k})_{jk}$. The matrix of $U_{\text{red}}(t)$ in the bases $(\ket{v_j(0)})_j$ and $(\ket{v_j(t)})_j$ is thus \cite{Kwon,Mostafazadeh1998}
\begin{align}
\rep U_{\text{red}}(t)=\mathcal T\expo^{\imag\int_{0}^t[\rep A(t')-\rep\Omega(t')]\diff t'},\label{melange}
\end{align} 
where $\rep A(t)=(\bk{v_j(t)}{\dot v_k(t)})_{jk}$ represents $\mathcal A$ with respect to the time-parameterization along $\mathcal C$. The presence of $\rep A(t)$ can be interpreted as an inertia effect which cancels out in a parallelly transported frame of reference. 

In the case $n=1$, the geometry (inertia) and the (remaining) dynamics give naturally rise to two distinct contributions since the reduced evolution operator is simply
\begin{align*}
U_{\text{red}}(t)=\Gamma[\mathcal C_t]\expo^{-\imag\int_0^t \bra{v(t')} H(t')\ket{v(t')}\diff t'},
\end{align*}
where $\mathcal C_t$ is the portion of $\mathcal C$ traced out from $t=0$ to $t=t$. The last phase factor contains the dynamical phase. When the path is closed, the geometric phase \eqref{MukundaSimon} that it induces is the Aharonov-Anandan phase of the loop in the ray space \cite{AA}. In the case $n>1$, the geometrical information appears somewhat \enquote{mixed} with the dynamical one in \eqref{melange}. To disentangle it, one needs to change the representation according to $\rep\Psi(t)\to\rep\Psi'(t)=\rep\Gamma^{-1}[\mathcal C_t]\rep\Psi(t)$ where $\rep\Gamma[\mathcal C_t]$ is the matrix of the transporter in the bases $\mathcal V(0)$ and $\mathcal V(t)$. The new representation $\rep\Psi'$ of $\ket\psi$ verifies the equation $(\diff +\imag\Omega'\diff t)\rep\Psi'=0$, where $\rep\Omega'(t)=\rep\Gamma^{-1}[\mathcal C_t]\rep\Omega(t)\rep\Gamma[\mathcal C_t]$, and one deduces
\begin{align*}
\rep U_{\text{red}}(t)=\rep\Gamma[\mathcal C_t]\Big[\mathcal T\expo^{-\imag\int_{0}^t\rep\Omega'(t')\diff t'}\Big].
\end{align*}
The last time-ordered exponential is the non-Abelian dynamical contribution of the torsion mechanism. The change of representation can be understood as a change of picture \cite{Kobe1985}.

In most cases, we do not have any invariant $n$-plane to our disposal but we can have a quasi-invariantone, i.e.\ an evolutive $n$-plane whose projection operator verifies approximately the Schr\"odinger-von Neumann equation. It yields approximative effective evolution operators \eqref{melange}. The paradigmatic example of such a situation is the adiabatic regime where $H$ varies slowly in time. If $P(t)$ is the projection operator into an $n$-fold degenerate level of $H(t)$ then, on the one hand, $P(t)$ obviously commutes with $H(t)$, and, on the other hand, $\dot P(t)\approx 0$ if the adiabatic assumptions are verified \cite{Teufel}. Hence, within these assumptions, $P(t)$ verifies approximately the Schr\"odinger-von Neumann equation and the interesting point is that $\rep\Omega(t)$ is simply the scalar matrix $E(t)\rep I_n$, where $E(t)$ is the energy of the eigenlevel:
\begin{align*}
U_{\text{ad}}(t)=\Gamma[\mathcal C_t]\expo^{-\imag\int_0^tE(t')\diff t'}.
\end{align*}
Therefore, the geometric and dynamical contributions are uncoupled. When the path is closed, $\Gamma[\mathcal C_t]$ is the Berry phase factor \cite{Berry1984} in the Abelian case $n=1$ and the Wilczek-Zee one \cite{WZ} in the non-Abelian case $n>1$.

\section{Application to a three-state system}\label{sec:3}

\subsection{The degeneracy conditions}\label{degeneracycondition}

Let us consider a quantum system whose state space is three-dimensional. In some fixed orthonormal basis $\mathscr B=\{\ket{e_1},\ket{e_2},\ket{e_3}\}$, its Hamiltonian $H$ is represented by a self-adjoint matrix having the general form
\begin{align}
\rep H=\begin{pmatrix}
a & \gamma & \beta^* \\ \gamma^* & b & \alpha \\
\beta & \alpha^* & c \\
\end{pmatrix},\label{Hmatrix}
\end{align}
where $a,b,c$ (resp.\ $\alpha,\beta,\gamma$) are real (resp.\ complex) numbers. Our first aim is to answer the following question: under which condition(s) on these coefficients does $H$ admit a degenerate spectrum? Translating rigidly the whole spectrum if necessary, we will first suppose that $H$ is traceless, i.e.\ that $a+b+c=0$. We will further assume that at least two of the numbers $\alpha,\beta,\gamma$ are nonzero otherwise the study is trivial. In particular, the sought spectrum will necessary contain an eigenvalue $E_1$ of multiplicity 1 and another one, $E_2$, of multiplicity 2. By the traceless assumption, they are such that $E_1+2E_2=0$.

Fundamentally, a twofold degenerate eigenvalue $E_2$ is the requirement that its associated eigenspace is of dimension 2, or that the kernel of $E_2\openone- H$ is of that dimension. Since the underlying vector space is three-dimensional, this property amounts to saying that $E_2\openone- H$ is of rank $3-2=1$, or equivalently that all its $2\times 2$ minors vanish \cite{Mostafazadeh1997}. By the Hermitianness of $E_2\openone-H$, the latter characterization is equivalent to the system of 6 equations
\begin{align}
\begin{aligned}
\alpha^*(E_2-a)+\beta\gamma&=0,\\
\beta^*(E_2-b)+\gamma\alpha&=0,\\
\gamma^*(E_2-c)+\alpha\beta&=0,\\
(E_2-a)(E_2-b)-\module{\gamma}^2&=0,\\
(E_2-b)(E_2-c)-\module{\alpha}^2&=0,\\
(E_2-c)(E_2-a)-\module{\beta}^2&=0.\label{system}
\end{aligned}
\end{align}
The three first lines show that none of the numbers $\alpha,\beta,\gamma$ could be zero otherwise another one would be so, in contradiction with our assumptions. Indeed, if for example $\alpha$ were zero then the first line would imply that $\beta$ or $\gamma$ vanishes as well, etc. The three first lines are thus equivalent to
\begin{align}
E_2=a-\frac{\beta\gamma}{\alpha^*}=b-\frac{\gamma\alpha}{\beta^*}=c-\frac{\alpha\beta}{\gamma^*}\,,\label{1-3}
\end{align}
and impose in particular the reality of the product $\alpha\beta\gamma$. Then, the fourth line becomes
\begin{align*}
\frac{\beta\gamma}{\alpha^*}\frac{\gamma\alpha}{\beta^*}=\module{\gamma}^2\iff \frac{\alpha\beta\gamma}{(\alpha\beta\gamma)^*}=1\iff\alpha\beta\gamma\in\mathbb R-\{0\},
\end{align*}
and imposes no new constraint. The two remaining lines yield the same result and the problem is solved: $H$ has a twofold degenerate eigenvalue $E_2$ iff the product $\alpha\beta\gamma$ has a nonzero real value whereas $a,b,c$ are given by
\begin{align}
 a=E_2+\frac{\beta\gamma}{\alpha^*}\;\;\;,\;\;\; b=E_2+\frac{\gamma\alpha}{\beta^*}\;\;\;,\;\;\; c=E_2+\frac{\alpha\beta}{\gamma^*}\,.\label{abc}
\end{align}
An alternative derivation of these conditions is presented in appendix \ref{app:alternative}. Summing the three expressions of $E_2$ in~\eqref{1-3}, and using the traceless assumption, one obtains the further expression
\begin{align}
E_2=-\frac13\bigg(\frac{\beta\gamma}{\alpha^*}+\frac{\gamma\alpha}{\beta^*}+\frac{\alpha\beta}{\gamma^*}\bigg)\label{E2}
\end{align}
which, in turn, allows to write $a,b,c$ as functions of $\alpha,\beta,\gamma$ alone. Finally, from
\begin{align}
E_1-E_2=\bigg(\frac{1}{\module{\alpha}^2}+\frac{1}{\module{\beta}^2}+\frac{1}{\module{\gamma}^2}\bigg)\alpha\beta\gamma,\label{DeltaE}
\end{align}
one deduces that $E_2$ is the lowest eigenvalue if $\alpha\beta\gamma>0$ and the highest one otherwise.

Under the degeneracy conditions derived above, it is clear from \eqref{abc} and \eqref{E2} that a homogeneous rescaling $(\alpha,\beta,\gamma)\to(\kappa\alpha,\kappa\beta,\kappa\gamma)$, with $\kappa\ne 0$, induces a rescaling $H\to\kappa H$ leaving invariant the eigenspaces of $H$. It is thus judicious to introduce the nonzero real coefficient $C$, the two positive real numbers $r,s$, and the two angles $\vartheta,\varphi$, such that
\begin{align*}
\alpha=C r\expo^{\imag \vartheta}\quad,\quad \beta=C s\expo^{-\imag\varphi}\quad,\quad \gamma=C \expo^{\imag( \varphi-\vartheta)}.
\end{align*}
One easily verifies that the Hamiltonian $H$ decomposes as $U_0(\vartheta,\varphi)H_0(r,s)U_0^\dag(\vartheta,\varphi)$ where $U_0(\vartheta,\varphi)$ is the unitary operator
\begin{align*}
U_0(\vartheta,\varphi)=\expo^{\imag\varphi}\kb{e_1}{e_1}+\expo^{\imag\vartheta}\kb{e_2}{e_2}+\kb{e_3}{e_3}
\end{align*}
and $H_0(r,s)$ is the self-adjoint operator whose matrix in the basis $\mathscr B$ takes the simple real form
\begin{align}
\rep H_0(r,s)=C\begin{pmatrix}
s/r & 1 & s \\ 1 & r/s & r \\ s & r & rs
\end{pmatrix}+E_2(r,s)\,\rep I_3\label{realform}
\end{align}
which remains true even if the traceless assumption is dropped. Hereafter, $C$ will be taken positive (so that the degenerate level is the ground one) and equal to 1 (unit of energy).\footnote{Recall that multiplying the Hamiltonian by a positive function of the time amounts to redefining the time.} After some basic algebra, one finds that the excited level of $H_0(r,s)$ is spanned by the unit vector
\begin{align}
\ket{1^0(r,s)}=\frac{rs}{\mathcal N_1(r,s)}\Big(r^{-1}\ket{e_1}+s^{-1}\ket{e_2}+\ket{e_3}\Big),\label{excitage}
\end{align}
where $\mathcal N_1(r,s)=\sqrt{r^2+s^2+r^2s^2}$, while the ground one is spanned by the orthonormal basis formed by
\begin{align}
\begin{aligned}
\ket{2^0_1(r,s)}&=\frac{r\ket{e_1}-\ket{e_3}}{\sqrt{1+r^2}}\,,\\
\ket{2_2^0(r,s)}&=\frac{r\ket{e_1}-(1+r^2)s\ket{e_2}+r^2\ket{e_3}}{\mathcal N_1(r,s)\sqrt{1+r^2}}\,.
\end{aligned}\label{basis}
\end{align}
Note that the Hamiltonian is invariant under a simultaneous exchange $r\leftrightarrow s$, $\vartheta\leftrightarrow\varphi$, $\ket{e_1}\leftrightarrow\ket{e_2}$. Its eigenspaces are invariant as well. 

In what follows, we will realize transports with the ground level. The whole parameter space of the problem is formed by the points $x$ coordinatized by the two positive quantities $r$, $s$ and the two angles $\vartheta$, $\varphi$. We will denote by $V_2(x)$ the ground eigenspace at $x=(r,s,\vartheta,\varphi)$ and by $P_2(x)$ the projection operator into $V_2(x)$. If $V_0(r,s)$ is the ground eigenspace of $H_0(r,s)$ then one has obviously $V(x)=U_0(\vartheta,\varphi)(V_0(r,s))$ and thus
\begin{align*}
P_2(x)=U_0(\vartheta,\varphi)P_0(r,s)U_0^\dag(\vartheta,\varphi),
\end{align*}
where $P_0(r,s)$ is the projection operator into $V_0(r,s)$. The excited eigenspace of $H(x)$ is spanned by $\ket{1(x)}=U_0(\vartheta,\varphi)\ket{1^0(r,s)}$ and the ground one by the basis $\ket{2_j(x)}=U_0(\vartheta,\varphi)\ket{2_j^0(r,s)}$ ($j=1,2$). Note that the operators $U_0(\vartheta,\varphi)$ verify 
\begin{align*}
U_0^\dag(\vartheta,\varphi)U_0(\vartheta',\varphi')=U_0(\vartheta'-\vartheta,\varphi'-\varphi).
\end{align*} 
To simplify the study, we will limit ourselves to adiabatic parallel transports at $r$ and $s$ constants, i.e.\ on a given torus $(r,s)$ coordinatized by $\vartheta$ and $\varphi$.

\subsection{Teleparallelism and geodesics with the ground degenerate level}

Let us consider two given points $x_0=(r,s,\vartheta_0,\varphi_0)$ and $x'_0=(r,s,\vartheta'_0,\varphi'_0)$ on the torus $(r,s)$. We will use the shortened notations $\ket{1}\equiv\ket{1(x_0)}$, $\ket{1'}=\ket{1(x'_0)}$, $\ket{2_j}=\ket{2_j(x_0)}$ and $\ket{2'_j}=\ket{2_j(x'_0)}$ ($j=1,2$). We will set $E_2(r,s)=0$ thus $E_1(r,s)=\tfrac{r}{s}+\frac{s}{r}+rs=\tfrac{\mathcal N_1^2}{rs}$. Now, let us determine the teletransporter from $V'_2=V_2(x'_0)$ to $V_2=V_2(x_0)$. Clearly, $V_2$ and $V'_2$ intersect each other and it follows that the projection operator $\Pi_{V_2V'_2}$ admits $\sigma_+=1$ as singular value. Besides $\sigma_+^2=1$, the map $\Pi_{V'_2V_2}\Pi_{V_2V'_2}$ admits the eigenvalue
\begin{align*}
\sigma_{-}^2&=\tr\big(P_2(x_0)P_2(x'_0)\big)-\sigma_+^2=\module{\bk{1}{1'}}^2
\end{align*}
(see appendix \ref{hyperplanes}). The two principal angles between $V_2(x_0)$ and $V_2(x'_0)$ are thus $\phi_+=0$ and
\begin{align*}
\phi_-=\arccos\module{\bk{1}{1'}}=\arcsin\bigg(\frac{rs\mathcal N_2}{\mathcal N_1^2}\bigg)
\end{align*}
with $\mathcal N_1=\mathcal N_1(r,s)$ and
\begin{align*}
\mathcal N_2=2\sqrt{r^2\sin^2\frac{\Delta\vartheta}{2}+s^2\sin^2\frac{\Delta\varphi}{2}+\sin^2\bigg(\frac{\Delta\vartheta-\Delta\varphi}{2}\bigg)},
\end{align*}
where $\Delta\varphi=\varphi'_0-\varphi_0$ and $\Delta\vartheta=\vartheta'_0-\vartheta_0$. The two spaces $V_2$ and $V'_2$ decompose as the orthogonal sums $V_2=V_+\oplus V_-$ and $V'_2=V'_+\oplus V'_-$ with $V_+=V_2\cap V'_2=V'_+$ spanned by the unit vector
\begin{align*}
\ket{+}=\frac{\bk{1'}{2_2}\ket{2_1}-\bk{1'}{2_1}\ket{2_2}}{\sin\phi_-}\,.
\end{align*}
Then, $V_-$ and $V_-'$ are respectively spanned by the unit vectors
\begin{align*}
\ket{-}&=\frac{P_2(x_0)\ket{1'}}{\|P_2(x_0)\ket{1'}\|}=\frac{\bk{2_1}{1'}\ket{2_1}+\bk{2_2}{1'}\ket{2_2}}{\sin\phi_-}\,,\\
\ket{-'}&=\frac{P_2(x'_0)\ket{1}}{\|P_2(x'_0)\ket{1}\|}=\frac{\bk{2'_1}{1}\ket{2'_1}+\bk{2'_2}{1}\ket{2'_2}}{\sin\phi_-}\,.
\end{align*}
It is shown in appendix \ref{app:complex} that $V_2$ and $V'_2$ are always anti-orthogonal except possibly for one or two \enquote{accidental} values (mod $2\pi$) of the couples $(\Delta\vartheta,\Delta\varphi)$. In the overwhelming majority of cases where $V_2$ and $V'_2$ are anti-orthogonal, $\sigma_-$ is the other singular value of $\Pi_{V_2V'_2}$ and the teletransporter from $V'_2$ to $V_2$ is
\begin{align}
\Gamma_{V_2V'_2}=\kb{+}{+}+\expo^{\imag\delta_-}\kb{-}{-'}\,,\label{GPP}
\end{align}
where $\delta_-$ is the Pancharatnam phase between $\ket{-'}$ and $\ket{-}$. By construction, one has $\bk{-}{-'}=-\bk{1'}{1}$ and thus
\begin{align*}
\delta_-=\pi+\arg\bk{1'}{1}=\pi-\arg\big(1+r^{-2}\expo^{\imag\Delta\varphi}+s^{-2}\expo^{\imag\Delta\vartheta}\big).
\end{align*}

\begin{figure}
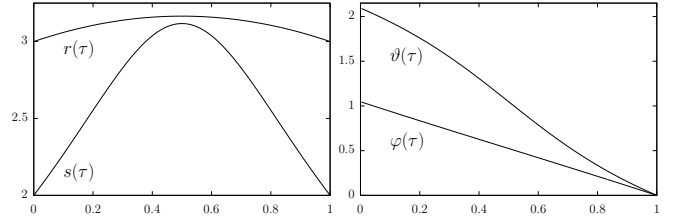

    \centering
    \resizebox{.49\columnwidth}{!}{\input{geodesicrs.tex}}
    \resizebox{.49\columnwidth}{!}{\input{geodesicpt.tex}}
    \caption{Plots of the coordinates of the shortest geodesic $V(x(\tau))$ from $V'_2=V(0)$ to $V_2=V(1)$ for $r=3$, $s=2$, $\vartheta_0=0=\varphi_0$, $\vartheta'_0=\tfrac{2\pi}{3}$ and $\varphi'_0=\tfrac{\pi}{3}$.}\label{geodesicsrspt}
\end{figure}

According to the general theory discussed in subsection \ref{trivialholo}, the shortest geodesic $\mathcal C_{\text{geo}}\colon\tau\mapsto V(\tau)$ in the Grassmann manifold of $2$-planes of $\mathcal H$, from $V'_2=V(0)$ to $V_2=V(1)$, is such that $V(\tau)=\spn\{\ket{+},\ket{-(\tau)}\}$ with 
\begin{align*}
\ket{-(\tau)}=\frac{\sin(\phi_-(1-\tau))\ket{-'}+\sin(\phi_-\tau)\expo^{\imag\delta_-}\ket{-}}{\sin\phi_-}\,.
\end{align*} 
Thanks to the equality $\bk{-}{1'}=\bk{-'}{1}$, it is clear that the vector
\begin{align*}
\ket{n(\tau)}=\sin(\phi_-(1-\tau))\ket{1'}-\sin(\phi_-\tau)\expo^{-\imag\delta_-}\ket{1}
\end{align*}
is orthogonal to $V(\tau)$. Its decomposition in the basis $\mathscr B$ reads 
\begin{align*}
\ket{n(\tau)}=\frac{A_1(\tau)}{r}\,\ket{e_1}+\frac{A_2(\tau)}{s}\,\ket{e_2}+A_3(\tau)\ket{e_3},
\end{align*}
with
\begin{align*}
A_1(\tau)&=\sin(\phi_-(1-\tau))\expo^{\imag\varphi'_0}+\sin(\phi_-\tau)\expo^{\imag(\varphi_0+\delta)},\\
A_2(\tau)&=\sin(\phi_-(1-\tau))\expo^{\imag\vartheta'_0}+\sin(\phi_-\tau)\expo^{\imag(\vartheta_0+\delta)},\\
A_3(\tau)&=\sin(\phi_-(1-\tau))+\sin(\phi_-\tau)\expo^{\imag\delta},
\end{align*}
and
\begin{align*}
\delta=\arg\bk{1}{1'}=\arg\big(1+r^{-2}\expo^{\imag\Delta\varphi}+s^{-2}\expo^{\imag\Delta\vartheta}\big).
\end{align*}
The quantities $A_i(\tau)$ are nonzero for $\tau\ne\tfrac{1}{2}$ whatever the constants $r$, $s$, $\vartheta_0$, $\vartheta'_0$, $\varphi_0$ and $\varphi'_0$ be. They are also all nonzero at $\tau=\tfrac{1}{2}$ iff $\delta$ is distinct from $0$, $\Delta\vartheta$, and $\Delta\varphi$ (mod $2\pi$). We will suppose that this is the case. Hence, it suffices to have a look at \eqref{excitage} to see that $V(\tau)$ is orthogonal to $\ket{1(x(\tau))}$ for $x(\tau)=(r(\tau),s(\tau),\vartheta(\tau),\varphi(\tau))$ with
\begin{align*}
r(\tau)=r\bigg\lvert\frac{A_3(\tau)}{ A_1(\tau)}\bigg\rvert\;,&& \varphi(\tau)&=\arg\bigg(\frac{A_1(\tau)}{A_3(\tau)}\bigg), \\ s(\tau)=s\bigg\lvert\frac{A_3(\tau)}{ A_2(\tau)}\bigg\rvert
\;,&&\vartheta(\tau)&=\arg\bigg(\frac{A_2(\tau)}{A_3(\tau)}\bigg).
\end{align*}
Consequently, if we suppose that $\delta\ne 0,\Delta\vartheta,\Delta\varphi$ (mod $2\pi$), the shortest geodesic $V(\tau)$ is realized by the above path $x(\tau)$ in the parameter space: $V(\tau)=V_2(x(\tau))$. We have shown that, in general, the shortest geodesic in the Grassmann manifold coincides with the shortest geodesic in the parameter space (but it is no more the case if we restrict the parameter space to the torus). Remark that $r(\tau)$ and $s(\tau)$ are even-symmetric while $\varphi(\tau)$ and $\vartheta(\tau)$ are odd-symmetric: 
\begin{align*}
r(\tau)=r(1-\tau)\;,&& \varphi(\tau)+\varphi(1-\tau)&=\varphi_0+\varphi'_0\,,\\
s(\tau)=s(1-\tau)\;,&& \vartheta(\tau)+\vartheta(1-\tau)&=\vartheta_0+\vartheta'_0\,,
\end{align*}
and that $r(\tau)$ and $s(\tau)$ reach at $\tau=\tfrac{1}{2}$ the extrema
\begin{align*}
r\big(\tfrac{1}{2}\big)=r\bigg\lvert\frac{\cos\big(\frac{\delta}{2}\big)}{\cos\big(\frac{\delta-\Delta\varphi}{2}\big)}\bigg\rvert\;,\quad s\big(\tfrac{1}{2}\big)=s\bigg\lvert\frac{\cos\big(\frac{\delta}{2}\big)}{\cos\big(\frac{\delta-\Delta\vartheta}{2}\big)}\bigg\rvert.
\end{align*}
As an illustration are plotted in figure \ref{geodesicsrspt} the coordinates of $x(\tau)$ for a given choice of the constants.

\subsection{Non-Abelian adiabatic holonomies}\label{abelianholo}

\begin{figure}
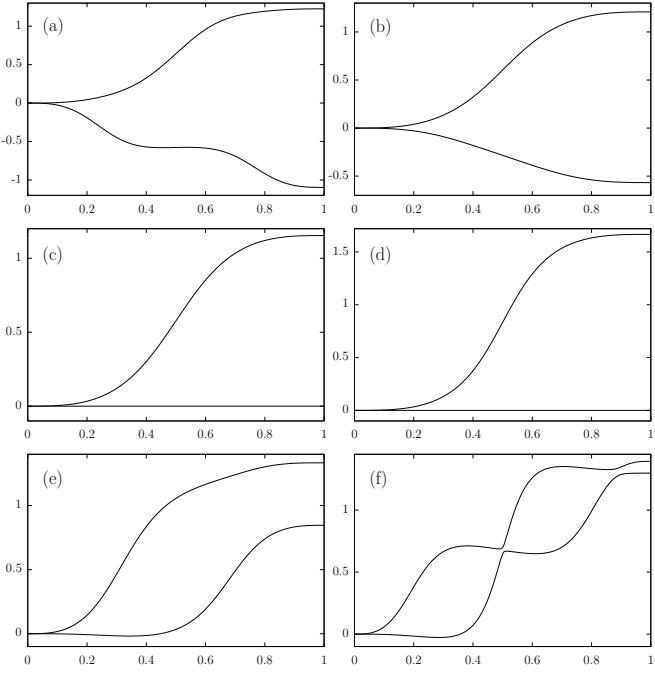

    \centering
    \resizebox{.49\columnwidth}{!}{\input{phases2m4.tex}}
    \resizebox{.49\columnwidth}{!}{\input{phases2m2.tex}}
    \resizebox{.49\columnwidth}{!}{\input{phases20.tex}}
    \resizebox{.49\columnwidth}{!}{\input{phases22.tex}}
    \resizebox{.49\columnwidth}{!}{\input{phases24.tex}}
    \resizebox{.49\columnwidth}{!}{\input{phases26.tex}}
    \caption{Plots of the two geometric phases of the holonomies associated with the degenerate eigenspace along the toroidal helix $(r,s,\omega_\vartheta t,\omega_\varphi t)$, $\omega_\vartheta>0$, as a function of $\frac{\omega_\vartheta t}{2\pi}\in[0,1]$, for $r=3$, $s=2$, and $\omega_\varphi=$ (a) $-2\omega_\vartheta$, (b) $-\omega_\vartheta$, (c) $0$, (d) $\omega_\vartheta$, (e) $2\omega_\vartheta$ and (f) $3\omega_\vartheta$. The holonomies are non-cyclic as long as $\omega_\vartheta t <2\pi$ and cyclic at $\omega_\vartheta t=2\pi$. Since $r^{-2}+s^{-2}<1$, the degenerate eigenspaces are all anti-orthogonal to each other (see the appendix \ref{app:complex}).}\label{plots}
\end{figure}

Now, let us realize continuous parallel transports of the ground degenerate eigenspace on the torus $(r,s)$ over which the gauge field $\mathcal A$ decomposes as $\rep A_\vartheta\diff\vartheta+\rep A_\varphi\diff\varphi$ with\footnote{Remark that $\imag\bk{1(x)}{\partial_\vartheta 1(x)}$ and $\imag\bk{1(x)}{\partial_\varphi 1(x)}$ are functions of $r$ and $s$ only. Therefore, on the torus, $\imag\bk{1(x)}{\diff 1(x)}$ is a pure gauge and cyclic continuous holonomies of the non-degenerate level are trivial (indeed, $\imag\bk{1(x)}{\diff 1(x)}$ is smoothly defined over the torus).}
\begin{align*}
\rep A_\vartheta(x)&=\imag\!\big[\bk{2_j(x)}{\partial_\vartheta 2_k(x)}\big]_{jk}=-\frac{(1+r^2)s^2}{\mathcal N_1^2}\begin{pmatrix}
0 & 0\\
0 & 1
\end{pmatrix},\\
\rep A_\varphi(x)&=\imag\big[\bk{2_j(x)}{\partial_\varphi 2_k(x)}\big]_{jk}\!=-\frac{r^2}{(1+r^2)}\!\begin{pmatrix}
1& \mathcal N_1^{-1}\\
\mathcal{N}_1^{-1} & \mathcal N_1^{-2}
\end{pmatrix}
\end{align*}
As a case study, consider a toroidal helix evolving in time according to $x(t)=(r,s,\vartheta_0+\omega_\vartheta t,\varphi_0+\omega_\varphi t)$ where $\omega_\varphi$ and $\omega_\vartheta$ are constant frequencies.\footnote{From $P_2(t)=U_0(t)P_0U_0^\dag(t)$ with $P_0\equiv P_0(r,s)$, one derives $\dot P_2(t)=\imag U_0(t)[H_1,P_0]U_0^\dag(t)$ where $H_1=-\imag U_0^\dag \dot U_0$ is a constant operator (that will be used in the subsequent subsection). It is clear that $\dot P_2^2(t)=-\imag U_0(t)[H_1,P_0]^2U_0^\dag(t)$ has a constant trace. Hence, $t$ is an affine parameter along the curve traced out in $\text{Gr}(2,\mathcal H)$.} It degenerates into a circular arc if one of the frequencies vanishes. Along the helix, the gauge field is represented by the constant matrix field $\rep A=\rep A_\vartheta\omega_\vartheta+\rep A_\varphi\omega_\varphi$. By the choice of the zero of the energies, the adiabatic evolution operator of the ground level coincides with the transporter along the curve $\mathcal C^{\text{ad}}_t$ developed in $\text{Gr}(2,\mathcal H)$ from $t=0$ to $t=t$. Their matrix in the bases $(\ket{2_1},\ket{2_2})$ and $(\ket{2_1(t)},\ket{2_2(t)})$, is
\begin{align}
\rep U^{\text{ad}}(t)=\rep \Gamma[\mathcal C^{\text{ad}}_t]=\expo^{\imag\rep A t}=\expo^{-\imag\zeta t} \rep R_{\mathbf{\hat u}}(\Omega t),\label{holonomyhelix}
\end{align}
where $\rep R_{\mathbf{\hat{u}}}(\Omega t)=\expo^{-\imag\frac{\Omega t}{2}\boldsymbol\sigma\cdot\mathbf{\hat u}}$ is the usual rotation matrix \cite{Cohen} of a spin $\tfrac{1}{2}$ by an angle $\Omega t$ around a unit three-vector $\mathbf{\hat u}$, with
\begin{align*}
\zeta&=\frac{s^2(1+r^2)\omega_\vartheta+r^2(1+s^2)\omega_\varphi}{2\mathcal N_1^2}\,,\\
\mathbf{\hat u}&=\frac{2r^2\mathcal N_1\omega_\varphi\,\mathbf{\hat i}+\big[r^2(\mathcal N^2-1)\omega_\varphi-(1+r^2)^2s^2\omega_\vartheta\big]\,\mathbf{\hat k}}{(1+r^2)\mathcal N_3},\\
\mathcal N_3&=\sqrt{\big[s^2(1+r^2)\omega_\vartheta-r^2(1+s^2)\omega_\varphi\big]^2+4r^2s^2\omega_\vartheta\omega_\varphi}\,,\\
\Omega&=\frac{\mathcal N_3}{\mathcal N_1^2}\,.
\end{align*}
If the helix is closed at $t=T$ then $\rep\Gamma[\mathcal C^{\text{ad}}_T]$ is the matrix of the holonomy in a unique basis: there is no twist since \eqref{basis} defines a smooth frame all over the torus ($\mathcal A$ is smoothly defined over it). In this case, the two geometric phases are clearly $(\pm\tfrac{\Omega}{2}-\zeta)T$. Otherwise, the non-cyclic holonomy at any instant $t$ is $\Gamma_{\text h}[\mathcal C^{\text{ad}}_t]=\Gamma_{V_2V_2(t)}\Gamma[\mathcal C^{\text{ad}}_t]$. The teletransporter $\Gamma_{V_2V_2(t)}$ is obtained from the previous subsection by taking $\vartheta'_0=\vartheta(t)$ and $\varphi'_0=\varphi(t)$, and the quantities $\ket{+}$, $\ket{-}$, $\ket{-'}$, $\delta_-$ are now time-dependent. To determine the two geometric phases $\gamma_\pm[\mathcal C^{\text{ad}}_t]$ associated with the holonomy, it suffices to find the determinant and the trace of $\Gamma_{\text h}[\mathcal C^{\text{ad}}_t]$. To this end, let us first render more symmetric the expression of the teletransporter term through the introduction of the unit vector
\begin{align*}
\ket{+'(t)}=\frac{\bk{1}{2_2(t)}\ket{2_1(t)}-\bk{1}{2_1(t)}\ket{2_2(t)}}{\sin\phi_-(t)}\,.
\end{align*}
It is obviously parallel to $\ket{+(t)}$ and one verifies that $\ket{+'(t)}=\expo^{\imag\delta_+(t)}\ket{+(t)}$ with
\begin{align*}
\delta_+(t)=\pi+(\omega_\vartheta+\omega_\varphi)t.
\end{align*}
Hence, the teletransporter from $V_2(t)$ to $V_2$ takes the form
\begin{align*}
\Gamma_{V_2V_2(t)}=\expo^{\imag\delta_+(t)}\kb{+(t)}{+'(t)}+\expo^{\imag\delta_-(t)}\kb{-(t)}{-'(t)}\,.
\end{align*}
Since, by construction, $\bk{2_j(t)}{{\pm'}(t)}=\bk{\pm(t)}{2_j}$ as well as $\bk{2_j(t)}{1}=\bk{1(t)}{2_j}$, the matrix of the teletransporter in the bases $(\ket{2_1(t)},\ket{2_2(t)})$ and $(\ket{2_1},\ket{2_2})$ admits the decomposition
\begin{align*}
\rep\Gamma_{V_2V_2(t)}=\rep S(t)\diag(\expo^{\imag\delta_+(t)},\expo^{\imag\delta_-(t)})\,\rep S^\top(t),
\end{align*}
where
\begin{align*}
\rep S(t)&=\begin{pmatrix}
\bk{2_1}{{+}(t)} & \bk{2_1}{{-}(t)} \\ \bk{2_2}{{+}(t)} & \bk{2_2}{{-}(t)}
\end{pmatrix}=\begin{pmatrix}
\mu_0^*(t) & \nu_0(t) \\ -\nu_0^*(t) & \mu_0(t)
\end{pmatrix}
\end{align*}
is a $SU(2)$ matrix whose elements are given by
\begin{align*}
\mu_0(t)&=\bk{2_2}{{-}(t)}=\frac{r^2-(1+r^2)\expo^{\imag\omega_\vartheta t}+\expo^{\imag\omega_\varphi t}}{\mathcal N_2(t)\sqrt{1+r^2}}\,,\\
\nu_0(t)&=\bk{2_1}{{-}(t)}=\frac{\mathcal N_1(\expo^{\imag\omega_\varphi t}-1)}{\mathcal N_2(t)\sqrt{1+r^2}}\,.
\end{align*}
Since $\rep S(t)$ and $\rep R_{\mathbf{\hat u}}(\Omega t)$ are $SU(2)$ matrices, the determinant of the holonomy is $\expo^{\imag(\delta_+(t)+\delta_-(t)-2\zeta t)}$. The geometric phases are thus
\begin{align}
\gamma_\pm[\mathcal C^{\text{ad}}_t]=\pi+\Big(\frac{\omega_\vartheta+\omega_\varphi}{2}-\zeta\Big)t-\xi(t)\pm\chi(t)\label{phasesphi}
\end{align}
where
\begin{align*}
\xi(t)=\frac12\,\arg\bk{1}{1(t)}=\frac12\,\arg\big(1+r^{-2}\expo^{\imag\omega_\varphi t}+s^{-2}\expo^{\imag\omega_\vartheta t}\big)
\end{align*}
and $\chi(t)$ is an angle such that
\begin{align*}
\cos\chi(t)&=\frac12\,\expo^{-\imag\big[\big(\frac{\omega_\vartheta+\omega_\varphi}{2}-\zeta\big) t-\xi(t)\big]}\tr\Gamma_{\text h}[\mathcal C_t].
\end{align*}
After some tedious calculations, one obtains an expression of the form
\begin{align}
\cos\chi(t)=\frac{1}{\mathcal N^2_2(t)}\bigg[C(t)\cos\bigg(\frac{\Omega t}{2}\bigg)+\frac{S(t)}{\mathcal N_3}\,\sin\bigg(\frac{\Omega t}{2}\bigg)\bigg].\label{formule}
\end{align}
The expressions of $C(t)$ and $S(t)$ are reported in appendix \ref{app:CS}. The geometric phases can be plotted by using formula \eqref{formule} but it is simpler to proceed numerically. The general principle of the algorithm is sketched in appendix \ref{app:num} and some cases are illustrated in figure \ref{plots}, the values of $r$ and $s$ being chosen so that all the degenerate eigenspaces over the torus are anti-orthogonal to each other.

We remark on figures \hyperref[plots]{\ref*{plots}(c)} and \hyperref[plots]{\ref*{plots}(d)} that one of the phases is identically zero. The reason for this in the plot \hyperref[plots]{\ref*{plots}(c)} lies in the fact that the state $\ket{2_1}$ remains fixed when $\omega_\varphi=0$, leading to holonomies
\begin{align}
\Gamma_{\text h}[\mathcal C^{\text{ad}}_t]=\kb{2_1}{2_1}+\expo^{\imag\gamma[\mathcal C^{\text{ad}}_t]}\kb{2_2(t)}{2_2}\,,\label{transportphinul}
\end{align}
where
\begin{align}
\gamma[\mathcal C^{\text{ad}}_t]&=\arg\bk{2_2}{2_2(t)}-\Im\int_0^\tau\bk{2_2(t')}{\dot 2_2(t')}\diff t' \label{gammaphinul}\\
&=\arg\Big[r^2+(1+r^2)s^2\expo^{\imag\omega_\vartheta t}\Big]-\frac{(1+r^2)s^2\omega_\vartheta t}{r^2+s^2+r^2s^2}\notag
\end{align}
is the geometric phase associated with the path traced out by $\spn(\ket{2_2(t)})$ in the ray space. It is shown in appendix \ref{app:berry} that one of the phases is identically zero iff $\omega_\varphi=0$ as in figure \hyperref[plots]{\ref*{plots}(c)} or $\omega_\vartheta=0$ or $\omega_\vartheta=\omega_\varphi$ as in \hyperref[plots]{\ref*{plots}(d)}.

Let me end this part of the study by contemplating the interesting case where $r=s$ and $\omega_\varphi+\omega_\vartheta=0$. Here, the geometric phases is reduced to $\gamma_\pm[\mathcal C^{\text{ad}}_t]=\pi-\xi(t)\pm\chi(t)$ where
\begin{align*}
\xi(t)=\frac12\,\arg\big[r^2+2\cos(\omega_\vartheta t)\big]\!=\!\begin{cases}
\frac{\pi}{2} & \text{if $\cos(\omega_0 t)<-\frac{r^2}{2}$}\\ 0 & \text{if $\cos(\omega_0 t)>-\frac{r^2}{2}$}
\end{cases}
\end{align*}
and $\omega_0=\module{\omega_\vartheta}=\module{\omega_\varphi}$. If $\xi(t)=\frac{\pi}{2}$, one finds on the basis of formula \eqref{phasesphi} and \eqref{formule} the geometric phases 
\begin{align}
\gamma_\pm[\mathcal C^{\text{ad}}_t]={}&\pm\arg\Big[1+(1+r^2)\cos(\omega_0 t)+\imag r\sqrt{2+r^2}\Big]\notag\\
&\mp\frac{r\omega_0t}{\sqrt{2+r^2}}\,.\label{geomphases}
\end{align}
If $\xi(t)=\tfrac{\pi}{2}$, a case which necessitates $r=s\leqslant\sqrt{2}$, the geometric phases are constrained to the values $0$ and $\pi$. Now, suppose $r=s<\sqrt 2$ and consider one cycle between 
$t=0$ and $t=\frac{2\pi}{\omega_0}$. There are exactly two values of $t$ in this interval such that $\cos(\omega_0 t)=-\tfrac{r^2}{2}$, viz.\
\begin{align*}
t_1=\frac{1}{\omega_0}\arccos\bigg(\!\!-\frac{r^2}{2}\bigg)\in\bigg(\frac{\pi}{2\omega_0},\frac{\pi}{\omega_0}\bigg)\;\;,\;\; t_2=\frac{2\pi}{\omega_0}-t_1\,.
\end{align*}
In the intervals $[0,t_1)$ and $(t_2,\tfrac{2\pi}{\omega_0}]$, the geometric phases evolve continuously according to \eqref{geomphases} whereas they are frozen to the values $0$ and $\pi$ on $(t_1,t_2)$. The discontinuity points $t_1$ and $t_2$ are precisely the values of $t$ for which $V(t)$ is not anti-orthogonal to $V$. The interval $(t_1,t_2)$ tends to $(\tfrac{\pi}{2\omega_0},\tfrac{3\pi}{2\omega_0})$ in the limit $r\to 0$ and decreases symmetrically until the singleton $\{\tfrac{\pi}{\omega_0}\}$ as $r$ grows up to $\sqrt 2$. Moreover, formula \eqref{geomphases} shows that the phases tend to zero in the limit $r\to 0$ outside of the frozen regime. Hence, in this limit, they are zero outside $(\tfrac{\pi}{2\omega_0},\tfrac{3\pi}{2\omega_0})$ and one of them jumps to the value $\pi$ inside.

\begin{figure}
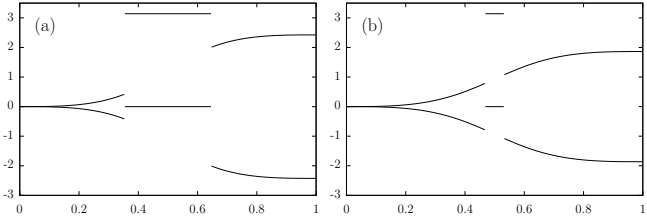

    \centering
    \resizebox{.49\columnwidth}{!}{\input{antiortho1.tex}}
    \resizebox{.49\columnwidth}{!}{\input{antiortho14.tex}}
    \caption{Plots of the two geometric phases of the holonomies associated with the degenerate eigenspace along the toroidal helix $(r,s,\omega_\vartheta t,\omega_\varphi t)$, $\omega_\vartheta>0$, as a function of $\frac{\omega_\vartheta t}{2\pi}\in[0,1]$, for $r=s$, $\omega_\varphi=-\omega_\vartheta$, and (a) $r=1.1$, (b) $r=1.4$. The discontinuities occur at the two values of $t$ for which $V(t)$ is not anti-orthogonal to $V(0)$. Between them, the intermediate geometric phases are blocked to the values $0$ and $\pi$. The central region decreases symmetrically with $r\in(0,\sqrt 2)$ from $(\tfrac{\pi}{2\omega_\vartheta},\tfrac{3\pi}{2\omega_\vartheta})$ to $\{\tfrac{\pi}{\omega_\vartheta}\}$.}\label{plots2}
\end{figure}

\subsection{Beyond the adiabatic limit}\label{beyond}

In this last part, we again consider the toroidal helices on the torus $(r,s)$ but we now drop the adiabatic hypothesis. Passing to the \enquote{rotating frame} through the change of picture
\begin{align*}
\ket{\psi(t)}&\longrightarrow\ket{\psi'(t)}=U_0^\dag(t)\ket{\psi(t)}\\
H(t)&\longrightarrow H'=\underbrace{U_0^\dag(t)H(t)U_0(t)}_{H_0}\underbrace{-\imag U_0^\dag(t)\dot U_0(t)}_{H_1}\\
\end{align*}
transforms the problem into a time-independent one ruled by the Hamiltonian $H'=H_0+H_1$ with
\begin{align*}
H_0\equiv H_0(r,s)\quad\text{and}\quad H_1=\omega_\varphi\kb{e_1}{e_1}+\omega_\vartheta\kb{e_2}{e_2}.
\end{align*}
Suppose that $\omega_\varphi$ and $\omega_\vartheta$ are treated as small parameters. Applying the standard Schr\"odinger-Rayleigh perturbation theory \cite{Ballentine}, we obtain the first order approximation of the eigenvalues of $H'$ and the zeroth order approximation of the corresponding eigenvectors. Then, we form the approximate evolution operator in the \enquote{rotating picture} and we go back to the initial one. The resulting evolution operator is precisely the adiabatic one, as it should be. Higher order expansions in the frequencies would bring more accurate superadiabatic approximations \cite{Berry1987}. However, since the dimension is 3, one can solve exactly the eigenproblem of $H'$. Let $\varepsilon_1,\varepsilon_2,\varepsilon_3$ be the eigenvalues of $H'$ and $(\ket{\varepsilon_1},\ket{\varepsilon_2},\ket{\varepsilon_3})$ be an eigenbasis such that
\begin{align*}
H'\ket{\varepsilon_k}=\varepsilon_k\ket{\varepsilon_k}\qquad(k=1,2,3).
\end{align*} 
The eigenvalues $\varepsilon_k$ are necessarily distinct as soon as at least one of the frequencies $\omega_\vartheta$ and $\omega_\varphi$ is nonzero. The solution of the eigenproblem of $H'$ yields three fundamental solutions of the Schr\"odinger equation in the initial picture:
\begin{align}
\ket{\alpha_k(t)}=\expo^{-\imag\varepsilon_k t}U_0(t)\ket{\varepsilon_k}.\label{lesalpha}
\end{align}
Applying formula \eqref{MukundaSimon}, each ray $\spn(\ket{\alpha_k(t)})$ accumulates a geometric phase $-\bra{\varepsilon_k}H_1\ket{\varepsilon_k}t$ along its motion in the ray space. In particular, if the evolution of the angles is cyclic on the torus, with a period $T$, then the states $U_0(t)\ket{\varepsilon_k}$ form a Floquet basis of the problem and the rays pick up an Aharonov-Anandan phase $-\bra{\varepsilon_k}H_1\ket{\varepsilon_k}T$ along a cycle in addition to a dynamical phase $-\bra{\varepsilon_k}H_0\ket{\varepsilon_k}T$. The sum of these two phases is $-\varepsilon_k T$ as it should be from \eqref{lesalpha}. 

Now, let us focus on the comparison between the exact treatment and the adiabatic one which was carried out in the previous subsection. To this end, we need to determine the two solutions $\ket{\psi_{21}(t)}$ and $\ket{\psi_{22}(t)}$ generated by the initial conditions $\ket{\psi_{2j}(0)}=U_0(0)\ket{2_j^0}=\ket{2_j}$ ($j=1,2$). Since we are mostly interested in the geometric quantities, it is then relevant to determine the Fubini-Study distance (see footnote \ref{Stiefel})
\begin{align*}
d_{\text{FS}}(t)=\arccos\left|\det\begin{pmatrix}
\bk{\psi_{21}(t)}{2_1(t)} & \bk{\psi_{21}(t)}{2_2(t)} \\
\bk{\psi_{22}(t)}{2_1(t)} & \bk{\psi_{22}(t)}{2_2(t)} 
\end{pmatrix}\right|
\end{align*} 
between the exact invariant 2-plane $V(t)$ spanned by $\ket{\psi_{21}(t)}$, $\ket{\psi_{22}(t)}$ and the adiabatic 2-plane $V_2(t)$. Other quantities of interest are the differences between the geometric phases accumulated along the evolutions of $V(t)$ and $V_2(t)$ in $\text{Gr}(2,\mathcal H)$.

\begin{figure}
\centering
\resizebox{.95\columnwidth}{!}{\input{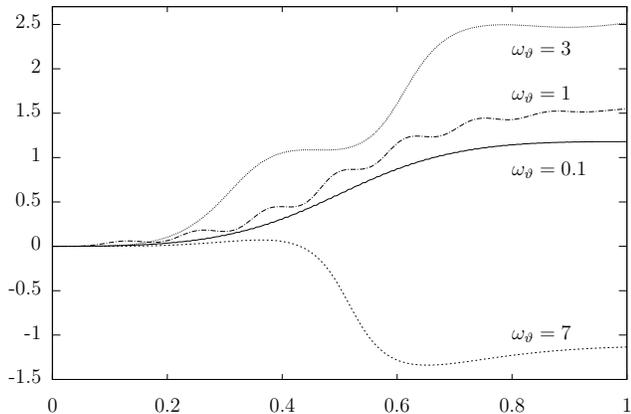}}
\caption{Plots of the geometric phase accumulated by $\ket{\psi_{22}(\tau)}$ as a function of $\tau=\frac{\omega_\vartheta t}{2\pi}\in[0,1]$ for some values of $\omega_\vartheta>0$, with $r=3$, $s=2$ and $\omega_\varphi=0$. In the limit $\omega_\vartheta\to 0$, it converges to the adiabatic geometric phase in figure \hyperref[plots]{\ref*{plots}(c)}.}\label{fig:phinul}
\end{figure}

\begin{figure}
\centering
\resizebox{.95\columnwidth}{!}{\input{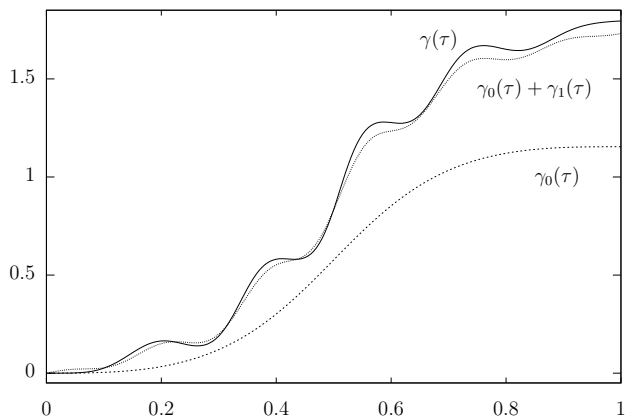}}
\caption{Comparison between the exact geometric phase $\gamma(\tau)$ accumulated by $\ket{\psi_{22}(\tau)}$ (solid line), the adiabatic phase $\gamma_0(\tau)$ and the first order superadiabatic phase $\gamma_0(\tau)+\gamma_1(\tau)$ for $\tau=\frac{\omega_\vartheta t}{2\pi}\in[0,1]$, $r=3$, $s=2$ and $\omega_\vartheta=1.5$. }\label{corrections}
\end{figure}

From the characteristic polynomial
\begin{align*}
\det(\lambda\openone-H')=\lambda^3-a_2\lambda^2+a_1\lambda-a_0
\end{align*}
of $H'$, with
\begin{align*}
a_2&=\frac{\mathcal N_1^2}{rs}+\omega_\vartheta+\omega_\varphi\,,\\
a_1&=\bigg(rs+\frac{s}{r}\bigg)\omega_\vartheta+\bigg(rs+\frac{r}{s}\bigg)\omega_\varphi+\omega_\vartheta\omega_\varphi\,,\\
a_0&=rs\omega_\vartheta\omega_\varphi\,,
\end{align*}
we see that 0 is an eigenvalue of $H'$ iff $\omega_\vartheta=0$ or $\omega_\varphi=0$. Furthermore, if $\omega_\vartheta$ and $\omega_\varphi$ are equal then $\omega_\vartheta=\omega_\varphi$ is an eigenvalue of $H'$. These three cases correspond to the existence of an obvious constant vector $\ket v$ such that $H(t)\ket{v}=E_2\ket v=0$ and allow to reduce the dimensionality of the problem by 1 (see the previous subsection and appendix \ref{app:berry}). To simplify, we will consider the case $\omega_\varphi=0$ for which we know that $\ket{2_1}$ is constant. Besides 0, the two other eigenvalues of $H'$ are
\begin{align*}
\varepsilon_{\pm}=\frac{\mathcal N_1^2}{2rs}+\frac12\,\omega_\vartheta\pm\omega,
\end{align*}
with
\begin{align*}
\omega=\frac12 \sqrt{\Big(\omega_\vartheta+\frac{r}{s}-\frac{s}{r}-rs\Big)^2+4(1+r^2)}.
\end{align*}
In the limit $\omega_\vartheta\to 0$ one has $\varepsilon_+\to\tfrac{\mathcal N_1^2}{rs}= E_1$ and $\varepsilon_-\to 0= E_2$. Corresponding unit eigenvectors of $H'$ are
\begin{align*}
\ket{\varepsilon_+}=\frac{A\ket{1^0}-B\ket{2_2^0}}{\sqrt{A^2+B^2}}\quad,\quad\ket{\varepsilon_-}=\frac{B\ket{1^0}+A\ket{2_2^0}}{\sqrt{A^2+B^2}}\,,
\end{align*}
with
\begin{align*}
\begin{aligned}
A&=\frac{\mathcal N_1^2}{2rs}+\bigg(\frac{r^2}{\mathcal N_1^2}-\frac12\bigg)\omega_\vartheta+\omega,\\
B&=\frac{rs}{\mathcal N_1^2}\sqrt{1+r^2}\,\omega_\vartheta\,.
\end{aligned}
\qquad(A^2+B^2=2A\omega)
\end{align*}
We have $\ket{\psi_{21}(t)}=\ket{2_1}=\text{cst.}$ and
\begin{align*}
\ket{\psi_{22}(t)}=\expo^{-\frac{\imag}{2}(\frac{\mathcal N_1^2}{rs}+\omega_\vartheta)t}&\bigg(\!\imag\frac{B}{\omega}\sin(\omega t)\ket{1(t)}\\ &+\bigg[\imag\frac{A}{\omega}\sin(\omega t)+\expo^{-\imag\omega t}\bigg]\ket{2_2(t)}\!\bigg).
\end{align*}
The transition probability to the excited level is as usually of the second order in $\omega_\vartheta$. It is easy to determine the first order superadiabatic approximation of $\ket{\psi_{22}(t)}$. Keeping in mind that $\omega_\vartheta t$ is finite, one must expand $\omega$ up to the second order in $\omega_\vartheta$ to obtain the correct first order expansion of $\omega t$, viz.
\begin{align*}
\ket{\psi_{22}(t)}=&\expo^{\imag\frac{rs}{8\mathcal N_1^2}\big[1+4\frac{r^2s^2(1+r^2)}{\mathcal N_1^4}\big]\omega_\vartheta^2t}U^{\text{ad}}(t)\ket{2_2(t)}\\
&+2\imag \expo^{-\frac{\imag}{2}(\frac{\mathcal N_1^2}{rs}+\omega_\vartheta)t}\frac{r^2s^2}{\mathcal N_1^4}\sqrt{1+r^2}\sin(\omega t)\omega_\vartheta\ket{1(t)}\\
&+\mathrm O(\omega_\vartheta^2).
\end{align*}
Of course, it is sufficient to limit ourselves to the first order approximation of $\omega$ in the sine. The distance between $V(t)$ and $V_2(t)$ is
\begin{align*}
d_{\text{FS}}(t)=\arccos\module{\bk{2_2(t)}{\psi_{22}(t)}}=\arcsin\left|\frac{B}{\omega}\sin(\omega t)\right|.
\end{align*}
It oscillates between $0$ and $\arcsin\module{\tfrac{B}{\omega}}\sim2\tfrac{r^2s^2}{\mathcal N_1^4}\sqrt{1+r^2}\omega_\vartheta$ with a period
\begin{align*}
\frac{\pi}{\omega}\sim\frac{2\pi rs}{\mathcal N_1^2}\bigg(1+\frac{rs}{\mathcal N_1^4}\big(r^2s^2-r^2+s^2\big)\omega_\vartheta\bigg).
\end{align*}
The expression of the geometric phase accumulated by $\ket{\psi_{22}(t)}$ derived from \eqref{MukundaSimon} is 
\begin{align*}
\gamma[\mathcal C_t]=&-\frac{\mathcal N_1^2}{2rs}\,t-\frac 12\,\omega_\vartheta t+\frac{rs(1+r^2)\omega_\vartheta t}{2\mathcal N_1^2\omega^2}\,\omega_\vartheta\\
&-\frac{rs(1+r^2)\sin(2\omega t)}{4\mathcal N_1^2\omega^3}\,\omega_\vartheta ^2\\
&+\arg\big[2\omega\, C(\omega_\vartheta t)\cos(\omega t)+\imag S(\omega_\vartheta t)\sin(\omega t)\big]
\end{align*}
with
\begin{align*}
C(\omega_\vartheta t)&=r^2+(1+r^2)s^2\expo^{\imag\omega_\vartheta t},\\
S(\omega_\vartheta t)&=\frac{\mathcal N_1^2}{rs}\,C(\omega_\vartheta t)+\big[r^2-(1+r^2)s^2\expo^{\imag\omega_\vartheta t}\big]\omega_\vartheta\,.
\end{align*}
In figure \ref{fig:phinul} are shown some plots of the phase for some choices of the parameters.
The geometric phase admits an expansion of the form  
\begin{align*}
\gamma[\mathcal C_t]=\gamma_0(t)+\gamma_1(t)+\text{O}(\omega_\vartheta^2)
\end{align*}
where $\gamma_0(t)$ is the adiabatic phase \eqref{gammaphinul} and $\gamma_1(t)$ is the first order superadiabatic correction. Explicitly:
\begin{widetext}
\begin{align*}
\gamma_1(t)=\frac{rs}{2\mathcal N_1^2}\bigg(\frac14+5\,\frac{r^2s^2}{\mathcal N_1^4}(1+r^2)\bigg)\omega_\vartheta^2 t-4\,\frac{r^3s^3}{\mathcal N_1^4}\sin(\omega t)\sin\bigg(\frac{\omega_\vartheta t}{2}\bigg)\frac{(1+r^2)s^2\sin\big(\frac{2\omega-\omega_\vartheta}{2}t\big)+r^2\sin\big(\frac{2\omega+\omega_\vartheta}{2}t\big)}{r^4+(1+r^2)^2s^4+2r^2s^2(1+r^2)\cos(\omega_\vartheta t)}\,\omega_\vartheta\,.
\end{align*}
\end{widetext}
In figure \ref{corrections} are plotted the exact geometric phase and its two lowest approximations for a value of $\omega_\vartheta$ sufficiently large to show a discrepancy between $\gamma[\mathcal C_t]$ and the first order approximation $\gamma_0(t)+\gamma_1(t)$.

\begin{figure}
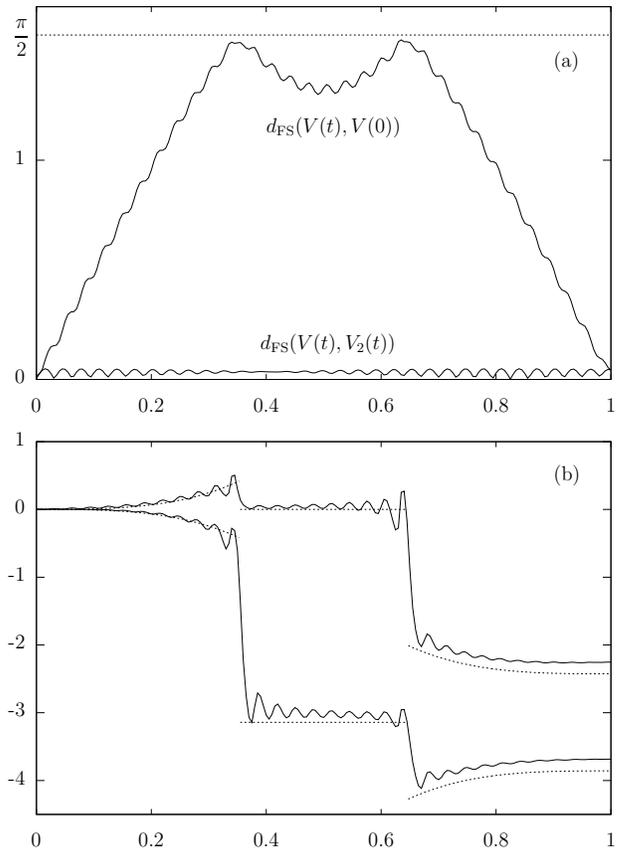

\centering
\resizebox{.95\columnwidth}{!}{\input{exactdist2.tex}}
\resizebox{.95\columnwidth}{!}{\input{exactphases2.tex}}
    \caption{(a) The Fubini-Study distances of the exact $2$-plane $V(t)$ to the adiabatic $2$-plane $V_2(t)$ and to the initial $2$-plane $V(0)=V_2$. (b) The exact geometric phases. The quantities are plotted as functions of $\frac{\omega_\vartheta t}{2\pi}\in[0,1]$ for $r=s=1.1$ and $\omega_\vartheta=0.1$. The dotted lines are the adiabatic phases.}\label{exactreqs}
\end{figure}

Let us end the study with an excursion in the general case where there does not exist a vector left constant by the dynamics ($\omega_\vartheta$, $\omega_\varphi$ and $\omega_\vartheta-\omega_\varphi$ are all nonzero). Here, we must apply the method of Tartaglia and Cardano \cite{Uspensky} to obtain exact expressions of the $\varepsilon_k$'s. Explicitly, they are
\begin{align}
\!\!\varepsilon_k=\frac{a_2}{3}+2\sqrt p\,\cos\bigg[\frac{1}{3}\arccos\bigg(\frac{q}{p^{3/2}}\bigg)+(k-1)\frac{2\pi}{3}\bigg]\!\label{exactev}
\end{align}
with
\begin{align*}
p=\frac{a_2^2}{9}-\frac{a_1}{3}\quad\text{and}\quad q=\frac{a_2^3}{27}-\frac{a_1a_2}{6}+\frac{a_0}{2}\,.
\end{align*} 
In the limit where the frequencies tend to zero, one has $\varepsilon_1\to E_1$ and $\varepsilon_2,\varepsilon_3\to E_2$. Then, as shown in the appendix \ref{app:colonnes}, one can choose
\begin{align*}
\ket{\varepsilon_k}={}& \varepsilon_k\ket{e_1}+\Big[\varepsilon_k^2-\Big(rs+\frac{s}{r}+\omega_\varphi\Big)\varepsilon_k+rs\omega_\varphi\Big]\ket{e_2}\\&+r(\varepsilon_k-\omega_\varphi)\ket{e_3}
\end{align*}
as eigenvectors of $H'$. All these expressions can be used to determine the invariant $2$-plane $V(t)$, its distance to $V_2(t)$, the holonomy associated with the path $t\mapsto V(t)$ and the corresponding geometric phases. However, the derived expressions are quite cumbersome and it is easier to proceed numerically. For example, in figure \ref{exactreqs} are plotted the various quantities of interest in a case where $r=s$ and $\omega_\vartheta=-\omega_\varphi$, a situation which was considered adiabatically at the end of the subsection \ref{abelianholo}. With the chosen values of the parameters, we remark that, along one cycle on the torus, the instantaneous $2$-plane $V(t)$ is always anti-orthogonal to $V(0)=V_2(0)$: the two anomalies in the adiabatic limit give rise to two instants where the distance $V(t)$ and $V(0)$ reach maxima which are slightly less than $\tfrac{\pi}{2}$.

\acknowledgments

This work is a contribution to the group \enquote{Connexions and Gauge Theories} which brings together mathematicians, physicists, and historians who meet regularly at Nancy. I have special thoughts for my longtime friend Simon Buron and --- since the French word \textit{transport} carries the poetic meaning of a passionate feeling --- Lucie.\\



\appendix

\section{On the polar decomposition}\label{PD}

In this appendix, I give a review of the polar decompositions \cite{Gallier} of linear maps between finite dimensional inner product spaces over the same field $\mathbb F=\mathbb R$ or $\mathbb C$. In the sequel, $V$, $W$, and $X$, will always be such spaces and all the inner products will be denoted by $\bk{\cdot}{\cdot}$. A typical example of dimension $n$ is the space $\mathbb F^{n\times 1}$ which is naturally equipped with the inner product $\bk{\rep u}{\rep v}=(\rep u^*)^\top\rep v$, where the asterisk denotes the complex conjugation. 

\subsection{Basic theory}\label{PDbasic}

Let me recall that the orthogonal of any vector subspace $X\subseteq V$ is the set $X^\perp$ formed by the vectors of $V$ which are orthogonal to $X$ (i.e.\ to all the vectors of $X$). It is in fact a supplement of $X$ in $V$, that is, a vector subspace of $V$ such that $X\oplus X^\perp=V$ (in particular, its dimension is $\dim V-\dim X$). Hence, the inner product structure allows to single out a natural supplement of $X$, namely its orthogonal supplement $X^\perp$. Actually, $X$ and $X^\perp$ are mutually orthogonal (i.e.\ $(X^{\perp})^{\perp}=X$). Direct sums of pairwise orthogonal vector subspaces are said orthogonal. If $V$ decomposes as a direct sum $X_1\oplus X_2\oplus\dots$ then an ordered basis of $V$ is said adapted to that decomposition if its $\dim X_1$ first elements form a basis of $X_1$, the $\dim X_2$ following ones a basis of $X_2$, etc.

Now, let $F\colon V\to W$ be a linear map. The inner product structures on $V$ and $W$ allow to decompose $V$ and $W$ into the orthogonal direct sums $\ker F\oplus(\ker F)^\perp$ and $F(V)\oplus F(V)^\perp$, respectively, and to extract canonically from $F$ an isomorphism $F'\colon(\ker F)^\perp\to F(V)$. By completing its inverse $F'^{-1}\colon F(V)\to(\ker F)^\perp$ with the null map on $F(V)^\perp$, one obtains the so-called Moore-Penrose pseudoinverse $F^+$ of $F$ (which coincides with the usual inverse if $F$ is bijective). The map $F$ also induces a unique map $F^\dag\colon W\to V$ such that $\bk{F^\dag w}{v}=\bk{w}{Fv}$ for all vectors $v\in V$ and $w\in W$. It is actually a linear map which is called the adjoint of $F$. More fundamentally, $F^\dag$ embodies the transpose of $F$ once $V$ and $W$ are identified with their duals thanks to the inner product structures. The adjoint of $F^\dag$ is $F$ itself, i.e.\ $F^{\dag\dag}=F$. If $G\colon W\to X$ is another linear map, the adjoint of $GF$ is $F^\dag G^\dag$. If $F$ is bijective, one has $(F^{-1})^\dag=(F^\dag)^{-1}$. For example, any $m\times n$ matrix $\rep A$ is naturally identified with the linear map $\rep u\mapsto \rep A\rep u$ from $\mathbb F^{n\times 1}$ to $\mathbb F^{m\times 1}$ whose adjoint $\rep A^\dag$ is the conjugate transpose of $\rep A$. Furthermore, if $\rep A$ is the matrix of a map $F\colon V\to W$ in some orthonormal basis $\mathcal V$ of $V$ and $\mathcal W$ of $W$, then the matrix of $F^\dag$ in $\mathcal W$ and $\mathcal V$ is $\rep A^\dag$.

A linear map $F\colon V\to W$ is said unitary if it preserves the inner product, i.e.\ if $\bk{Fv}{Fv'}=\bk{v}{v'}$ for all vectors $v,v'\in V$. The following properties are equivalent: (i) $F$ is unitary, (ii) $F$ preserves the norm, i.e.\ $\|Fv\|=\|v\|$ for all $v\in V$, (iii) $F$ is bijective and such that $F^\dag=F^{-1}$, (iv) there exists an orthonormal basis of $V$ and an orthonormal basis of $W$ in which the matrix $\rep A$ of $F$ is unitary, i.e.\ invertible and such that $\rep A^\dag=\rep A^{-1}$, (v) in any orthonormal bases of $V$ and $W$ the matrix of $F$ is unitary, (vi) $F^\dag$ is unitary. Obviously the set $ U(V,W)$ of unitary maps $V\to W$ is nonempty iff $V$ and $W$ have the same dimension.

An endomorphism $F\colon V\to V$ is self-adjoint if $F^\dag=F$ and positive semidefinite (resp.\ positive definite) if $\bk{v}{Fv}$ is nonnegative (resp.\ positive) for all $v\in V$. It has one of these properties iff there exists an orthonormal basis of $V$ in which the matrix of $F$ has the same property, and this is equivalent to say that any matrix of $F$ in an orthonormal basis of $V$ has that property. A self-adjoint endomorphism $H$ of $V$ verifies two interesting properties: (i) it can only admit real eigenvalues\footnote{If $Hv=\lambda v$ with $v\ne 0$ then $\lambda\bk{v}{v}=\bk{v}{\lambda v}=\bk{v}{Hv}=\bk{H^\dag v}{v}=\bk{Hv}{v}=\bk{\lambda v}{v}=\lambda^*\bk{v}{v}$ hence $\lambda=\lambda^*\in\mathbb R$.} 
and (ii) if $H$ leaves invariant a vector subspace $X$ of $V$, i.e.\ if $H(X)\subseteq X$, then it leaves invariant $X^\perp$ as well.\footnote{For any $y\in X^\perp$ one has, for all $x\in X$, $Hx\in X$ and thus $\bk{x}{Hy}=\bk{Hx}{y}=0$, that is, $Hy\in X^\perp$.} 
In particular, applying (ii) to the obvious invariant space $\ker H$ leads to $(\ker H)^\perp=H(V)$.\footnote{One has $H(V)=H((\ker H)^\perp)\subseteq(\ker H)^\perp$ hence $H(V)$ and $(\ker H)^\perp$ coincide since they have the same dimension.} The two properties (i) and (ii) are the cornerstones of a spectral theorem (in finite dimension) which states that the self-adjoint endomorphisms are precisely the endomorphisms which are diagonalizable in an orthogonal eigenbasis, with real eigenvalues. To put it another way, an endomorphism $F$ of $V$ is self-adjoint iff it can be decomposed into a sum $\lambda_1P_{V_1}+\dots+\lambda_kP_{V_k}$ where the $\lambda_i$ are distinct real numbers and the $P_{V_i}$ are (orthogonal) projections into nonzero vector subspaces $V_i$ such that $V$ is the orthogonal direct sum $V_1\oplus\dots\oplus V_k$. In this case, the sum $\lambda_1P_{V_1}+\dots+\lambda_kP_{V_k}$ is unique (up to a reordering) and called the spectral decomposition of $F$, the $\lambda_i$ are the eigenvalues of $F$ and the $V_i$ are the associated eigenspaces $\ker(F-\lambda_i\id_V)$. From there, it is clear that $F$ is further positive semidefinite (resp.\ positive definite) iff the $\lambda_i$ are all nonnegative (resp.\ positive).

\refstepcounter{prop}
\textbf{Proposition \theprop\label{prop1}.} \textsl{Let $F\colon V\to W$ be a linear map. One has the following properties:
\setlist{nolistsep}
\begin{enumerate}[noitemsep,align=left,leftmargin=*,label=(\roman*)]
\item $F^\dag F$ and $FF^\dag$ are positive semidefinite self-adjoint endomorphisms of $V$ and $W$ respectively. 
\item $\ker(F^\dag F)=\ker F=F^\dag(W)^\perp$.
\item $\ker(F F^\dag)=\ker F^\dag=F(V)^\perp$. 
\item $F$, $F^\dag$, $F^\dag F$, and $FF^\dag$ have the same rank.
\end{enumerate}}

\textbf{Proof.} (i) One has $(F^\dag F)^\dag=F^\dag F^{\dag\dag}=F^\dag F$ thus $F^\dag F$ is self-adjoint. It is also positive semidefinite since $\bk{v}{F^\dag Fv}=\bk{Fv}{Fv}$ for all $v\in V$. The same thing for $FF^\dag$. (ii) One has automatically $\ker F\subseteq\ker(F^\dag F)$. However, if $F^\dag Fv=0$ one has $0=\bk{v}{F^\dag Fv}=\bk{Fv}{Fv}$ thus $Fv=0$. This establishes the reverse inclusion. The second equality comes from the fact that $Fv=0$ iff $\bk{w}{Fv}=0$ for all $w\in W$ i.e.\ iff $\bk{F^\dag w}{v}=0$ for all $w\in W$. (iii) The same thing than for (ii). (iv) The isomorphism $(\ker F)^\perp\simeq F(V)$ and the equality $(\ker F)^\perp=F^\dag(W)$ provide $\rank F=\rank F^\dag$. Then, $F$ and $F^\dag F$ (resp.\ $F^\dag$ and $FF^\dag$) being defined over the same finite dimensional vector space $V$ (resp.\ $W$) and having kernels of the same dimension, their ranks are equal according to the rank-nullity theorem.\hfill$\square$

\refstepcounter{theorem}
\textbf{Theorem \thetheorem\label{thm1}.} \textsl{For any semidefinite self-adjoint endomorphism $H$ of $V$ there exists a unique semidefinite self-adjoint endomorphism $R$ of $V$ such that $H=R^2$. More precisely, if the spectral decomposition of $H$ is $\lambda_1^2P_1+\dots+\lambda_k^2P_k$ ($\lambda_i\geqslant0$) then $R=\lambda_1P_1+\dots+\lambda_kP_k$. In particular, $R$ and $H$ have the same kernel and the same rank, and $R$ is positive definite iff $H$ is so. The map $R$ is called the square root of $H$ and it can be written $\sqrt H$. }

\textbf{Proof.} Suppose first that such a map $R$ exists. Since $H=R^2=R^\dag R$ one has $\ker R=\ker H$ by the point (ii) of proposition \ref{prop1}. In addition, for any $\lambda>0$, one can factorize the map $H-\lambda^2\id_V$ as $(R+\lambda\id_V)(R-\lambda\id_V)$. Since $R$ is semidefinite, it cannot admit $-\lambda$ as eigenvalue. Hence, $R+\lambda\id_V$ is injective and the factorization implies the equality of the kernels of $H-\lambda^2\id_V$ and $R-\lambda\id_V$. What has been shown is that the kernels of $R-\lambda\id_V$ and $H-\lambda^2\id_V$ coincide for any nonnegative real $\lambda$. Therefore, $R$ must be the map $\lambda_1P_1+\dots+\lambda_kP_k$ which is obviously semidefinite, self-adjoint, and such that $H=R^2$.\hfill$\square$

\refstepcounter{lem}
\textbf{Lemma \thelem\label{lem1}.} \textsl{Let $A\colon Y\to Z$, $B\colon Z\to Y$ be linear maps between any vector spaces $Y$ and $Z$ over a same field. The compositions $BA$ and $AB$ have the same nonzero eigenvalues (if any), and the eigenspaces $Y'$ of $BA$ and $Z'$ of $AB$ associated with a same nonzero eigenvalue are isomorphic. More precisely, $A$ (resp.\ $B$) realizes an isomorphism $Y'\to Z'$ (resp.\ $Z'\to Y'$).}

\textbf{Proof.} Let $\lambda$ be a nonzero eigenvalue of $BA$ and $Y'$ be the associated eigenspace. For all nonzero eigenvectors $y\in Y'$ one has $B(Ay)=BAy=\lambda y\ne 0$ thus $Ay\ne 0$ and $AB(Ay)=A(BAy)=A(\lambda y)=\lambda(Ay)$. One deduces that $\lambda$ is also an eigenvalue of $AB$ and that $A$ maps injectively $Y'$ into the eigenspace $Z'$ of $AB$ associated with $\lambda$. In particular, one has the inclusion $A(Y')\subseteq Z'$, and also $B(Z')\subseteq Y'$ since $BA(Bz)=B(ABz)=B(\lambda z)=\lambda(Bz)$ for all $z\in Z'$. These two inclusions lead to $AB(Z')=A(B(Z'))\subseteq A(Y')$. However, since $Z'$ is an eigenspace of $AB$ associated with a nonzero eigenvalue, one has $AB(Z')=Z'$. It follows the reverse inclusion $ A(Y')\supseteq Z'$ thus the equality $A(Y')=Z'$. In summary, $A$ realizes a map $Y'\to Z'$ which is both injective and surjective, i.e.\ bijective. The lemma is finally established after exchanging the roles of $A$ and $B$.\hfill$\square$

\refstepcounter{theorem}
\textbf{Theorem \thetheorem\label{thm2}.} \textsl{Let $F\colon V\to W$ be a linear map. The maps $F^\dag F$ and $FF^\dag$ are diagonalizable and have the same nonzero eigenvalues $\sigma_1^2,\dots,\sigma_p^2$ ($\sigma_i>0$) with the same multiplities. The $\sigma_i$ are called the singular values of $F$. Furthermore, if $V_i$ (resp.\ $W_i$) is the eigenspace of $F^\dag F$ (resp.\ $FF^\dag$) associated with $\sigma_i$ then $\sigma_i^{-1}F$ realizes a unitary isomorphism $V_i\to W_i$.}

\textbf{Proof.} The diagonalizability of $F^\dag F$ and $FF^\dag$ stems from their self-adjointness while the nonnegativity of their eigenvalues comes from their positive semidefiniteness. Since $V_i$ and $W_i$ are isomorphic according to the lemma, they have the same dimension, i.e.\ the multiplicities of $\sigma_i^2$ as an eigenvalue of $F^\dag F$ and $FF^\dag$ coincide. Moreover, one knows from the lemma that $F$ realizes an isomorphism $V_i\to W_i$ and obviously so does $\sigma_i^{-1}F$. The latter is unitary since, for all $v,v'\in V_i$, one has $\bk{\sigma_i^{-1}Fv}{\sigma_i^{-1}Fv'}=\sigma_i^{-2}\bk{F^\dag Fv}{v'}=\sigma_i^{-2}\bk{\sigma_i^2v}{v'}=\bk{v}{v'}$.\hfill$\square$

Note that if $V$ and $W$ have the same dimension then $F^\dag F$ and $FF^\dag$ have the same eigenvalues with the same multiplicities since, according to the last point of the proposition \ref{prop1}, the kernels of $F^\dag F$ and $FF^\dag$ have in this case the same dimension.

\refstepcounter{theorem}
\textbf{Theorem \thetheorem\label{thm3}. (polar decompositions).} \textsl{Suppose that $V$ and $W$ have the same dimension, and let $F\colon V\to W$ be a linear map. There exist positive semidefinite self-adjoint endomorphisms $R$ of $V$ and $R'$ of $W$, as well as unitary isomorphisms $U$, $U'\colon V\to W$, such that $F=UR=R'U'$. Such factorizations $UR$ and $R'U'$ are called right and left polar decompositions of $F$, respectively. Furthermore, (i) $R$ and $R'$ are unique, $R$ (resp.\ $R'$) being the square root of $F^\dag F$ (resp.\ $FF^\dag$) and (ii) the set of maps $U$ coincides with the set of maps $U'$, all these unitary maps realizing the same unitary isomorphism $U_*\colon(\ker F)^\perp\to F(V)$ and an arbitrary one from $\ker F$ to $F(V)^\perp$. In particular, if $F$ is bijective then $R$ and $R'$ are positive definite while $U$ and $U'$ are unique and equal.} 

\textbf{Proof.} Let us first assume the existence of such maps $R$, $R'$, $U$, and $U'$. One has $F^\dag F=RU^{-1}UR=R^2$ and $FF^\dag=R'U'U'^{-1}R'=R'^2$. Hence, $R$ and $R'$ are necessarily the square roots of the semidefinite self-adjoint maps $F^\dag F$ and $FF^\dag$, respectively. The map $F$ realizes an isomorphism $F_*\colon(\ker F)^\perp\to F(V)$ and the null map $F_0\colon \ker F\to F(V)^\perp$ while $R$ (resp.\ $R'$) realizes an automorphism $R_*$ of $R(V)=(\ker F)^\perp$ (resp.\ $R'_*$ of $R'(V)=F(V)$) and the null endomorphism $r$ of $\ker F$ (resp.\ $R'_0$ of $F(V)^\perp$). To sum up, one has the decompositions
\begin{align*}
R&\colon\;\;\left|\;\begin{aligned}
(\ker F)^\perp&\xrightarrow[R_*]{\quad\sim\quad}(\ker F)^\perp\\
\ker F&\xrightarrow[r=0]{\quad\phantom{\sim}\quad}\ker F
\end{aligned}\right.\\[10pt]
F&\colon\;\;\left|\; \begin{aligned}
(\ker F)^\perp&\xrightarrow[F_*]{\quad\sim\quad}F(V)\\
\ker F&\xrightarrow[F_0=0]{\quad\phantom{\sim}\quad}F(V)^\perp
\end{aligned}\right.\\[10pt]
R'&\colon\;\;\left|\;\hspace{.245cm} \begin{aligned}
F(V)&\xrightarrow[R'_*]{\quad\sim\quad}F(V)\\
F(V)^\perp&\xrightarrow[R'_0=0]{\quad\phantom{\sim}\quad}F(V)^\perp
\end{aligned}\right.\\
\end{align*}
The equalities $F=UR=R'U'$ thus imply that $U,U'$ realize unitary isomorphisms $U_*,U'_*\colon(\ker F)^\perp\to F(V)$ and $U_0,U'_0\colon\ker F\to F(V)^\perp$ such that $F_*=U_*R_*=R'_*U'_*$ and $F_0=U_0r=R'_0U'_0$. Hence, $U_*$ and $U'_*$ are necessarily given by $U_*=F_*R_*^{-1}$ and $U'_*=R'^{-1}_*F_*$ while $U_0$ and $U'_0$ are arbitrary. Now, let $\sigma_1,\dots,\sigma_p$ be the distinct singular values of $F$. One has the orthogonal direct sums $(\ker F)^\perp= V_1\oplus\dots\oplus V_p$ and $F(V)^\perp= W_1\oplus\dots\oplus W_p$ with $V_i$ (resp.\ $W_i$) the eigenspace of $R$ (resp.\ $R'$) associated with $\sigma_i$. One knows from the lemma that $F$ realizes isomorphisms $F_i\colon V_i\to W_i$ and from theorem \ref{thm1} that the square root $R$ of $F^\dag F$ (resp.\ $R'$ of $FF^\dag$) realizes scaling transformations $R_i=\sigma_i\id_{V_i}$ (resp.\ $R'_i=\sigma_i\id_{W_i}$). Hence, $U_*$ and $U_*'$ must decompose into unitary isomorphisms $U_i,U'_i\colon V_i\to W_i$ which are characterized by $F_i=U_iR_i=R'_iU'_i$. One has $F_i=U_i\sigma_i=\sigma_iU'_i$ thus $U_i$ and $U'_i$ are unique and both equal to $\sigma_i^{-1}F_i$, i.e.\ to the unitary isomorphism brought to light in theorem \ref{thm2}. The deduced equality $U_*=U'_*$ combined with the arbitrariness of $U_0$ and $U'_0$ proves that unitary isomorphisms entering left polar decompositions enters also right polar ones, and vice versa. To establish the existence of polar decompositions of $F$, it remains to verify two points: (i) the unitariness of $U_*$ and (ii) the equality of the dimensions of $\ker F$ and $F(V)^\perp$ without which $U_0,U'_0$ could not exist. The first point is evident since the $U_i$ are unitary while the $V_i$ are in an orthogonal direct sum as well as the $W_i$. The second point is obvious from the equality of the dimensions of $V=(\ker F)^\perp\oplus\ker F$ and $W=F(V)\oplus F(V)^\perp$. Finally, if $F$ is bijective, $R$ and $R'$ are obviously positive definite and $U=U_*=U'$.\hfill$\square$

The proof above is constructive and allows to characterize the set $\mathcal U(F)$ of unitary isomorphisms entering the polar decompositions of $F$. Keeping the same notations, this set is formed by the elements of $U(V,W)$ realized by $U_*\colon(\ker F)^\perp\to F(V)$ and by an arbitrary unitary isomorphism $U_0\colon\ker F\to F(V)^\perp$. The whole work consists in the determination of $U_*$. It entails the computation of the inverse of one of the maps $R_*$ or $R'_*$ which are the respective square roots of $F^\dag_*F_*$ and $F_*F^\dag_*$. It is achieved in a systematic way by first solving the eigenproblem of $F^\dag F$ or $FF^\dag$ then by proceeding as described in the proof. Finally, one deduces $U_*=F_*R_*^{-1}=R'^{-1}_*F_*$ which can be completed to a map $\Gamma(F)\colon V\to W$ vanishing on $\ker F$. The latter is a partial unitary map\footnote{We say that a map $G\colon V\to W$ verifies partially a property if its restriction $(\ker G)^\perp\to G(V)$ has that property.}  of the same rank than $F$ and is given by $\Gamma(F)=FR^+=R'^+F$.

\refstepcounter{prop}
\textbf{Proposition \theprop\label{prop2}.} \textsl{Suppose that $V$, $W$, and $X$ have the same dimension. Let $F\colon V\to W$ be a linear map and $T_1\colon X\to V$, $T_2\colon W\to X$ be unitary maps. One has $\Gamma(FT_1)=\Gamma(F)T_1$ and $\Gamma(T_2F)=T_2\Gamma(F)$.}

\textbf{Proof.} Since $(FT_1)(FT_1)^\dag=FF^\dag$, the positive semidefinite self-adjoint endomorphisms of $W$ which enter the left polar decompositions of $F$ and $FT_1$ are identical. Denoting it by $R'$ one has $\Gamma(FT_1)=R'^+FT_1=\Gamma(F)T_1$. One proceeds in the same way for the second equality.\hfill$\square$

\subsection{Some matrix characterizations}\label{subsec:matchar}

Hereafter, we keep the hypothesis and notations of theorems \ref{thm2} and \ref{thm3}, and $n$ will be the dimension shared by $V$ and $W$. Our aim is to give various characterizations, in matrix terms, of the set $\mathcal U(F)$ of unitary isomorphisms $V\to W$ involved in the polar decompositions of $F$. Let me first precise some elements of notation. If $\mathcal Y$ and $\mathcal Z$ are any (ordered) bases of two finite dimensional vector spaces $Y$ and $Z$, respectively, then the matrix of a linear map $A\colon Y\to Z$ in these bases will be generically denoted by $\mat(A;\mathcal Z,\mathcal Y)$ or simply $\mat(A;\mathcal Y)$ if $Y=Z$ and $\mathcal Y=\mathcal Z$. Moreover, if $\mathcal Y=(y_1,y_2,\dots)$ then $A(\mathcal Y)$ will denote the family $(Ay_1,Ay_2,\dots)$. Recall that, once an orthonormal basis $\mathcal V$ of $V$ is fixed, $U(\mathcal V)$ is an orthonormal basis of $W$ for any $U\in U(V,W)$ and there exists, for any orthonormal basis $\mathcal W$ of $W$, a unique element $U\in U(V,W)$ such that $\mathcal W=U(\mathcal V)$. In other words, $\mathcal V$ being fixed, there is a biunivocal correspondence between $U(V,W)$ and the set of orthonormal bases of $\mathcal W$ allowing to reason in terms of the former set instead of the latter one. Moreover, one can also \enquote{rotate} any orthonormal basis $\mathcal X=(x_1,\dots,x_n)$ of $X$ ($=V$ or $W$) via $n\times n$ unitary matrices $\rep T$ through
\begin{align*}
x_i\longmapsto x'_i=\sum_{j=1}^n x_j\rep T_{ji}\,.
\end{align*} 
The resulting orthonormal basis $\mathcal X'=(x'_1,\dots,x'_n)$ of $X$ will be simply denoted by $\mathcal X\rep T$. Here again, once an orthonormal basis $\mathcal X$ of $X$ is fixed, there exists, for any orthonormal basis $\mathcal X'$ of $X$, a unique rotation matrix $\rep T$ such that $\mathcal X'=\mathcal X\rep T$. 

\refstepcounter{theorem}
\textbf{Theorem \thetheorem\label{thm4}.} \textsl{For any orthonormal basis $\mathcal V$ of $V$, $\mat(F;U(\mathcal V),\mathcal V)$ is a positive semidefinite self-adjoint matrix iff $U\in \mathcal U(F)$, in which case $\mat(F;U(\mathcal V),\mathcal V)=\mat(R;\mathcal V)$.}

\textbf{Proof.} Let $\mathcal V$ be an orthonormal basis of $V$. If $U\in \mathcal U(F)$ then one has $F=UR$ and therefore $\mat(F;U(\mathcal V),\mathcal V)=\mat(U;U(\mathcal V),\mathcal V)\mat(R;\mathcal V)=\mat(R;\mathcal V)$ is a positive semidefinite self-adjoint matrix. Reciprocally, let $U\in U(V,W)$ and suppose that $\mat(F;U(\mathcal V),\mathcal V)$ is positive semidefinite and self-adjoint. Introducing an element $U_\bullet\in \mathcal U(F)$, one has $F=U_\bullet R$ thus $\mat(F;U(\mathcal V),\mathcal V)=\mat(U_\bullet;U(\mathcal V),\mathcal V)\mat(R;\mathcal V)$. It is a right polar decomposition of the matrix in the left-hand side. However, since $\mat(F;U(\mathcal V),\mathcal V)$ is positive semidefinite and self-adjoint, it also admits the right polar decomposition $\rep I_n\mat(F;U(\mathcal V),\mathcal V)$. By the uniqueness of the positive semidefinite self-adjoint factor involved in right polar decompositions of a matrix, one deduces $\mat(F;U(\mathcal V),\mathcal V)=\mat(R;\mathcal V)=\mat(U;U(\mathcal V),\mathcal V)\mat(R;\mathcal V)$, that is, $F=UR$.\hfill$\square$

\refstepcounter{cor}
\textbf{Corollary \thecor\label{cor1}.} \textsl{For any orthonormal basis $\mathcal V$ of $V$, the matrix $\rep M(F;U(\mathcal V),\mathcal V)$ is diagonal and nonnegative iff $U\in \mathcal U(F)$ and $\mathcal V$ is adapted to the decomposition $V_1\oplus\dots\oplus V_p\oplus\ker F$ (up to a reordering of its elements). In this case, the matrix takes the form}
\begin{align*}
\rep M(F;U(\mathcal V),\mathcal V)=\begin{pmatrix}
\sigma_1\rep I_{n_1} & & &\\
 & \ddots & &\\
 & & \sigma_p \rep I_{n_p} & \\
 & & & 0
\end{pmatrix},\quad n_i=\dim V_i\,.
\end{align*}

\textbf{Proof.} If $U\notin\mathcal U(F)$ then, by theorem \ref{thm4}, the matrix $\rep M(F;U(\mathcal V),\mathcal V)$ cannot be diagonal and nonnegative. Consider the case where $U\in\mathcal U(F)$. Let $v$ be an element of $\mathcal V$. It decomposes as a sum $v_1+\dots+v_p+v_0$ with $v_i\in V_i$, $v_0\in\ker F$, and one has $Fv=\sigma_1Uv_1+\dots+\sigma_pUv_p$. Clearly, $Fv$ belongs to the span of $Uv$ iff $v$ belongs to one of the $V_i$ or to $\ker F$. Hence, $\rep M(F;U(\mathcal V),\mathcal V)$ is diagonal iff $\mathcal V$ is (up to a reordering of its elements if necessary) adapted to the decomposition $V_1\oplus\dots\oplus V_p\oplus\ker F$, in which case the matrix takes the above form.\hfill$\square$

The following theorem gives a characterization of $\mathcal U(F)$ based on a minimization problem about the trace of the matrices of $F$ in orthonormal bases. 

\refstepcounter{theorem}
\textbf{Theorem \thetheorem\label{thm5}.} \textsl{Let $\mathcal V$ be an orthonormal basis of $V$. Over the set of unitary isomorphisms $U\in U(V,W)$, the real part of the trace of $\mat(F;U(\mathcal V),\mathcal V)$ admits a maximum which is attained iff $U\in \mathcal U(F)$. Alternatively stated, one has
\begin{align*}
\mathcal U(F)=\argmax_{U\in U(V,W)}\Big\{\Re\big[\tr\mat(F;U(\mathcal V),\mathcal V)\big]\Big\}.
\end{align*}
Furthermore, the maximum does not depend on $\mathcal V$.}

\textbf{Proof.} Let $U\in U(V,W)$. If $U\in \mathcal U(F)$, one has $\mat(F;U(\mathcal V),\mathcal V)=\mat(R;\mathcal V)$ by theorem \ref{thm4} and the trace of $\mat(F;U(\mathcal V),\mathcal V)$ has for value the invariant $\tr R$. Now, suppose $U\notin \mathcal U(F)$ and introduce an orthonormal basis $\mathcal V'=(v'_1,\dots,v'_n)$ of $V$ such that $(v'_1,\dots,v'_p)$ spans $R(V)=(\ker R)^\perp=(\ker F)^\perp$. There exists a rotation matrix $\rep T$ such that $\mathcal V=\mathcal V'\rep T$. Since $\mat(F;U(\mathcal V'\rep T),\mathcal V'\rep T)=\mat(F;U(\mathcal V')\rep T,\mathcal V'\rep T)=\rep T^{-1}\mat(F;U(\mathcal V),\mathcal V)\rep T$, the matrix $\mat(F;,U(\mathcal V'),\mathcal V')$ has the same trace than $\mat(F;U(\mathcal V),\mathcal V)$. Then, introduce an element $U_\bullet\in\mathcal U(F)$. One has $\mat(F;U(\mathcal V'),\mathcal V')=\mat(U_\bullet;U(\mathcal V'),\mathcal V')\mat(R;\mathcal V')$ and thus
\begin{align*}
\Re\big[\tr\mat(F;U(\mathcal V),\mathcal V)\big]=\sum_{i=1}^p\Re\big[\bk{Uv'_i}{U_\bullet v'_i}\big]\bk{v'_i}{Rv'_i}.
\end{align*}
By construction, one has $\bk{v'_i}{Rv'_i}>0$, and the unitariness of $U$ and $U_\bullet$ implies $\Re\bk{Uv'_i}{U_\bullet v'_i}\leqslant 1$ with equality iff $Uv'_i=U_\bullet v'_i$. Recalling that an element of $U(V,W)$ belongs to $\mathcal U(F)$ iff it realizes the unitary isomorphism $U_*\colon(\ker F)^\perp\to F(V)$ brought to light in theorem \ref{thm3}, the fact that $U\notin \mathcal U(F)$ implies the existence of an element $v'_j$ of $(v'_1,\dots,v'_p)$ such that $Uv'_j\ne U_\bullet v'_j$, inducing $\Re\bk{Uv'_j}{U_\bullet v'_j}<1$. One finally concludes
\begin{align*}
\Re\big[\tr\mat(F;U(\mathcal V),\mathcal V)\big]<\sum_{i=1}^p\bk{v'_i}{Rv'_i}=\tr R.
\end{align*}
This achieves the proof.\hfill$\square$

Let us give a last characterization of $\mathcal U(F)$ as the solution of an equivalent matrix optimization problem. To this end, we introduce the Frobenius norm $\|\rep A\|_{\text{Fro}}=\tr(\rep A^\dag\rep A)^{1/2}$ in the vector space of $n\times n$ complex matrices $\rep A$. The Frobenius distance between two such matrices $\rep A$ and $\rep B$ is thus $\|\rep A-\rep B\|_{\text{Fro}}$. After choosing a basis $\mathcal V$ of $V$ and an element $U\in U(V,W)$, let $\rep F$ be the matrix of $F$ in the bases $\mathcal V$ and $U(\mathcal V)$. The squared Frobenius distance between $\rep F$ and the $n\times n$ identity matrix $\rep I_n$ is thus
\begin{align*}
\|\rep F-\rep I_n\|_{\text{Fro}}^2&=\tr((\rep F-\rep I_n)^\dag(\rep F-\rep I_n))\\
&=n+\tr(F^\dag F)-2\Re\big[\tr \rep F\big].\\
\end{align*}
Therefore, one infers at once from theorem \ref{thm5} the following statement.

\refstepcounter{cor}
\textbf{Corollary \thecor\label{cor2}.} \textsl{Let $\mathcal V$ be a basis of $V$. Over the set of unitary isomorphisms $U\in U(V,W)$, the Frobenius distance between $\mat(F;U(\mathcal V),\mathcal V)$ and the $n\times n$ identity matrix admits a minimum which is attained iff $U\in \mathcal U(F)$. Alternatively stated, one has
\begin{align*}
\mathcal U(F)=\argmin_{U\in U(V,W)}\Big\{\|\mat(F;U(\mathcal V),\mathcal V)-\rep I_n\|_{\mathrm{Fro}}\Big\}.
\end{align*}
Furthermore, the minimum does not depend on $\mathcal V$.}

The fact that the extrema in the theorem \ref{thm5} and its corollary do not depend on the basis says that these quantities are intrinsic to the map $F$.

\section{Dynamical invariants and invariant planes}\label{annexe:DI}

A dynamical invariant is commonly defined as a self-adjoint linear operator $I(t)$ verifying the equation \cite{Fock,Lewis,Kwon}
\begin{align}
\imag \dot I+[I,H]=0.\label{FI}
\end{align}
The expectation values of such an operator remain constant in time. Indeed, for any motion $\ket{\psi(t)}$, the Ehrenfest theorem \cite{Messiah1} gives
\begin{align*}
\imag\frac{\diff}{\diff t}\Big[\bra{\psi}I\ket{\psi}\Big]=\bra{\psi}\Big[\imag \dot I+[I,H]\Big]\ket{\psi}=0.
\end{align*}
One has thus $\bra{\psi(t)}I(t)\ket{\psi(t)}=\bra{\psi(0)}I(0)\ket{\psi(0)}$ for any motion $\ket{\psi(t)}$, i.e.\
\begin{align*}
\bra{\psi(0)}U(t)^{-1}I(t)U(t)\ket{\psi(0)}=\bra{\psi(0)}I(0)\ket{\psi(0)}.
\end{align*}
The initial state being arbitrary, one deduces the equality $U(t)^{-1}I(t)U(t)=I(0)$ which means that the Heisenberg representation of $I(t)$ remains constant in time \cite{Messiah1}. It is equivalent to $I(t)U(t)=U(t)I(0)$. Reciprocally, if $I(t)$ verifies $I(t)U(t)=U(t)I(0)$ then, multiplying this equality by $\imag$ and taking its total derivative with respect to $t$ gives $[\imag\dot I(t)+H(t)]U(t)$ for the left-hand side, $H(t)U(t)I(0)=H(t)I(t)U(t)$ for the right-hand side, and one obtains \eqref{FI}. This paragraph has brought equivalent characterizations of a dynamical invariant. According to the terminology of most authors, a dynamical invariant is furthermore a constant of motion if it does not depend on time, i.e.\ if it is a time-independent self-adjoint operator commuting with the Hamiltonian \cite{Messiah1,Schiff,Cohen,Merzbacher,Ballentine,Weinberg}.

Now, let $\alpha$ be an eigenvalue of $I(0)$, $V(0)$ be the associated eigenspace, and $\ket{\psi(t)}$ be a solution generated by some initial state $\ket{\psi(0)}$. If $\ket{\psi(0)}$ belongs to $V(0)$ then $I(t)\ket{\psi(t)}=I(t)U(t)\ket{\psi(0)}=U(t)I(0)\ket{\psi(0)}=\alpha\, U(t)\ket{\psi(0)}=\alpha\,\ket{\psi(t)}$. Otherwise, if $\ket{\psi(0)}$ is orthogonal to $V(0)$ then $0=U(t)I(0)\ket{\psi(0)}=I(t)U(t)\ket{\psi(0)}=I(t)\ket{\psi(t)}$ i.e.\ $\ket{\psi(t)}\in V(t)^\perp$. One has shown that $\alpha$ is still an eigenvalue of $I(t)$ and that $U(t)$ realizes a unitary isomorphism from $V(0)$ to the eigenspace $V(t)$ of $I(t)$ associated with $\alpha$. Consequently, the time-dependent eigenspace $V(t)$ associated with the constant eigenvalue $\alpha$ is an invariant $n$-plane. This is how a dynamical invariant generates invariant planes of $\mathcal H$. Reciprocally, if $V(t)$ is an invariant $n$-plane and $P(t)$ the projection operator into $V(t)$ then $V(t)$ is the eigenspace of the dynamical invariant $P(t)$ associated with the constant eigenvalue 1. To sum up, seeking invariant planes amounts to seeking dynamical invariants. Once such an operator $I(t)$ is found, the dynamics becomes orthogonally \enquote{block-diagonalized}. If it happens that $I$ is furthermore a genuine constant of motion, all the \enquote{blocks} are time-independent. The latter case reflects a common symmetry of all the instantaneous Hamiltonians $H(t)$ and is of great interest to reduce the dynamics. However, it is helpless to generate nontrivial geometry effects.

\section{An alternative derivation of the degeneracy conditions}\label{app:alternative}

Faced with the problem of finding the degeneracy conditions in subsection \ref{degeneracycondition}, a large number of students in physics start by focusing themselves on the algebraic side of the multiplicities. They transcribe the sought property in terms of the existence of a root of degree 2 admitted by the characteristic polynomial $P_{\text c}(\lambda)=\det(\lambda \openone- H)$, where $\lambda$ is the indeterminate. Using~\eqref{Hmatrix} and the traceless hypothesis, one has explicitly
\begin{align}
P_{\text c}(\lambda)=\det(\lambda \openone- H)=\lambda^3+p\,\lambda+q \label{Pcharexp}
\end{align}
with
\begin{align*}
p&=ab+bc+ca-\module{\alpha}^2-\module{\beta}^2-\module{\gamma}^2,\\
q&=a\module{\alpha}^2+b\module{\beta}^2+c\module{\gamma}^2-abc-2\Re(\alpha\beta\gamma).
\end{align*}
On the other hand, the characteristic polynomial must read
\begin{align*}
P_{\text c}(\lambda)=(\lambda-E_1)(\lambda-E_2)^2=(\lambda+2E_2)(\lambda-E_2)^2.
\end{align*}
Expanding this last expression and identifying the result with~\eqref{Pcharexp} yields the necessary and sufficient conditions $3E_2^2=-p\quad\text{and}\quad 2E_2^3=q$ from which is deduced the necessary condition $4p^3+27q^2=0$ on the entries of $\rep H$ (in particular, $p<0$). The latter is actually also sufficient since it amounts to the vanishing of the discriminant of $P_{\text c}$ (one can refer to the method of Tartaglia and Cardano \cite{Uspensky} to solve the roots of polynomials of the third degree). However, one is last with a quite complicated characterization and the students who have followed this route conclude that the problem is too difficult.

The problem can nevertheless be solved by using a polynomial method. Let me first recall that, according to the Hamilton-Cayley theorem \cite{Lang1}, $P_{\text c}$ is a polynomial verifying $P_{\text c}(H)=0$. Actually, the diagonalizability of $H$ makes this property obvious. To verify it, let $E_1,E_2$, and $E_3$ be the (not necessarily distinct) eigenvalues of $H$, so that
\begin{align*}
P_{\text c}(H)=(H-E_1 \openone)(H-E_2\openone)(H-E_3\openone).
\end{align*}
Obviously, the three factors in the right-hand side mutually commute.
Now, let $\{\ket{u_1},\ket{u_2},\ket{u_3}\}$ be an eigenbasis of $H$ such that $H\ket{u_i}=E_i\ket{u_i}$. Since $(H-E_i\openone)\ket{u_i}=0$, one has $P_{\text c}(H)\ket{u_i}=0$. The evaluations of $P_{\text c}(H)$ on the vectors of the eigenbasis being zero, one deduces the sought property $P_{\text c}(H)=0$. 

Amongst the set $\mathscr I$ of polynomials $P$ verifying $P(H)=0$, $P_{\text c}$ is not of the lowest degree if the spectrum of $H$ is degenerate. Indeed, if we suppose for example that $E_2=E_3$ then, clearly, the polynomial
\begin{align}
P_{\text m}(\lambda)=(\lambda-E_1)(\lambda-E_2)\label{Pmin}
\end{align}  
verifies $P_{\text m}(H)=0$ as well (it suffices to evaluate $P_{\text m}$ on the eigenbasis as above). Reciprocally, if there exist two values $E_1$ and $E_2$ such that~\eqref{Pmin} belongs to $\mathscr I$ then $E_1$ and $E_2$ form the set of eigenvalues of $H$. Indeed, any eigenvalue $\mu$ of $H$ brings an eigenvalue $P_{\text m}(\mu)$ of $P_{\text m}(H)$ and the assumption whereby $P_{\text m}(H)$ is the null operator imposes $P_{\text m}(\mu)=0$, i.e.\ the fact that $\mu$ is equal to either $E_1$ or $E_2$. Since by assumption $\rep H$ has two distinct eigenvalues, they are necessarily $E_1$ and $E_2$.

One deduces from the above considerations and the traceless hypothesis that $H$ admits a degenerate eigenvalue $E_2$ iff the polynomial
\begin{align*}
P_{\text m}(\lambda)=(\lambda+2E_2)(\lambda-E_2)
\end{align*}
belongs to $\mathscr I$. In this case, since $H$ is not proportional to the identity operator, $P_{\text m}$ is the monic polynomial of the lowest degree in $\mathscr I$; it is the so-called \emph{minimal polynomial} of $H$. Writing down the system of six equations corresponding to $P_{\text m}(\rep H)=0$, one easily verifies that they are equivalent to the system \eqref{system}. 

As a conclusion, let me mention that the minimal polynomial of a linear operator $H$ (in a vector space of finite dimension), whether diagonalizable or not, is a key concept of algebra which is closely related to ring theory \cite{Lang2}. It has been for example implicitly used in reference \onlinecite{LeoneMonjou} to determine the degeneracy conditions of a family of Hamiltonians in a four-state quantum system. 

\section{On hyperplanes}\label{hyperplanes}

Let $\mathcal H$ be a $d$-dimensional Hilbert space and $V$, $W$ be two distinct hyperplanes of $\mathcal H$. For obvious dimensional reasons, the dimension of the intersection $V\cap W$ is $d-2$. Hence, $\sigma_+=1$ is a $(d-2)$-fold singular value of the mutual projections $\Pi_{WV}$ and $\Pi_{VW}$. Besides $\sigma^2_+=1$, let $\sigma_-^2$ ($\sigma_-\geqslant 0$) be the other eigenvalue of $\Pi_{WV}\Pi_{VW}$ and $\Pi_{VW}\Pi_{WV}$. One has
\begin{align*}
\tr(P_WP_V)=\sigma^2_-+(d-2)\sigma^2_+=\sigma^2_-+d-2.
\end{align*}
On the other hand, one has
\begin{align*}
\tr(P_WP_V)&=\tr\big[(\openone-P_{W^\perp})(\openone-P_{V^\perp})\big]\\
&=\tr\big[\openone-P_{W^\perp}-P_{V^\perp}+P_{W^\perp}P_{V^\perp}\big]\\
&=d-2+\tr\big[P_{W^\perp}P_{V^\perp}\big].
\end{align*}
Hence, if $\ket{\tilde v}$ and $\ket{\tilde w}$ are unit vectors orthogonal to $V$ and $W$, respectively, one deduces
\begin{align*}
\sigma_-=\sqrt{\tr\big[P_{W^\perp}P_{V^\perp}\big]}=\module{\bk{\tilde w}{\tilde v}}.
\end{align*}
The principal angles between $V$ and $W$ are thus $\phi_+=\arccos(\sigma_+)=0$ and $\phi_-=\arccos(\sigma_-)=\arccos\module{\bk{\tilde w}{\tilde v}}$. Finally, the Fubini-Study distance between $V$ and $W$ is
\begin{align*}
d_{\text{FS}}(V,W)=\arccos(\sigma_+^{d-2}\sigma_-)=\phi_-=d_{\text{FS}}(V^\perp,W^\perp).
\end{align*}

\section{About the sum of three nonzero complex numbers}\label{app:complex}

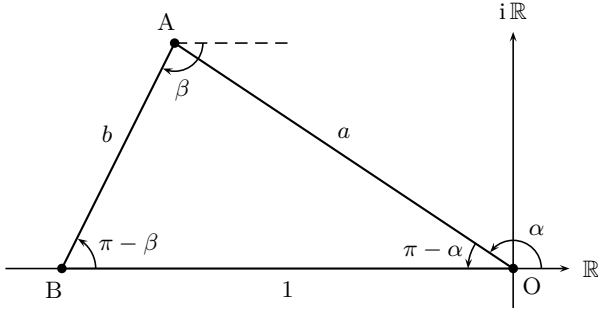
\begin{figure}
\centering
\psset{unit=.75cm}
\begin{pspicture*}(-9,-.7)(1.5,4.8)
\psline[linewidth=.6pt]{->}(0,0)(1,0)
\psline[linewidth=.6pt](-9,0)(-8,0)
\psline[linewidth=.6pt]{->}(0,-.7)(0,4.2)
\psline[linestyle=dashed,linewidth=.6pt](-6,4)(-4,4)
\psline(0,0)(-6,4)(-8,0)(0,0)
\psdots(-6.,4.)(-8,0)(0,0)
\uput[-45](0,0){O}
\uput[0](1,0){$\mathbb R$}
\uput[90](0,4.2){$\imag\mathbb R$}
\psarc[linewidth=.6pt]{->}(0,0){.5}{0}{145}
\uput{.65}[60](0,0){$\alpha$}
\rput(-3,2.4){$a$}
\rput(-7.2,2.4){$b$}
\psarc[linewidth=.6pt]{->}(0,0){.8}{146}{180}
\uput{.9}[162](0,0){$\pi-\alpha$}
\uput[-90](-4,0){1}
\psarc[linewidth=.6pt]{<-}(-6,4){.5}{245}{0}
\uput{.65}[-80](-6,4){$\beta$}
\uput[110](-6,4){A}
\uput[-110](-8,0){B}
\psarc[linewidth=.6pt]{->}(-8,0){.6}{0}{62}
\uput{.7}[27](-8,0){$\pi-\beta$}
 \end{pspicture*}
 \caption{Seeking the conditions on $a$, $b$, $\alpha$ and $\beta$ for which the equality \eqref{sum-1} is verified amounts to seeking the points A in the complex plane located at a distance $a$ from the origin O and $b$ from the point B$(-1,0)$. Such a point A is characterized by the values of $\alpha$ and $\beta$ and the inequalities \eqref{inegalite} are the \textit{condicio sine qua non} for its existence. When they are strict, A is not on the real axis and there are two solutions symmetric with respect to that axis. They degenerate into a single solution on the real axis if either $\module{a-b}=1$ (A outside the segment OB) or $a+b=1$ (A inside OB). The relation \eqref{condangles} expresses the fact that if \enquote{OA goes upwards} then \enquote{AB goes downwards} and vice versa. Finally, the law of cosines (also known in France as Al-Kashi theorem) applied to the angles $\angle\text{AOB}$ and $\angle\text{OBA}$ gives formulas \eqref{cosinus}.}\label{alkashi}
\end{figure}

In this appendix, we seek the conditions for which the sum of three nonzero complex numbers vanishes. Dividing the sum by one of the terms and using the exponential form, it amounts to seeking the conditions on the two positive numbers $a$, $b$ and the two angles $\alpha$, $\beta$, for which one has
\begin{align}
a\expo^{\imag\alpha}+b\expo^{\imag\beta}+1=0.\label{sum-1}
\end{align}
Rewriting this sum as $a\expo^{\imag\alpha}+b\expo^{\imag\beta}=-1$, and applying the triangle inequalities, one obtains the necessary condition
\begin{align}
\module{a-b} \leqslant 1\leqslant a+b \label{inegalite}
\end{align}
on $a$ and $b$. Then, taking the imaginary parts of \eqref{sum-1}, one deduces the necessary condition
\begin{align}
\sin\alpha\sin\beta\leqslant 0\label{condangles}
\end{align}
on the angles. Now, multiplying \eqref{sum-1} by $\expo^{-\imag\alpha}$, rewriting it as $a+\expo^{-\imag\alpha}=-b\expo^{\imag(\beta-\alpha)}$, and taking the modulus of the two sides, give $\cos\alpha$ as a function of $a$ and $b$. A similar manipulation allows to isolate $\cos\beta$ and one has
\begin{align}
\cos\alpha=\frac{b^2-a^2-1}{2a}\;,\quad \cos\beta=\frac{a^2-b^2-1}{2b}\,.\label{cosinus}
\end{align}
It is easily seen that \eqref{cosinus} makes sense iff \eqref{inegalite} is verified and that the relations \eqref{inegalite}, \eqref{condangles} and \eqref{cosinus} are also sufficient for having \eqref{sum-1}. If the inequalities are strict in \eqref{inegalite}, the cosines belong to $(-1,1)$ and there are (mod $2\pi$) exactly two couples $(\alpha,\beta)$ verifying \eqref{condangles} and \eqref{cosinus}. In the case where $\lvert a-b\rvert=1$ or $a+b=1$, these two couples degenerate into a single one. The results of this appendix could also have been derived geometrically (see figure \ref{alkashi}).

As an application of this basic study, one can determine the conditions on the parameters for which two degenerate eigenspaces $V(x)$ and $V(x')$ are not anti-orthogonal in the context of section \ref{sec:3}. Such a non anti-orthogonality amounts to the orthogonality of $\ket{1(x)}$ and $\ket{1(x')}$. However, one has
\begin{align*}
\bk{1(x)}{1(x')}=\frac{rr'ss'}{\mathcal N_1(r,s)\mathcal N_1(r',s')}\bigg(\frac{\expo^{\imag\Delta\varphi}}{rr'}+\frac{\expo^{\imag\Delta\vartheta}}{ss'}+1\bigg),
\end{align*}
with $\Delta\varphi=\varphi'-\varphi$ and $\Delta\vartheta=\vartheta'-\vartheta$. Therefore, the condition of non anti-orthogonality of $V(x)$ and $V(x')$ takes the form \eqref{sum-1} with $a=(rr')^{-1}$, $b=(ss')^{-1}$, $\alpha=\Delta\varphi$ and $\beta=\Delta\vartheta$.

\section{The functions $\boldsymbol{C(t)}$ and $\boldsymbol{S(t)}$}\label{app:CS}

The functions $C(t)$ and $S(t)$ in formula \eqref{formule} are

\begin{widetext}
\begin{align*}
C(t)={}&r^2\cos\bigg(\frac{\omega_\vartheta +\omega_\varphi}{2}\,t+\xi(t)\bigg)-2r^2\cos\bigg(\frac{\omega_\vartheta -\omega_\varphi}{2}\,t-\xi(t)\bigg)-\cos\bigg(\frac{\omega_\vartheta +\omega_\varphi}{2}\,t-\xi(t)\bigg)\\
&+(1+r^2)\cos\bigg(\frac{3\omega_\vartheta -\omega_\varphi}{2}\,t-\xi(t)\bigg)+\big[\text{ exchange } (r,\omega_\vartheta)\leftrightarrow(s,\omega_\varphi) \text{ in all the terms }\big],\\
S(t)={}&s^2(\mathcal N_1^2-r^4)\omega_\vartheta\sin\bigg(\frac{\omega_\vartheta +\omega_\varphi}{2}\,t+\xi(t)\bigg)-2r^2\big[(\mathcal N_1^2-r^2)\omega_\vartheta-(\mathcal N_1^2+s^2)\omega_\varphi\big]\sin\bigg(\frac{\omega_\vartheta-\omega_\varphi}{2}\,t-\xi(t)\bigg)\\
&-2(\mathcal N_1^2-r^2)\omega_\vartheta\sin\bigg(\frac{\omega_\vartheta +\omega_\varphi}{2}\,t-\xi(t)\bigg)+\big[s^2(1+r^2)^2\omega_\vartheta-r^2(\mathcal N_1^2-1)\omega_\varphi\big]\sin\bigg(\frac{3\omega_\vartheta -\omega_\varphi}{2}\,t-\xi(t)\bigg)\\
&+\big[\text{ exchange } (r,\omega_\vartheta)\leftrightarrow(s,\omega_\varphi) \text{ in all the terms }\big].
\end{align*}
\end{widetext}

\section{Numerical holonomies}\label{app:num}

\lstset{frameshape={RYRYNYYYY}{yny}{yny}{RYRYNYYYY},language=Octave,showstringspaces=false,tabsize=4,
morekeywords={arg,i},basicstyle=\footnotesize,numberstyle=\tiny,numbers=left,framexleftmargin=20pt,xleftmargin=2em}
\begin{figure*}
\begin{lstlisting}
clear ;
datafile = fopen("Phases2pi-4pi.txt","w")
			
# Determination of the degenerate eigenspace as a function of r, s, theta and phi :

function [deg1,deg2] = deg(r,s,theta,phi) 
	deg1 = [ r*exp(i*phi) ; 0 ; -1 ] / sqrt(1+r^2) ;
	deg2 = [ r*exp(i*phi) ; -(1+r^2)*s*exp(i*theta) ; r^2 ] / sqrt((1+r^2)*(r^2+s^2+(r*s)^2)) ;	
endfunction

# Definition of the constants :

r = 3 ; s = 2 ; wtheta = 2*pi ; wphi = -4*pi ; n_iter = 200 ;

# Initialization :

[deg1,deg2] = deg(r,s,0,0) ; degi1 = deg1 ; degi2 = deg2 ; S = eye(2) ; 
fprintf(datafile,"%d, %d, %d\n",0,0,0) ;

# Computation of the (intermediate) holonomies :

for j = 1 : n_iter
	t = j/n_iter ; theta = wtheta*t ; phi = wphi*t ;
	degb1 = deg1 ; degb2 = deg2 ; [deg1,deg2] = deg(r,s,theta,phi) ;
	S = [ dot(deg1,degb1) , dot(deg1,degb2) ; dot(deg2,degb1) , dot(deg2,degb2) ] * S ;
	B = [ dot(degi1,deg1) , dot(degi1,deg2) ; dot(degi2,deg1) , dot(degi2,deg2) ] * S ;
	phases = sort(arg(eig(inv(sqrtm(B*B'))*B))) ;
    fprintf(datafile,"%d, %d, %d\n",t,phases)
endfor

fclose(datafile)
\end{lstlisting}
\caption{The content of the \texttt{Octave} script file used to produce the data plotted in figure \hyperref[plots]{\ref*{plots}(a)} with the terminal \texttt{epslatex} of \texttt{gnuplot}. The values of $r$ and $s$ are chosen so that one knows a priori that the intermediate holonomies are all non-partial (indeed, $r^{-2}+s^{-2}<1$). Hence, no pseudoinversion is needed (in line 27, the inverse function \texttt{inv} is used instead of the pseudoinverse \texttt{pinv}). Furthermore, as can be seen in figure \hyperref[plots]{\ref*{plots}(a)}, the eigenvalues of $\Gamma_j$ never cross the negative real axis: sorting the principal value of their argument suffices to obtain continuous plots of the phases on $\lambda\in[0,1]$. Otherwise, if such crossings occur and if one wants continuous plots, one needs to (i) sort the eigenvalues in such a way that the eigenvalues are continuous and (ii) unwrap their phases, but it is an elementary exercise.}\label{Octave}
\end{figure*}

Assume that we have a path $\mathcal C\colon\lambda'\mapsto \text{V}(\lambda')$ in the Grassmann manifold $\text{Gr}(n,\mathcal H)$. Our aim is to give a systematic scheme to compute the holonomies along the portions $\mathcal C_\lambda$ traced out from $\lambda'=0$ to $\lambda'=\lambda$ as a function of $\lambda\in[0,1]$. Let $N$ be some positive integer, $\lambda_j=\tfrac{j}{N}$ ($0\leqslant j\leqslant N$) be a homogeneous discretization of $[0,1]$, $\mathcal C_j$ be the portion $\mathcal C_{\lambda_j}$, $\mathcal V_j$ be an orthonormal basis of $V_j=V(\lambda_j)$, and $\Pi_{k,j}$ be the projection $V_j\to V_k$ represented by the overlap matrix $\rep S_{k,j}=\rep S(\mathcal V_k,\mathcal V_j)$ in the bases $\mathcal V_j$ and $\mathcal V_k$. In the limit $N\to\infty$, each product $\Pi_j=\Pi_{j,j-1}\dots\Pi_{2,1}\Pi_{1,0}$ tends to the transporter $\Gamma[\mathcal C_j]$. Its matrix in the bases $\mathcal V_0$ and $\mathcal V_j$ is the product $\rep S_j=\rep S_{j,j-1}\dots\rep S_{2,1}\rep S_{1,0}$. Then, the (possibly partial) holonomy along $\mathcal C_j$ is in this limit
\begin{align}
\Gamma_j=\Gamma_{V_0V_j}\Gamma[\mathcal C_j]={\sqrt{\Pi_{0,j}\Pi_{0,j}^\dag}\,}^+\Pi_{0,j}\Pi_j\,.\label{forme1}
\end{align}
However, since $\Pi_j$ is unitary, proposition \ref{prop2} of appendix \ref{PD} guarantees that $\Gamma_j$ is also given by
\begin{align}
\Gamma_j={\sqrt{(\Pi_{0,j}\Pi_j)(\Pi_{0,j}\Pi_j)^\dag}\,}^+\Pi_{0,j}\Pi_j\,.\label{forme2}
\end{align}
For computations, $N$ is finite and $\Pi_j$ is less and less unitary as $j$ grows (its operator norm decreases with $j$). Hence, formula \eqref{forme2} is more adapted than \eqref{forme1} since it compensates that loss and forces $\rep\Gamma_j$ to be unitary (one might also force $\Pi_j$ to be so). Expression \eqref{forme2} in matrix form in the basis $\mathcal V_0$ is
\begin{align*}
\rep\Gamma_j={\sqrt{\rep B_j\rep B_j^\dag}\,}^+\rep B_j\,,
\end{align*}   
where $\rep B_j$ is the generalized Bargmann invariant $\rep S_{0,j}\rep S_j$. The square root and the pseudoinversion can be computed numerically thanks to pre-existing routines (e.g. by the functions \texttt{sqrtm} and \texttt{pinv} in \texttt{Octave}). Otherwise, one can proceed by performing a unitary diagonalization of $\rep B_j\rep B_j^\dag$. If $r_j$ is the rank of $\Pi_{0,j}$ and if one sorts the eingenvalues of $\rep B_j\rep B_j^\dag$ in the decreasing order then one obtains a decomposition of the form $\rep B_j\rep B_j^\dag=\rep U_j\rep D_j\rep U_j^\dag$ where $\rep U_j$ is a unitary matrix and $\rep D_j=\diag(\sigma_1^2,\dots,\sigma_{r_j}^2,0,\dots,0)$, with $\sigma_i>0$. It follows that
\begin{align*}
{\sqrt{\rep B_j\rep B_j^\dag}\,}^+=\rep U_j\diag(\sigma_i^{-1},\dots,\sigma_{r_j}^{-1},0,\dots,0)\rep U_j^\dag.
\end{align*}
If $r_j=n$, the holonomy $\Gamma_j$ is non-partial. Hence, it is a unitary automorphism of $V_0$ and we can diagonalize it to extract the geometric phases. In figure \ref{Octave} is given the minimal algorithm written in \texttt{Octave} that was used to produce the plots of figure \ref{plots}.

\section{On some adapted bases}\label{app:berry}

In this last appendix, we establish the conditions on $\Delta\vartheta$ and $\Delta\varphi$ for which one of the phases is identically zero along the toroidal helices considered in subsection \ref{abelianholo}. It is the case iff there exists a vector of the degenerate eigenspace which remains invariant. This goes for $\Delta\varphi=0$ since $\ket{2_1(x)}$ does not depend on $\vartheta$ in the choice of orthonormal frame of reference \eqref{basis}. When $\Delta\vartheta=0$, it suffices to perform the change of frame
\begin{align*}
\ket{2_1(x)}&\longrightarrow\frac{s\expo^{\imag\vartheta}\ket{e_2}-\ket{e_3}}{\sqrt{1+s^2}}\,,\\
\ket{2_2(x)}&\longrightarrow\frac{r(1+s^2)\expo^{\imag\varphi}\ket{e_1}-s\expo^{\imag\vartheta}\ket{e_2}-s^2\ket{e_3}}{\mathcal N_1(r,s)\sqrt{1+s^2}}\,.
\end{align*}  
For any other combination $\Delta\vartheta+q\Delta\varphi=0$ ($q\in\mathbb R-\{0\}$), this will again be possible iff there exists a vector field  belonging to the degenerate eigenspaces which depends on $\sigma=\vartheta+q\varphi$ but not on $\tau=\vartheta-q\varphi$. However, the degenerate eigenspace is spanned by $\ket{v(x)}=r\expo^{\imag\varphi}\ket{e_1}-\ket{e_3}$ and $\ket{w(x)}=s\expo^{\imag\vartheta}\ket{e_2}-\ket{e_3}$. Hence, a linear combination
\begin{align*}
\ket{u(x)}&=\xi(x)\ket{v(x)}+\eta(x)\ket{w(x)}\\
&=\xi(x) r\expo^{\imag\varphi}\ket{e_1}+\eta(x) s\expo^{\imag\vartheta}\ket{e_2}-\big[\xi(x)+\eta(x)\big]\ket{e_3}
\end{align*}
is independent of $\tau$ iff
\begin{align*}
\partial_\tau\xi-\imag\,\frac{\xi}{2q}=\partial_\tau\eta+\imag\,\frac{\eta}{2}=\partial_\tau(\xi+\eta)=0,
\end{align*}
i.e.\ iff there exists two complex coefficients $C(r,s,\sigma)$ and $D(r,s,\sigma)$ such that
\begin{align*}
\xi=C\expo^{\imag\frac{\tau}{2q}}\;,\;\;\eta=D\expo^{-\imag\frac{\tau}{2}}\;,\;\;\frac{C}{q}\,\expo^{\imag\frac{\tau}{2q}}-D\,\expo^{-\imag\frac{\tau}{2}}=0.
\end{align*}
The last equality imposes $q=-1$ and it suffices to take opposite constants for $C$ and $D$. Consequently, the intermediate holonomies admit a form \eqref{transportphinul}-\eqref{gammaphinul} iff one has $\Delta\varphi=0$ or $\Delta\vartheta=0$ or $\Delta\vartheta-\Delta\varphi=0$. In the last case, an adapted orthonormal frame of reference is obtained through the replacement
\begin{align*}
\ket{2_1(x)}&\longrightarrow\frac{r\ket{e_1}-s\expo^{\imag(\vartheta-\varphi)}\ket{e_2}}{\sqrt{r^2+s^2}}\,,\\
\ket{2_2(x)}&\longrightarrow\frac{rs^2\expo^{\imag\varphi}\ket{e_1}+r^2s\expo^{\imag\vartheta}\ket{e_2}-(r^2+s^2)\ket{e_3}}{\mathcal N_1(r,s)\sqrt{r^2+s^2}}\,.
\end{align*} 

\section{On the eigenvectors of $H'$}\label{app:colonnes}

Let $\varepsilon$ be a nonzero eigenvalue \eqref{exactev} of the operator $H'$ defined in subsection \ref{beyond}, and let $A=H'-\varepsilon\openone$. In the basis $\mathscr B$, the matrix of $A$ is
\begin{align*}
\mat(A;\mathscr B,\mathscr B)=\begin{pmatrix}
\frac{s}{r}+\omega_\varphi-\varepsilon & 1 & s \\
1 & \frac{r}{s}+\omega_\vartheta-\varepsilon & r \\
s & r & rs-\varepsilon
\end{pmatrix}.
\end{align*}
Determining an eigenvector of $H'$ associated with $\varepsilon$ by substituting $\varepsilon$ with its expression is not a good strategy. Instead, let us operate on the columns of the above matrix. Fixing the second column, one can nullify the upper corners to obtain a matrix of the form
\begin{align*}
\mat(A;\mathscr B,\mathscr B')=\begin{pmatrix}
0 & 1 & 0 \\
Q(\varepsilon) & \frac{r}{s}+\omega_\vartheta-\varepsilon & s(\varepsilon-\omega_\vartheta) \\
r(\varepsilon-\omega_\varphi) & r& -\varepsilon
\end{pmatrix}
\end{align*}
where the source basis is now $\mathscr B'=(\ket{e'_1},\ket{e_2},\ket{e'_3})$, with
\begin{align*}
\ket{e'_1}&=\ket{e_1}+\Big(\varepsilon-\frac{s}{r}-\omega_\varphi\Big)\ket{e_2}\;,\quad\ket{e'_3}=\ket{e_3}-s\ket{e_2}.
\end{align*}
Since $\varepsilon\ne 0$, another column operation reduces the matrix of $A$ to the form
\begin{align*}
\mat(A;\mathscr B,\mathscr B'')=\begin{pmatrix}
0 & 1 & 0 \\
Q'(\varepsilon) & \frac{r}{s}+\omega_\vartheta-\varepsilon & s(\varepsilon-\omega_\vartheta) \\
0 & r & -\varepsilon
\end{pmatrix}
\end{align*}
where the source basis is $\mathscr B''=(\ket{e''_1},\ket{e'_2},\ket{e'_3})$ with
\begin{align}
\ket{e''_1}={}&\varepsilon\ket{e'_1}+r(\varepsilon-\omega_\varphi)\ket{e'_3}\notag\\
={}&\varepsilon\ket{e_1}+\Big[\varepsilon^2-\Big(rs+\frac{s}{r}+\omega_\varphi\Big)\varepsilon+rs\omega_\varphi\Big]\ket{e_2}\notag\\&+r(\varepsilon-\omega_\varphi)\ket{e_3}.\label{nonzeroeig}
\end{align}   
Since $\varepsilon$ is an eigenvalue of $H'$, the operator $A$ is singular. Hence, one has necessarily $Q'(\varepsilon)=0$. Consequently, $A\ket{e''_1}=0$ and $\ket{\varepsilon}=\ket{e''_1}$ is an eigenvector of $H'$ associated with $\varepsilon$. Its squared norm is a polynomial expression of the fourth order in $\varepsilon$ whose coefficients depend on $r,s,\omega_\varphi$. It can be reduced to a second-order one of the form $f(r,s,\omega_\varphi)\varepsilon^2+g(r,s,\omega_\varphi)\varepsilon$ by using the fact that $\varepsilon$ is a root of the characteristic polynomial of $H'$. Dividing $\ket{\varepsilon}$ by the square root of that expression, one obtains a unit vector. 

Obviously, the route followed to derive an eigenvector of $H'$ associated with $\varepsilon$ is not unique. One could have operated on the columns in a different way and the resulting eigenvector would have had a different expression. For example, since the matrix of $A$ in $\mathscr B$ is invariant under the simultaneous exchanges $r\leftrightarrow s$, $\omega_\vartheta\leftrightarrow\omega_\varphi$, $\ket{e_1}\leftrightarrow\ket{e_2}$, these exchanges realized in the last member of \eqref{nonzeroeig} leads to another possible choice of eigenvector (whose squared norm is $f(s,r,\omega_\vartheta)\varepsilon^2+g(s,r,\omega_\vartheta)\varepsilon$). .

\bibliography{Holonomy3D}

\end{document}